\documentclass[11pt]{article}

\usepackage[a4paper,margin=0.9in]{geometry}
\usepackage[authoryear,round]{natbib}
\usepackage{amsmath,amssymb,amsbsy,amsfonts,amsthm,mathtools,mathrsfs}
\usepackage[T1]{fontenc}
\usepackage[utf8]{inputenc}
\usepackage{microtype}
\usepackage{graphicx}
\usepackage{booktabs,multirow,array,calc}
\usepackage{caption}
\usepackage{subcaption}
\usepackage{algorithm}
\usepackage{algorithmic}
\usepackage[hidelinks]{hyperref}
\usepackage{xurl}

\hypersetup{
  pdftitle={Scalable Estimation and Inference for Varying Coefficient Expectile Regression with Streaming Data},
  pdfauthor={Haojie Dong; Xinyuan Song; Xinghui Wang},
  pdfkeywords={expectile regression; functional stochastic gradient descent; multiplier bootstrap; online learning; reproducing kernel Hilbert spaces}
}

\allowdisplaybreaks

\ifPDFTeX
\DeclareUnicodeCharacter{2011}{-}
\DeclareUnicodeCharacter{2013}{--}
\DeclareUnicodeCharacter{2014}{---}
\DeclareUnicodeCharacter{2018}{`}
\DeclareUnicodeCharacter{2019}{'}
\DeclareUnicodeCharacter{201C}{``}
\DeclareUnicodeCharacter{201D}{''}
\fi

\newtheorem{theorem}{Theorem}
\newtheorem{lemma}[theorem]{Lemma}

\newtheorem{assumption}{Assumption}
\theoremstyle{definition}

\theoremstyle{remark}
\newtheorem{remark}{Remark}

\title{Scalable Estimation and Inference for Varying Coefficient Expectile Regression with Streaming Data}
\author{
Haojie Dong\textsuperscript{1},
Xinyuan Song\textsuperscript{2}, and
Xinghui Wang\textsuperscript{1}\thanks{Corresponding author: \texttt{wangxinghui@ahu.edu.cn}.}\\[1ex]
\small \textsuperscript{1}School of Big Data and Statistics, Anhui University,
Hefei, Anhui, China\\
\small \textsuperscript{2}Department of Statistics and Data Science,
The Chinese University of Hong Kong, Hong Kong, China
}
\date{}

\begin{document}
\maketitle

\begin{abstract}
Nonparametric expectile regression methods face new challenges from streaming data, where retaining the full dataset for batch processing becomes increasingly impractical. In this paper, we develop a scalable nonparametric expectile regression framework for varying coefficient models. We estimate the coefficient functions online via functional stochastic gradient descent with Polyak--Ruppert averaging in a product reproducing kernel Hilbert space, requiring only gradient information and one pass over the data. For this online estimator, we establish non-asymptotic error bounds, prove its optimal nonparametric convergence rate under the proposed step-size schemes, and derive its Berry--Esseen bound for pointwise asymptotic normality. To perform online inference for the varying coefficient expectile regression models, we propose a multiplier bootstrap procedure that perturbs the stochastic gradients with independent random weights and recursively constructs bootstrap trajectories. We establish the bootstrap consistency of the online inference procedure, yielding asymptotically valid pointwise confidence intervals. Simulation studies and a real data application confirm the performance of the proposed methods.
\end{abstract}

\vspace{0.5em}

\section{Introduction}

Expectile regression (ER, \cite{newey1987asymmetric}) is a powerful tool for revealing complex dependence structures between responses and covariates, and an important extension of traditional regression frameworks. Classical least squares focuses solely on the conditional mean, a measure of central tendency that often masks underlying heterogeneity and obscures tail characteristics. Although the seminal quantile regression (QR, \cite{koenker1978regression}) overcomes this limitation by capturing conditional quantiles, the non-differentiability of its asymmetric absolute loss introduces inherent optimization challenges. Furthermore, QR is relatively inefficient for near-Gaussian error distributions. In contrast, ER minimizes an asymmetric squared loss function:
\begin{equation}
\Psi_\tau (u) = |\tau - \mathbb{I}(u < 0)|\cdot u^2,
\label{loss}
\end{equation}
where $\mathbb{I}(\cdot)$ is the indicator function. The asymmetric loss \eqref{loss} confers several advantages: it produces a family of expectiles that summarizes the conditional distribution; it is more efficient than QR for near-Gaussian errors; and its differentiability avoids the optimization difficulties. Analogous to how QR summarizes the conditional distribution via quantiles, ER admits a natural interpretation through ``expectiles.'' For further discussion, we refer readers to \cite{efron1991regression}, \cite{de2009quantiles}, and \cite{daouia2020tail}.

Existing ER research encompasses parametric frameworks (e.g., \cite{kuan2009assessing}, \cite{gu2016high}, and \cite{chen2020efficient}) for interpretable risk modeling, and nonparametric approaches (e.g., \cite{xie2014varying}, \cite{gu2019ae}, and \cite{zhao2026minimax}) designed to capture complex, nonlinear tail dependencies. However, these studies remain largely confined to offline settings that assume full data access and sufficient computational resources. In modern data environments, the high-speed sequential arrival of data streams, such as those from sensor networks (\cite{gama2013evaluating} and \cite{wang2017distributed}) and financial transaction logs (\cite{glaeser2017extrapolative} and \cite{wang2022renewable}), poses significant challenges to offline methods. First, continuous and potentially infinite data streams often make storing all historical samples infeasible due to hardware memory limits. Second, high-speed data arrival imposes stringent timeliness requirements on model updates. To address these challenges, online learning (e.g., \cite{robbins1951stochastic} and \cite{duchi2011adaptive}) offers a natural solution. By processing data streams in one pass, it operates under resource constraints and enables low-latency updates.

Despite the advantages of online learning for streaming data, existing online methods are limited to linear ER (e.g., \cite{pan2024renewable} and \cite{cao2026renewable}). To the best of our knowledge, online nonparametric ER remains an open challenge. To this end, we develop an online estimation and inference framework for nonparametric ER within reproducing kernel Hilbert spaces (RKHS, \cite{aronszajn1950theory}). To balance flexibility with interpretability, we adopt a varying coefficient structure. Let $Y \in \mathbb{R}$ be a response variable, $\mathbf{Z} = (Z_0, Z_1, \ldots, Z_p)^{\top} \in \mathbb{R}^{p+1}$ be covariates with $Z_0 \equiv 1$ and fixed $p$, and $X \in \mathcal{X}$ be a random-design variable. For a fixed expectile level $\tau \in (0,1)$, we consider the following varying coefficient ER model:
\begin{equation}
e_{Y\mid (X,\mathbf{Z})}(\tau) = \mathbf{Z}^{\top} \boldsymbol{\beta}^*(X,\tau) = \beta_0^*(X,\tau) + \sum_{k=1}^p Z_{k} \beta_k^*(X,\tau),
\label{model}
\end{equation}
where $\boldsymbol{\beta}^*(\cdot,\tau) = (\beta_0^*(\cdot,\tau), \ldots, \beta_p^*(\cdot,\tau))^{\top}$ is the unknown coefficient function vector over $\mathcal{X}$. For notational simplicity, we take $\mathcal{X} = [0,1]^d$, the unit cube in $\mathbb{R}^d$. Our goal is to perform online estimation and inference of $\boldsymbol{\beta}^*$ in a product RKHS.

Online nonparametric estimation has a rich methodological literature, including local polynomial estimation (\cite{kong2019efficiency}), kernel smoothing (\cite{huang2013recursive}), dynamic penalized splines (\cite{xue2022dynamic}), RKHS-based projection estimation (\cite{zhang2023online}), local linear approximation (\cite{fang2023online}), and penalized orthogonal basis expansion (\cite{quan2024optimal}). However, these online estimation methods, designed for least squares, rely on closed-form recursive updates that the asymmetric ER loss \eqref{loss} generally precludes when $\tau\neq 0.5$. Although \cite{yang2024online} proposed an online nonparametric optimization method, it requires both gradient and Hessian information; convolution-based smoothing (\cite{fernandes2021smoothing}) could make the ER objective twice differentiable to fit this framework, but it introduces a hard-to-tune bandwidth and unnecessarily complicates online optimization. Fundamentally, the ER loss \eqref{loss} is first-order smooth, allowing parameter updates using only gradient information.

Therefore, stochastic gradient descent (SGD, \cite{robbins1951stochastic}) provides a highly scalable alternative. Along this line, \cite{tarres2014online}, \cite{dieuleveut2016nonparametric}, and \cite{liu2026scalable} explored functional stochastic gradient descent (FSGD) algorithms in RKHS for online nonparametric least squares. In this paper, we propose a product RKHS-based FSGD estimation method for coefficient functions under asymmetric least squares. Specifically, we derive the functional gradient under the ER loss in the product RKHS, then construct the FSGD estimator via Polyak--Ruppert averaging (\cite{ruppert1988efficient} and \cite{polyak1992acceleration}).

To achieve online inference, we employ resampling method. The bootstrap is widely used to estimate the sampling distribution of a statistic for uncertainty quantification; see \cite{tibshirani1993introduction}. However, traditional resampling-based bootstrap requires storing all historical data and repeatedly scanning a fixed dataset, which contradicts the low-storage and one-pass goals of online learning. To address this, \cite{fang2018online} proposed the online multiplier bootstrap for linear models by injecting multiplicative perturbation into the stochastic gradient to mimic the random fluctuation of online estimates. \cite{xie2026scalable} extended this method to functional linear models, and \cite{liu2026scalable} further generalized it to nonparametric least squares. Inspired by these works, we extend the multiplier bootstrap to online varying coefficient ER. We simultaneously maintain a set of perturbed FSGD trajectories by independent random weights. These bootstrap estimators are updated recursively and provide an empirical approximation to the sampling distribution of the original estimator, requiring only one pass over the data. We prove that the conditional distribution of the normalized bootstrap estimator captures the random fluctuation of the original estimator around the true value. Further, we construct valid pointwise confidence intervals (PCIs) for the true coefficient function.

We summarize our main contributions as follows.
\begin{itemize}
    \item \textbf{Online estimation in product RKHS.} We propose an online nonparametric ER framework for varying coefficient models. By embedding the coefficient functions in a product RKHS and establishing FSGD with Polyak--Ruppert averaging, the method processes streaming data in one pass and requires only the first-order gradient of the asymmetric squared loss.

    \item \textbf{Non-asymptotic analysis and pointwise asymptotic normality.} We conduct a rigorous non-asymptotic analysis and prove that the FSGD estimator attains the optimal nonparametric convergence rate under the proposed step-size schemes. Moreover, we establish its Berry--Esseen bound for pointwise asymptotic normality.

    \item \textbf{Online multiplier bootstrap inference.} We extend the multiplier bootstrap procedure to varying coefficient ER model. By recursively updating perturbed FSGD trajectories with random weights and updating them alongside the original estimator, these trajectories approximate its sampling distribution. We prove bootstrap consistency, ensuring that the constructed PCIs are asymptotically valid.
\end{itemize}

The remainder of this paper is organized as follows. Section \ref{Sec2} develops the online FSGD estimator and establishes its theoretical guarantees. Section \ref{Sec3} constructs the multiplier bootstrap procedure for online inference. Section \ref{Sec4} reports simulation studies that evaluate the finite-sample performance of the proposed methods and provides implementation details. Section \ref{Sec5} presents an application to cryptocurrency data from Binance, and Section \ref{Sec6} concludes the paper.

\section{Online Estimation via FSGD in RKHS}\label{Sec2}

Throughout this paper, we consider the varying coefficient ER model \eqref{model} at a fixed expectile level $\tau\in(0,1)$, so we omit $\tau$ in some subsequent notation. For any function vector $\boldsymbol{f}=(f_0,\ldots,f_p)^{\top}$ with $f_k$ over $\mathcal{X}$, denote $\Vert \boldsymbol{f}\Vert_{\infty}=\max_{0\leq k\leq p}\Vert f_k\Vert_{\infty}=\max_{0\leq k\leq p}\sup_{x\in\mathcal{X}}|f_k(x)|$ as the supremum norm. For any two sequences $a_n$ and $b_n$ depending on $n$, denote $a_n \asymp b_n$ if there exist positive constants $c$ and $C$ such that $c|a_n|\leq |b_n|\leq C|a_n|$. For any matrix $\mathbf{A}$, let $\lambda_{\min}(\mathbf{A})$ and $\lambda_{\max}(\mathbf{A})$ be the minimum and maximum eigenvalues of $\mathbf{A}$, respectively.

\subsection{Preliminaries on RKHS}

In this subsection, we briefly review background on RKHS. Let $F_X$ denote the marginal distribution of covariate $X\in\mathcal{X}$, and $L_2(F_X)$ is the space of all square-integrable functions with respect to $F_X$ over $\mathcal{X}$. Denote by $\|\cdot\|_{L_2}$ the $L_2$-norm of the space $L_2(F_X)$ induced by the inner product $\left\langle f, g \right\rangle_{L_2} = \int_{\mathcal{X}} f(x)g(x) dF_X(x)$. An RKHS is a Hilbert space $\mathcal{H}\subset L_2(F_X)$ of functions defined on a set $\mathcal{X}$ and equipped with an inner product $\left\langle \cdot, \cdot \right\rangle_{\mathcal{H}}$. Each $\mathcal{H}$ is uniquely associated with a reproducing kernel function $K(\cdot, \cdot): \mathcal{X} \times \mathcal{X} \to \mathbb{R}$.

\begin{assumption}\label{A1}
$\mathcal{X}$ is a compact topological space and $\mathcal{H}$ is an RKHS associated with a positive definite, symmetric, and continuous kernel $K$ on the set $\mathcal{X}$.
\end{assumption}

The RKHS satisfies the following properties:
\begin{itemize}
\item[(1)] For all $x \in \mathcal{X}$, $K(x, \cdot) : t \mapsto K(x, t) \in \mathcal{H}$.
\item[(2)] $(\mathcal{H}, \left\langle \cdot, \cdot \right\rangle_{\mathcal{H}})$ is a separable Hilbert space of functions.
\item[(3)] For all $f \in \mathcal{H}$ and $x \in \mathcal{X}$, the reproducing property holds: $f(x) = \left\langle f, K(x, \cdot) \right\rangle_{\mathcal{H}}$.
\end{itemize}
From the above facts, we can show that for any $x \in \mathcal{X}$, the evaluation functional at $x$ defined by $L_x(f) = f(x)$ is a continuous linear functional on the RKHS. Under Assumption \ref{A1}, Mercer’s theorem states that the $K$ has the following spectral decomposition:
\begin{equation}
K(x, x') = \sum_{m=1}^\infty \mu_m \phi_m(x) \phi_m(x'), \quad \forall x, x' \in \mathcal{X},
\label{mercer}
\end{equation}
where $\mu_1 \geq \mu_2 \geq \cdots \geq 0$ is the sequence of eigenvalues, and $\{\phi_m\}_{m=1}^\infty$ are the corresponding eigenfunctions forming an orthonormal basis in $L_2(F_X)$. We next introduce some technical assumptions regarding the spectral decomposition \eqref{mercer}.

\begin{assumption}\label{A2}
The eigenfunctions $\{ \phi_m \}_{m=1}^{\infty}$ of $K$ are uniformly bounded on $\mathcal{X}$, i.e., there exists a constant $c_\phi>0$ such that $\sup_{m \geq 1} \|\phi_m\|_{\infty} \leq c_\phi.$
\end{assumption}

\begin{assumption}\label{A3}
There exists a constant $C_\phi>0$ such that eigenfunctions satisfy the Lipschitz condition $|\phi_m(s) - \phi_m(t)| \leq C_\phi m |s - t|$ for any $s, t \in \mathcal{X}$.
\end{assumption}

\begin{assumption}\label{A4}
The eigenvalues $\{ \mu_m \}_{m=1}^{\infty}$ of $K$ satisfy $\mu_m \asymp m^{-\nu}$ for some $\nu > 1$.
\end{assumption}
\begin{remark}
Assumptions \ref{A2}--\ref{A4} are commonly adopted in the RKHS-based literature, such as \cite{mendelson2010regularization} and \cite{shang2013local}. Assumption \ref{A2} requires that the eigenfunctions are uniformly bounded, and Assumption \ref{A3} characterizes the its smoothness. Assumption \ref{A4} assumes $K$ has polynomially decaying eigenvalues.
%Assumptions \ref{A2}--\ref{A4} can be easily satisfied. For example, translation-invariant kernels $K(t, s) = g(t - s)$, where $g$ is an even function with period one. By employing the Fourier series expansion, the eigenfunctions of $K$ on the domain $\mathcal{X} = [0, 1]$ are given by the trigonometric functions $\phi_{2m-1}(x) = \sin(\pi mx)$ and $\phi_{2m}(x) = \cos(\pi mx)$ for $m \in\mathbb{N}_+$. It is straightforward to verify that we can take $c_\phi = 1$ and $C_\phi = \pi$ to satisfy the above assumptions.
\end{remark}

\begin{lemma}\label{Kbound}
Under Assumptions \ref{A1}-\ref{A4}, we have $\sup_{x\in\mathcal{X}} K(x, x) \leq c_\phi^2 \sum_{m=1}^{\infty} m^{-\nu} <\infty$.
\end{lemma}
The lemma \ref{Kbound} establishes the boundedness of the reproducing kernel function $K(\cdot,\cdot)$. Further details on RKHS can be found in \cite{Berlinet2011}.

\subsection{FSGD in RKHS}

Define the expected loss functional $\mathcal{E}(\boldsymbol{\beta}) = \mathbb{E}\left\{ \Psi_\tau\left(Y - \mathbf{Z}^{\top}\boldsymbol{\beta}(X)\right)\right\}$ over the product RKHS space $\mathcal{H}^{p+1}$ equipped with the inner product $\left\langle \boldsymbol{\beta}, \boldsymbol{\theta} \right\rangle_{\mathcal{H}^{p+1}} = \sum_{k=0}^p \left\langle \beta_k, \theta_k \right\rangle_{\mathcal{H}}$ for any function vector $\boldsymbol{\beta}(\cdot)=(\beta_0(\cdot),\ldots,\beta_p(\cdot))^{\top}$ and $\boldsymbol{\theta}(\cdot)=(\theta_0(\cdot),\ldots,\theta_p(\cdot))^{\top}$ with $\beta_k(\cdot),\theta_k(\cdot)\in\mathcal{H}$. Let $\nabla \mathcal{E}(\boldsymbol{\beta}) \in \mathcal{H}^{p+1}$ denote the functional gradient of $\mathcal{E}(\cdot)$ at $\boldsymbol{\beta}$. Before deriving the functional gradient, we impose the following regularity assumptions on model \eqref{model}.
\begin{assumption}\label{A5}
Let $\boldsymbol{\beta}^* = (\beta_0^*, \ldots, \beta_p^*)^\top$ denote the true coefficient function vector and assume that $\boldsymbol{\beta}^* \in \mathcal{H}^{p+1}$.
Given $(X, \mathbf{Z})$, the noise $\varepsilon := Y - \mathbf{Z}^{\top}\boldsymbol{\beta}^*(X)$ satisfies $\mathbb{E}\{|\tau - \mathbb{I}(\varepsilon < 0)|\cdot \varepsilon \mid X,\mathbf{Z} \} = 0$ and $\mathbb{E}(\varepsilon^2) < \infty$. Moreover, its conditional density $f_{\varepsilon}$ is bounded and satisfies $0 < f_{\varepsilon}(\cdot) \leq C_\varepsilon < \infty$ for some constant $C_\varepsilon > 0$.
\end{assumption}
\begin{assumption}\label{A6}
Define the matrix $\mathbf{\Sigma}_Z = \mathbb{E}(\mathbf{Z}\mathbf{Z}^{\top})$, and there exist positive constants $c_\Sigma$, $C_\Sigma$, $C_Z$ such that $c_\Sigma \leq \lambda_{\min}(\mathbf{\Sigma}_Z) \leq \lambda_{\max}(\mathbf{\Sigma}_Z) \leq C_\Sigma$ and $\|\mathbf{Z}\|_2 \leq C_Z$.
\end{assumption}
\begin{remark}
Assumptions \ref{A5} and \ref{A6} impose standard regularity conditions on the covariates and the error term, respectively, which are commonly adopted in the literature focusing on ER (e.g., \cite{xie2014varying}, \cite{liu2023functional}, and \cite{cao2026renewable}). Assumption \ref{A5} characterizes the error distribution via a zero conditional $\tau$-expectile, thereby ensuring identifiability of $\boldsymbol{\beta}^*$. The boundedness of the conditional density and the finite second moment are imposed to establish the consistency of the FSGD estimator. Assumption \ref{A6} requires that the explanatory variables have no multicollinearity and are bounded.
\end{remark}

Define $\psi_\tau: \mathbb{R} \to \mathbb{R}$ as the derivative of $\Psi_\tau(u)$, given by $\psi_\tau(u) =2|\tau - \mathbb{I}(u < 0)|u$. By the first-order condition for convexity, for any $u, v \in \mathbb{R}$, we have $\Psi_\tau(v) - \Psi_\tau(u) \geq \psi_\tau(u) (v - u)$. Let $u = Y - \mathbf{Z}^{\top}\boldsymbol{\beta}(X)$ and $v = Y - \mathbf{Z}^{\top}\boldsymbol{\theta}(X)$, we then obtain
\begin{equation}
\Psi_\tau(Y - \mathbf{Z}^{\top}\boldsymbol{\theta}(X)) - \Psi_\tau(Y - \mathbf{Z}^{\top}\boldsymbol{\beta}(X)) \geq \psi_\tau(Y - \mathbf{Z}^{\top}\boldsymbol{\beta}(X))\mathbf{Z}^{\top}(\boldsymbol{\beta}-\boldsymbol{\theta})(X).
\label{subgineq1}
\end{equation}
By the reproducing property of RKHS, denoting $K_X=K(X,\cdot)$, we have $(\beta_k - \theta_k)(X) = \left\langle \beta_k - \theta_k, K_X \right\rangle_{\mathcal{H}}$ for $k=0,\ldots,p$, which implies that
\begin{equation}
\mathbf{Z}^{\top}(\boldsymbol{\beta} - \boldsymbol{\theta})(X)=\sum_{k=0}^p Z_k (\beta_k - \theta_k)(X) = \sum_{k=0}^p Z_k \left\langle \beta_k - \theta_k, K_X \right\rangle_{\mathcal{H}} = \left\langle \boldsymbol{\beta} - \boldsymbol{\theta}, \mathbf{Z} K_X \right\rangle_{\mathcal{H}^{p+1}}.
\label{subgineq2}
\end{equation}
From \eqref{subgineq1}--\eqref{subgineq2}, we can get
\begin{align}
\Psi_\tau(Y - \mathbf{Z}^{\top}\boldsymbol{\theta}(X)) - \Psi_\tau(Y - \mathbf{Z}^{\top}\boldsymbol{\beta}(X)) \geq \left\langle \boldsymbol{\beta} - \boldsymbol{\theta}, \psi_\tau(Y - \mathbf{Z}^{\top}\boldsymbol{\beta}(X)) \mathbf{Z} K_X \right\rangle_{\mathcal{H}^{p+1}}.
\label{subgineq3}
\end{align}
Taking the expectation of \eqref{subgineq3}, we obtain
\begin{align}
\mathcal{E}(\boldsymbol{\theta})-\mathcal{E}(\boldsymbol{\beta}) &\geq \mathbb{E}\left\langle \boldsymbol{\theta}-\boldsymbol{\beta}, -\psi_\tau(Y - \mathbf{Z}^{\top}\boldsymbol{\beta}(X)) \mathbf{Z} K_X \right\rangle_{\mathcal{H}^{p+1}}\notag\\
&= \left\langle \boldsymbol{\theta}-\boldsymbol{\beta}, -\mathbb{E}\left\{\psi_\tau(Y - \mathbf{Z}^{\top}\boldsymbol{\beta}(X)) \mathbf{Z} K_X\right\} \right\rangle_{\mathcal{H}^{p+1}}.
\label{subgineq4}
\end{align}
where the last equality holds under Lemma \ref{Kbound} and Assumption \ref{A6}. Consequently, from \eqref{subgineq4}, we can identify that $\nabla \mathcal{E}(\boldsymbol{\beta})=-\mathbb{E}\left\{\psi_\tau\left(Y - \mathbf{Z}^{\top}\boldsymbol{\beta}(X)\right) \mathbf{Z} K_X\right\}\in\mathcal{H}^{p+1}$ is the functional gradient of $\mathcal{E}(\cdot)$ at $\boldsymbol{\beta}$. In stochastic approximation, the FSGD algorithm employs a more efficient strategy by replacing $\nabla \mathcal{E}(\boldsymbol{\beta})$ with an unbiased estimator constructed from one observation $(Y_i, X_i, \mathbf{Z}_i)$:
\begin{equation}
\boldsymbol{g}_i(\boldsymbol{\beta}) = -\psi_\tau\left(Y_i - \mathbf{Z}_i^{\top}\boldsymbol{\beta}(X_i)\right)\mathbf{Z}_i K_{X_i},
\label{fsubg}
\end{equation}
where $\mathbf{Z}_i = (Z_{i,0}, Z_{i,1}, \ldots, Z_{i,p})^{\top}$. Then, \eqref{fsubg} is used in the online estimation.

\subsection{Online FSGD Estimator}

Let $\left\{(Y_i, X_i, \mathbf{Z}_i)\right\}_{i=1,2,\ldots}$ be streaming data sequentially and independently sampled from model \eqref{model}. Upon receiving the $i$-th observation $(Y_i, X_i, \mathbf{Z}_i)$, we incur the instantaneous empirical loss $\mathcal{E}_i(\boldsymbol{\beta}) = \Psi_\tau (Y_i - \mathbf{Z}_i^{\top}\boldsymbol{\beta}(X_i))$. Given an initial estimate $\widehat{\boldsymbol{\beta}}_0\in\mathcal{H}^{p+1}$, according to \eqref{fsubg}, we construct $\boldsymbol{g}_i(\widehat{\boldsymbol{\beta}}_{i-1}) \in\mathcal{H}^{p+1}$ as the functional gradient of $\mathcal{E}_i(\cdot)$ at the current estimate $\widehat{\boldsymbol{\beta}}_{i-1}$. Then, the FSGD recursively updates the estimate along the negative gradient direction:
\begin{equation}
\widehat{\boldsymbol{\beta}}_i = \widehat{\boldsymbol{\beta}}_{i-1} - \gamma_i \boldsymbol{g}_i(\widehat{\boldsymbol{\beta}}_{i-1}) = \widehat{\boldsymbol{\beta}}_{i-1} + \gamma_i \psi_\tau\left(Y_i - \mathbf{Z}_i^{\top}\widehat{\boldsymbol{\beta}}_{i-1}(X_i)\right) \mathbf{Z}_i K_{X_i}, \quad i = 1, 2, \ldots,
\label{update}
\end{equation}
where $\{\gamma_i\}_{i=1,2,\ldots}$ is a sequence of positive step sizes, also referred to as learning rates.

Motivated by \cite{ruppert1988efficient} and \cite{polyak1992acceleration}, we employ the Polyak--Ruppert averaging for the FSGD estimator \eqref{update} to improve estimation accuracy by averaging over the entire update trajectory. The averaged FSGD estimator of $\boldsymbol{\beta}$ based on the accumulated dataset $\mathcal{D}_n=\{(Y_i, X_i, \mathbf{Z}_i)\}_{i=1}^n$ is defined as
\begin{equation}
\overline{\boldsymbol{\beta}}_n = \frac{1}{n} \sum_{i=1}^n \widehat{\boldsymbol{\beta}}_i.
\label{festi}
\end{equation}
This averaged estimator can be computed efficiently in an online fashion, via the recursive update $\overline{\boldsymbol{\beta}}_i = (i - 1)\overline{\boldsymbol{\beta}}_{i-1}/i + \widehat{\boldsymbol{\beta}}_i/i$ for $i=1,\ldots,n$. We summarize the above recursive estimation procedure in Algorithm \ref{alg1}, and refer to the estimator obtained from \eqref{festi} as the \emph{online FSGD estimator}.

\begin{algorithm}[htbp]
\def\baselinestretch{1}
\caption{Online FSGD Estimation for Varying Coefficient ER}
\label{alg1}
\begin{algorithmic}
\REQUIRE step-sizes $\{\gamma_i\}_{i=1,\ldots,n} > 0$, initial estimate $\widehat{\boldsymbol{\beta}}_0$.
\ENSURE online FSGD estimator $\overline{\boldsymbol{\beta}}_n$.
\FOR{$i = 1,\ldots,n $}
\STATE Update $\widehat{\boldsymbol{\beta}}_i = \widehat{\boldsymbol{\beta}}_{i-1} + \gamma_i \psi_\tau\left(Y_i - \mathbf{Z}_i^{\top}\widehat{\boldsymbol{\beta}}_{i-1}(X_i)\right) \mathbf{Z}_i K_{X_i}$;
\STATE Update $\overline{\boldsymbol{\beta}}_i = (i-1)\overline{\boldsymbol{\beta}}_{i-1}/i + \widehat{\boldsymbol{\beta}}_i/i$;
\ENDFOR
\RETURN $\overline{\boldsymbol{\beta}}_n$.
\end{algorithmic}
\end{algorithm}

For the FSGD algorithm, the step-size scheme $\{\gamma_i\}_{i=1,\ldots,n}$ constitutes the unique tunable component. We focus on two common settings for the step-size (e.g., \cite{bottou2007tradeoffs} and \cite{bottou2010large}) to identify the convergence rate of $\overline{\boldsymbol{\beta}}_n$ as follows:
\begin{itemize}
\item [(1)] the constant step-size scheme, where $\gamma_i \equiv \gamma = \Gamma(n)$ for all $1\leq i \leq n$, with $\Gamma(\cdot)$ being some positive function of the total sample size $n$;
\item [(2)] the non-constant step-size scheme, where the step-size decays polynomially, taking the form $\gamma_i = i^{-\kappa}$ for some $\kappa > 0$.
\end{itemize}

%\begin{remark}
%Although the constant step-size scheme is formally simpler, it suffers from two prominent limitations. First, it requires prior knowledge of the sample size $n$, which is typically unavailable in streaming data scenarios. Second, the optimal estimation error is achieved only at the final iteration. Therefore, for online scenarios, adopting an adaptive step-size mechanism is generally more practical. Despite introducing substantial theoretical complexity, such adaptive schemes with non-constant step-sizes successfully bypass the aforementioned limitations, yielding an algorithm intrinsically tailored to the streaming nature of the data.
%\end{remark}

\subsection{Non-asymptotic Analysis of Online FSGD Estimator}\label{Sec2.4}

We first introduce some necessary notation. Let $\otimes$ denote the tensor product operator on $\mathcal{H}^{p+1}$ defined through $(\boldsymbol{U} \otimes \boldsymbol{V})\boldsymbol{\beta} = \left\langle \boldsymbol{\beta}, \boldsymbol{V} \right\rangle_{\mathcal{H}^{p+1}} \boldsymbol{U},$ for any $\boldsymbol{U}, \boldsymbol{V}, \boldsymbol{\beta} \in \mathcal{H}^{p+1}$. In particular, we have $(\mathbf{Z}_i K_{X_i} \otimes \mathbf{Z}_i K_{X_i})\boldsymbol{\beta} = \left\langle \boldsymbol{\beta}, \mathbf{Z}_i K_{X_i} \right\rangle_{\mathcal{H}^{p+1}} \mathbf{Z}_i K_{X_i}$. By utilizing the reproducing property, the iterative updating formula \eqref{update} can be rewritten as
\begin{align}
\widehat{\boldsymbol{\beta}}_i &= \widehat{\boldsymbol{\beta}}_{i-1} + \gamma_i \psi_\tau\left(Y_i - \left\langle \widehat{\boldsymbol{\beta}}_{i-1}, \mathbf{Z}_i K_{X_i} \right\rangle_{\mathcal{H}^{p+1}}\right) \mathbf{Z}_i K_{X_i}\notag\\
%&=\widehat{\boldsymbol{\beta}}_{i-1} + 2\gamma_i \eta_{i-1,i}^\tau\left(Y_i - \left\langle \widehat{\boldsymbol{\beta}}_{i-1}, \mathbf{Z}_i K_{X_i} \right\rangle_{\mathcal{H}^{p+1}}\right) \mathbf{Z}_i K_{X_i}\notag\\
&=\left\{I - 2\gamma_i \eta_{i-1,i}^\tau \left(\mathbf{Z}_i K_{X_i} \otimes \mathbf{Z}_i K_{X_i}\right)\right\} \widehat{\boldsymbol{\beta}}_{i-1} + 2\gamma_i \eta_{i-1,i}^\tau Y_i \mathbf{Z}_i K_{X_i},
\label{update2}
\end{align}
where $I$ denotes the identity map on $\mathcal{H}^{p+1}$ and $\eta_{i-1,i}^\tau=\left|\tau - \mathbb{I}\left(Y_i - \mathbf{Z}_i^{\top}\widehat{\boldsymbol{\beta}}_{i-1}(X_i) < 0\right)\right|$. Under Assumption \ref{A5}, substituting $Y_i=\mathbf{Z}_i^{\top}\boldsymbol{\beta}^*(X_i)+\varepsilon_i$ into \eqref{update2} yields:
\begin{equation}
\boldsymbol{\zeta}_i:=\widehat{\boldsymbol{\beta}}_i - \boldsymbol{\beta}^* = \left\{I - 2\gamma_i \eta_{i-1,i}^\tau \left( \mathbf{Z}_i K_{X_i} \otimes \mathbf{Z}_i K_{X_i}\right)\right\} \left( \widehat{\boldsymbol{\beta}}_{i-1} - \boldsymbol{\beta}^*\right) + 2\gamma_i \eta_{i-1,i}^\tau \varepsilon_i \mathbf{Z}_i K_{X_i}.\label{update3}
\end{equation}
For simplicity, we denote $\boldsymbol{\xi}_i:=\mathbf{Z}_i K_{X_i} \otimes \mathbf{Z}_i K_{X_i}$ below. Note that $\boldsymbol{\xi}_i$ is a random operator on $\mathcal{H}^{p+1}$. We then define $\mathbf{\Xi} := \mathbb{E}(\mathbf{Z} K_{X} \otimes \mathbf{Z} K_{X}): \mathcal{H}^{p+1} \to \mathcal{H}^{p+1}$ as the unique linear operator such that, for any $\boldsymbol{\beta}, \boldsymbol{\theta} \in \mathcal{H}^{p+1}$, $\left\langle \boldsymbol{\beta}, \mathbf{\Xi} \boldsymbol{\theta} \right\rangle_{\mathcal{H}^{p+1}} = \mathbb{E}\left[\left\{\mathbf{Z}^\top \boldsymbol{\theta}(X)\right\} \left\{\mathbf{Z}^\top \boldsymbol{\beta}(X)\right\}\right]$.
%This population-level weighted covariance operator $\mathbf{\Xi}$ captures the covariance structure of the random elements in $\mathcal{H}^{p+1}$ induced by the joint distribution of $(X, \mathbf{Z})$.

Next, we introduce the error decomposition of the FSGD estimator $\overline{\boldsymbol{\beta}}_i$. This error decomposition allows us to distinguish between the deterministic term responsible for bias and the stochastic fluctuation term contributing to variance. Without loss of generality, we consider a zero initialization $\widehat{\boldsymbol{\beta}}_0 = \mathbf{0}$ in this paper. We decompose the recursion of $\boldsymbol{\zeta}_i$ into two finer recursions $\boldsymbol{\zeta}_i=\boldsymbol{\zeta}_i^{\text{bias}}+\boldsymbol{\zeta}_i^{\text{noise}}$, where bias recursion term is
\begin{equation}
\boldsymbol{\zeta}_i^{\text{bias}} = (I - 2\gamma_i \eta_{i-1,i}^\tau \boldsymbol{\xi}_i)\boldsymbol{\zeta}_{i-1}^{\text{bias}} \quad \text{with} \quad \boldsymbol{\zeta}_0^{\text{bias}} = \widehat{\boldsymbol{\beta}}_0 - \boldsymbol{\beta}^* = -\boldsymbol{\beta}^*,
\label{bias}
\end{equation}
and the noise recursion term is
\begin{equation}
\boldsymbol{\zeta}_i^{\text{noise}} = (I - 2\gamma_i \eta_{i-1,i}^\tau \boldsymbol{\xi}_i)\boldsymbol{\zeta}_{i-1}^{\text{noise}} + 2\gamma_i \eta_{i-1,i}^\tau \varepsilon_i \mathbf{Z}_iK_{X_i} \quad \text{with} \quad \boldsymbol{\zeta}_0^{\text{noise}} = \boldsymbol{0}.
\label{noise}
\end{equation}
Let $\eta_i^{\tau,*} := |\tau - \mathbb{I}(\varepsilon_i < 0)|$, we further decompose the recursive update into three parts:
\begin{enumerate}
\item [(1)] the leading bias recursion term
\begin{equation}
\boldsymbol{\zeta}_i^{\text{bias},*} = (I - 2\gamma_i \eta_i^{\tau,*} \mathbf{\Xi})\boldsymbol{\zeta}_{i-1}^{\text{bias},*}, \quad \text{with} \quad \boldsymbol{\zeta}_0^{\text{bias},*} = \boldsymbol{\zeta}_0^{\text{bias}},
\label{leadbias}
\end{equation}
which captures the leading deterministic component of the recursion in \eqref{update3}.

\item [(2)] the leading noise recursion term
\begin{equation}
\boldsymbol{\zeta}_i^{\text{noise},*} = (I - 2\gamma_i \eta_i^{\tau,*} \mathbf{\Xi})\boldsymbol{\zeta}_{i-1}^{\text{noise},*} + 2\gamma_i \eta_i^{\tau,*} \varepsilon_i \mathbf{Z}_iK_{X_i}, \quad \text{with} \quad \boldsymbol{\zeta}_0^{\text{noise},*} = \boldsymbol{\zeta}_0^{\text{noise}},
\label{leadnoise}
\end{equation}
which captures the leading stochastic fluctuation component in \eqref{update3}.

\item [(3)] the remainder recursion terms $R_i^{\text{bias}}=\boldsymbol{\zeta}_i^{\text{bias}} -\boldsymbol{\zeta}_i^{\text{bias},*}$ and $R_i^{\text{noise}}=\boldsymbol{\zeta}_i^{\text{noise}} -\boldsymbol{\zeta}_i^{\text{noise},*}$.
\end{enumerate}

Furthermore, we define averaged decomposition $\overline{\boldsymbol{\zeta}}_i^{\text{bias},*} = i^{-1} \sum_{j=1}^i \boldsymbol{\zeta}_j^{\text{bias},*}$ as the leading bias term, $\overline{\boldsymbol{\zeta}}_i^{\text{noise},*} = i^{-1} \sum_{j=1}^i \boldsymbol{\zeta}_j^{\text{noise},*}$ as the leading noise term, $\overline{R}_i^{\text{bias}}=i^{-1} \sum_{j=1}^i R_j^{\text{bias}}$ as the bias remainder term and $\overline{R}_i^{\text{noise}}=i^{-1} \sum_{j=1}^i R_j^{\text{noise}}$ as the noise remainder term in $\overline{\boldsymbol{\zeta}}_i:=\overline{\boldsymbol{\beta}}_i - \boldsymbol{\beta}^*$. Based on the error decomposition for the accumulated dataset $\mathcal{D}_n$, establishing a non-asymptotic error bound for $\overline{\boldsymbol{\beta}}_n - \boldsymbol{\beta}^*$ is equivalent to bounding the decomposed components: $\overline{\boldsymbol{\zeta}}_n=\overline{\boldsymbol{\zeta}}_n^{\text{bias},*}+\overline{\boldsymbol{\zeta}}_n^{\text{noise},*}+\overline{R}_n^{\text{bias}}+\overline{R}_n^{\text{noise}}$.

Theorem \ref{TH1} establishes non-asymptotic error bounds for the proposed online estimator under the supremum norm. The results indicate that the remainder terms are of strictly higher order in $n$ than the two leading terms associated with $\overline{\boldsymbol{\zeta}}_n$.
%Notably, the constant step-size scheme attains the optimal nonparametric convergence rate only at the final iteration. In contrast, the non-constant step-size scheme guarantees that the FSGD estimator achieves this optimal rate at any intermediate iteration, thereby making it more suitable for online scenarios.

\begin{theorem}[Non-asymptotic error bound under supremum norm]\label{TH1}
Under Lemma \ref{Kbound} and Assumptions \ref{A5}--\ref{A6}. Let $\underline{\tau} := \min\{\tau, 1-\tau\}$ and $\overline{\tau} := \max\{\tau, 1-\tau\}$. Then there exist constants such that the following statements hold.

\begin{enumerate}
\item[(1)] \textbf{Constant step-size} $(\gamma_i \equiv \gamma=\Gamma(n))$. The leading bias term satisfies that $\left\|\overline{\boldsymbol{\zeta}}_n^{\text{bias},*}\right\|_{\infty} \leq C_1/\sqrt{\underline{\tau}n\gamma}$. Assume that $0<\gamma< 1/(2\mu_1)$ and $\overline{\tau}/\underline{\tau} <\infty$, then the leading noise term satisfies
$\left\|\mathrm{Var}\left(\overline{\boldsymbol{\zeta}}_n^{\text{noise},*}\right)\right\|_{\infty} \leq C_2\gamma^{1/\nu}n^{-1+1/\nu}$. Furthermore, the remainder terms satisfy
$$
\mathbb{P}\left( \left\|\overline{R}_n^{\text{bias}}\right\|_\infty^2 \geq \gamma^{1/2}\frac{1}{\underline{\tau} n \gamma} \right) \leq \frac{C_3\overline{\tau}^2\gamma^{1/2}}{\underline{\tau}},
$$
and
$$
\mathbb{P}\left(\left\|\overline{R}_n^{\text{noise}}\right\|_\infty^2 \geq \gamma^{1/2}\frac{\gamma^{1/\nu}}{n^{1-1/\nu}} \right) \leq \frac{C_4\overline{\tau}^4\gamma^{1/2}}{\underline{\tau}}.
$$

\item[(2)] \textbf{Non-constant step-size} $(\gamma_i = i^{-\kappa})$. The leading bias term satisfies that $\left\|\overline{\boldsymbol{\zeta}}_n^{\text{bias},*}\right\|_{\infty} \leq C'_1/\sqrt{\underline{\tau}n\gamma_n}$. Assume that $0<\kappa< 1/2$ and $\overline{\tau}^2/\underline{\tau}^{(2-\kappa)/(1-\kappa)} <\infty$, then the leading noise term satisfies $\left\|\mathrm{Var}\left(\overline{\boldsymbol{\zeta}}_n^{\text{noise},*}\right)\right\|_{\infty} \leq C'_2\gamma_n^{1/\nu}n^{-1+1/\nu}.$ Moreover, the remainder terms satisfy that
$$
\mathbb{P}\left(\left\|\overline{R}_n^{\text{bias}}\right\|_\infty^2 \geq \gamma_n^{1/2}\frac{1}{\underline{\tau} n \gamma_n}\right) \leq \frac{C'_3 \overline{\tau}^2\gamma_n^{1/2} }{\underline{\tau}^2(1-\kappa)^2},
$$
and
$$
\mathbb{P}\left( \left\|\overline{R}_n^{\mathrm{noise}}\right\|_\infty^2 \geq \gamma_n^{1/2} \frac{\gamma_n^{1/\nu}}{n^{1-1/\nu}} \right)\leq \frac{C'_4 \overline{\tau}^4}{\underline{\tau} (1-\kappa)^2} \gamma_n^{1/2}.
$$
\end{enumerate}
\end{theorem}

\begin{remark}[Bias-variance trade-off and implicit regularization]
In FSGD, the step-size controls a bias-variance trade-off: a larger step-size reduces bias but inflates variance, whereas a smaller one increases bias but suppresses variance. Theorem \ref{TH1} shows that this trade-off is achieved by taking $\Gamma(n)=n^{-1/(\nu+1)}$ for constant step-sizes and $\gamma_i=i^{-1/(\nu+1)}$ for non-constant step-sizes, yielding the optimal nonparametric rate $n^{-\nu/(\nu+1)}$. This mechanism mirrors the penalty parameter in RKHS-penalized methods (Remark \ref{offline}): a larger step-size corresponds to weaker penalization, allowing greater flexibility, while a smaller step-size imposes stronger shrinkage. The cumulative step-size $\sum_{i=1}^n \gamma_i$ can be viewed as the total path length in the RKHS; since the functional gradient is defined with respect to the RKHS norm, this path length serves as an early-stopping mechanism that controls model complexity and prevents overfitting in one pass over the data.
\end{remark}

%\begin{remark}[Tail vulnerability of extreme expectiles]\label{R5}
%Theorem \ref{TH1} shows that in extreme tail regimes ($\tau \to 0$ or $1$), the non-asymptotic error bound of the FSGD estimator inflates: the bias grows as the tail probability decays, due to the intrinsic difficulty of tail function approximation, while the variance diverges at an even faster polynomial rate because the effective sample size shrinks drastically. Thus, the ER estimator converges increasingly slowly as the tail becomes more extreme—a classic difficulty in ER stemming from the asymptotic behavior of extreme tail probabilities; see \cite{girard2021extreme} and \cite{daouia2024extreme}.
%\end{remark}

\begin{remark}[Computational complexity]\label{offline}
Given a static dataset $\{(Y_i, X_i, \mathbf{Z}_i)\}_{i=1}^n$, classical estimation procedures minimize the empirical loss $n^{-1}\sum_{i=1}^{n}\mathcal{E}_i(\boldsymbol{\beta})$. A commonly used offline approach is RKHS-penalized ER:
$$\widehat{\boldsymbol{\beta}} = \underset{\boldsymbol{\beta}\in\mathcal{H}^{p+1}}{\mathrm{argmin}} \left\{\frac{1}{n}\sum_{i=1}^{n}\Psi_\tau (Y_i - \mathbf{Z}_i^{\top}\boldsymbol{\beta}(X_i))+\lambda\Vert\boldsymbol{\beta}\Vert_{\mathcal{H}^{p+1}}^2\right\},$$
where $\lambda > 0$ is the penalty parameter. Since this objective has no closed-form solution in general, it is usually solved by iteratively reweighted least squares (IRLS). Each IRLS iteration solves a weighted kernel ridge regression subproblem with time complexity $O(n^3)$ and space complexity $O(n^2)$. Consequently, offline estimation requires $O(Mn^3)$ time and $O(n^2)$ space, where $M$ is the number of IRLS iterations. In contrast, the proposed FSGD algorithm requires $O(n^2)$ time and $O(n)$ space, making it well-suited for online learning.
\end{remark}

To construct online inference for $\boldsymbol{\beta}^*$, we study the distributional convergence of $\overline{\boldsymbol{\beta}}_n(x) - \boldsymbol{\beta}^*(x)$ for any $x\in\mathcal{X}$. In Theorem \ref{TH2}, we analyze the leading noise term $\overline{\boldsymbol{\zeta}}_n^{\mathrm{noise},*}(x_0)$ to establish the pointwise limiting distribution of $\overline{\boldsymbol{\beta}}_n(x_0)-\boldsymbol{\beta}^*(x_0)$ for any fixed point $x_0\in\mathcal{X}$, and provide a finite-sample upper bound on the Kolmogorov distance between its sampling distribution and the multivariate standard normal distribution.

\begin{theorem}[Berry--Esseen bound for the pointwise asymptotic normality]\label{TH2}
Under Lemma \ref{Kbound} and Assumptions \ref{A5}--\ref{A6}, we further suppose that $\mathbb{E}(|\varepsilon|^3)<\infty$. Let $\Phi_{p+1}(\cdot)$ denote the cumulative distribution function of the $(p+1)$-dimensional standard normal distribution and let $\mathbf{u} = (u_1, \ldots, u_{p+1})^\top \in \mathbb{R}^{p+1}$. For any fixed point $x_0\in\mathcal{X}$, there exist constants such that the following statements hold.
\begin{enumerate}
\item[(1)] \textbf{Constant step-size} $(\gamma_i \equiv \gamma=\Gamma(n))$. If $0<\gamma< 1/(2\mu_1)$, we have
\begin{align*}
\sup_{\mathbf{u} \in \mathbb{R}^{p+1}} & \Big| \mathbb{P}\left( \left[\mathrm{Var}\left\{\overline{\boldsymbol{\zeta}}_n^{\mathrm{noise},*}(x_0)\right\}\right]^{-1/2} \left\{ \overline{\boldsymbol{\beta}}_{n}(x_0) - \boldsymbol{\beta}^*(x_0) - \overline{\boldsymbol{\zeta}}_{n}^{\mathrm{bias},*}(x_0) \right\} \leq \mathbf{u} \right) \\
& - \Phi_{p+1}(\mathbf{u})\Big| \leq C \frac{\overline{\tau}^3}{\underline{\tau}^{3 - 2/\nu}} \frac{(n\gamma)^{1/(2\nu)}}{n^{1/2}}.
\end{align*}

\item[(2)] \textbf{Non-constant step-size} $(\gamma_i = i^{-\kappa})$. If $0<\kappa< 1/2$, we have
\begin{align*}
\sup_{\mathbf{u} \in \mathbb{R}^{p+1}} & \Big|\mathbb{P}\left(\left[\mathrm{Var}\left\{\overline{\boldsymbol{\zeta}}_n^{\mathrm{noise},*}(x_0)\right\}\right]^{-1/2} \left\{ \overline{\boldsymbol{\beta}}_{n}(x_0) - \boldsymbol{\beta}^*(x_0) - \overline{\boldsymbol{\zeta}}_{n}^{\mathrm{bias},*}(x_0) \right\}\leq \mathbf{u} \right) \\
& - \Phi_{p+1}(\mathbf{u}) \Big| \leq C'\frac{\overline{\tau}^3}{\underline{\tau}^{3/(1-\kappa)}} \left(n^{-1/2}+n^{3\kappa-3/2}\right)(n\gamma_n)^{1/(2\nu)}.
\end{align*}
\end{enumerate}
\end{theorem}

\begin{remark}[Undersmoothing for asymptotically valid inference]\label{rem:undersmooth}
Theorem \ref{TH2} shows that the bias-corrected, standardized estimator $\overline{\boldsymbol{\beta}}_n(x_0)-\boldsymbol{\beta}^*(x_0)$ converges in distribution to a multivariate standard normal for any fixed $x_0\in\mathcal{X}$. By Theorem \ref{TH1}, the leading bias is $O(1/\sqrt{n\gamma_n})$, while the leading noise standard deviation is $O\{\gamma_n^{1/(2\nu)}n^{-1/2+1/(2\nu)}\}$. At the optimal step-size, i.e., $\gamma=n^{-1/(\nu+1)}$ (constant) or $\gamma_i=i^{-1/(\nu+1)}$ (non-constant), these two terms are of the same order under the scaling in Theorem \ref{TH2}. Hence the bias does not vanish asymptotically and, if uncorrected, distorts coverage. For inference, undersmoothing can suppress it by choosing a slightly slower decaying step-size, e.g., $\gamma=n^{-1/(\nu+1+\delta)}$ with small $\delta>0$ (or $\gamma_i=i^{-1/(\nu+1+\delta)}$), such that $\left\|\left\{\mathrm{Var}\left(\overline{\boldsymbol{\zeta}}_n^{\mathrm{noise},*}\right)\right\}^{-1/2} \overline{\boldsymbol{\zeta}}_n^{\mathrm{bias},*}\right\|_\infty =o(1).$ Consequently, the bias becomes negligible relative to the noise standard deviation, so Theorem \ref{TH2} can be used for inference without explicit bias correction.
\end{remark}

\begin{remark}[Challenges for online inference]
Although Theorem \ref{TH2} gives the pointwise limiting distribution of the FSGD estimator $\overline{\boldsymbol{\beta}}_n$, its limiting covariance matrix depends on the unknown eigenvalues $\{\mu_m\}_{m=1}^{\infty}$ and eigenfunctions $\{\phi_m\}_{m=1}^{\infty}$ of the reproducing kernel $K$, the joint distribution of $(X,\mathbf{Z})$, and the weighted noise second moment. Consequently, an effective online inference procedure cannot be established.
\end{remark}

\section{Online Statistical Inference via Multiplier Bootstrap}\label{Sec3}

In this section, we develop online inference for $\boldsymbol{\beta}^*$ based on the FSGD estimator $\overline{\boldsymbol{\beta}}_n$. Specifically, for each component $\beta_k^*$, we aim to construct a nominal level $1-\alpha$ PCI as $CI_{n,k}(x_0; \alpha) = \left[U_{n,k}(x_0; \alpha), V_{n,k}(x_0; \alpha)\right]$ for any fixed $x_0 \in \mathcal{X}$ and for $\alpha\in (0,1)$. We require the PCI to be asymptotically valid, meaning that the coverage probabilities approach the nominal level $1-\alpha$. Formally, this translates to $\mathbb{P}\left( \beta_k^*(x_0) \in CI_{n,k}(x_0; \alpha) \right) \to 1 - \alpha$.

To this end, we propose a multiplier bootstrap procedure for online inference based on the FSGD estimator. We then establish its consistency by proving that the introduced perturbation accurately mimics the randomness of the online estimation process, so that the conditional distribution of the bootstrapped FSGD estimator asymptotically coincides with the sampling distribution of the original FSGD estimator. 

\subsection{Online Bootstrapped FSGD Estimator}

Upon receiving the $i$-th observation $(Y_i, X_i, \mathbf{Z}_i)$, we generate a set of independent random perturbation weights $w_{i,b} \sim \mathrm{W}$ for $b=1,\ldots,B$, where $B$ denotes the number of bootstrap replicates. Given any initial estimates $\widehat{\boldsymbol{\beta}}_0^b \in \mathcal{H}^{p+1}$ and following equations \eqref{update} and \eqref{update2}, we construct the randomly perturbed FSGD update as follows:
\begin{align*}
\widehat{\boldsymbol{\beta}}_i^b &= \widehat{\boldsymbol{\beta}}_{i-1}^b - w_{i,b}\gamma_i \boldsymbol{g}_i\left(\widehat{\boldsymbol{\beta}}_{i-1}^b\right) \\
&=\left\{I - 2w_{i,b}\gamma_i \eta_{i-1,i}^{\tau,b} \left(\mathbf{Z}_i K_{X_i} \otimes \mathbf{Z}_i K_{X_i}\right)\right\} \widehat{\boldsymbol{\beta}}_{i-1}^b + 2w_{i,b}\gamma_i \eta_{i-1,i}^{\tau,b} Y_i \mathbf{Z}_i K_{X_i},
\end{align*}
for $i = 1, 2, \ldots$ and $b = 1, \ldots, B$, where $\eta_{i-1,i}^{\tau,b} = \left|\tau - \mathbb{I}\left(Y_i - \mathbf{Z}_i^{\top}\widehat{\boldsymbol{\beta}}_{i-1}^b(X_i) < 0\right)\right|$. We then obtain the averaged bootstrapped FSGD estimator over the accumulated dataset $\mathcal{D}_n$ as $\overline{\boldsymbol{\beta}}_n^b = n^{-1} \sum_{i=1}^n \widehat{\boldsymbol{\beta}}_i^b$. We summarize the above recursive estimation procedure in Algorithm \ref{alg2}, and refer to the estimator $\overline{\boldsymbol{\beta}}_n^b$ as the \emph{online bootstrapped FSGD estimator}.

\begin{algorithm}[htbp]
\def\baselinestretch{1}
\caption{Online Bootstrapped FSGD Estimation for Varying Coefficient ER}
\label{alg2}
\begin{algorithmic}
\REQUIRE step-sizes $\{\gamma_i\}_{i=1,\ldots,n} > 0$, initial estimates $\widehat{\boldsymbol{\beta}}_0^b$, the number of bootstrap replicates $B$.
\ENSURE Online bootstrapped FSGD estimators $\overline{\boldsymbol{\beta}}_n^b$.
\FOR{$i = 1,\ldots,n $}
\FOR{$b=1,\ldots,B$}
\STATE Update $\widehat{\boldsymbol{\beta}}_i^b = \widehat{\boldsymbol{\beta}}_{i-1}^b + w_{i,b}\gamma_i \psi_\tau\left(Y_i - \mathbf{Z}_i^{\top}\widehat{\boldsymbol{\beta}}_{i-1}^b(X_i)\right) \mathbf{Z}_i K_{X_i}$;
\STATE Update $\overline{\boldsymbol{\beta}}_i^b = (i-1)\overline{\boldsymbol{\beta}}_{i-1}^b/i + \widehat{\boldsymbol{\beta}}_i^b/i$;
\ENDFOR
\ENDFOR
\RETURN $\overline{\boldsymbol{\beta}}_n^b$ for $b=1,\ldots,B$.
\end{algorithmic}
\end{algorithm}

Algorithm \ref{alg2} produces a sequence of online bootstrapped FSGD estimators. By collecting $B$ such trajectories, we can approximate the sampling distribution of the original FSGD estimator and construct PCIs for $\beta_k^*$, $k=0,\ldots,p$, in a fully online fashion.

\subsection{Bootstrap Consistency for Online Inference}

To obtain asymptotically valid online PCI for a fixed component index $k$ and a given point $x_0 \in \mathcal{X}$, we approximate the sampling distribution of $\overline{\beta}_{n,k}(x_0) - \beta_k^*(x_0)$ by the conditional empirical distribution of the bootstrap differences $\overline{\beta}_{n,k}^b(x_0) - \overline{\beta}_{n,k}(x_0)$ given dataset $\mathcal{D}_n$. We impose the following conditions on the bootstrap multipliers.

\begin{assumption}\label{A7}
The multipliers $\{w_{i,b}\}_{i,b=1}^{n,B}$ are i.i.d. copies of a random variable $W$ that satisfies $\mathbb{E}(W)=1$, $\mathrm{Var}(W)=1$, and $\mathbb{P}(|W|\geq t) \leq 2\exp(-t^2/C)$ for all $t\geq 0$ with a constant $C>0$.
\end{assumption}
\begin{assumption}\label{A8}
The error $\varepsilon$ is sub-exponential, i.e., there exist constants $c_e, C_e>0$ such that
$\mathbb{P}(|\varepsilon|\geq t)\leq C_e\exp(-t/c_e)$ for all $t\geq 0$.
\end{assumption}
\begin{remark}
Assumption \ref{A7} is standard in the multiplier bootstrap literature to establish bootstrap consistency. Assumption \ref{A8} imposes a sub-exponential tail condition on the random error $\varepsilon$. This tail assumption is used to control the Kolmogorov-type error bound. Compared with the sub-Gaussian error assumptions commonly imposed in the ER literature (\cite{gu2016high}), the sub-exponential requirement is substantially weaker.
\end{remark}

We derive an error decomposition for the online bootstrapped FSGD estimator analogous to Section \ref{Sec2.4}. The bootstrapped leading noise recursion follows as
\begin{equation*}
\boldsymbol{\zeta}_i^{\mathrm{noise},*,b} = \left(I - 2\gamma_i \eta_i^{\tau,*} \mathbf{\Xi}\right) \boldsymbol{\zeta}_{i-1}^{\mathrm{noise},*,b} + 2w_{i,b}\gamma_i \eta_i^{\tau,*} \varepsilon_i \mathbf{Z}_i K_{X_i}, \quad \boldsymbol{\zeta}_0^{\mathrm{noise},*,b} = \mathbf{0},
\end{equation*}
with the bootstrapped leading noise term $\overline{\boldsymbol{\zeta}}_n^{\mathrm{noise},*,b} = n^{-1}\sum_{i=1}^n \boldsymbol{\zeta}_i^{\mathrm{noise},*,b}$. The bootstrapped leading bias term coincides with the original leading bias term. Theorem \ref{TH3} establishes the consistency for online inference of the bootstrapped leading noise term.

\begin{theorem}[Consistency of bootstrapped leading noise term]\label{TH3}
Under Lemma \ref{Kbound} and Assumptions \ref{A5}--\ref{A8}, for any fixed point $x_0\in\mathcal{X}$, there exist constants such that the following statements hold.

\begin{enumerate}
\item[(1)] \textbf{Constant step-size} $(\gamma_i \equiv \gamma=\Gamma(n))$. If $0<\gamma< 1/(2\mu_1)$, with probability at least $1 - C_1 n^{-1}$, we have
\begin{align*}
\sup_{\mathbf{u}\in\mathbb{R}^{p+1}} & \left|\mathbb{P}\left( \sqrt{n(n\gamma)^{-1/\nu}}\left\{\overline{\boldsymbol{\zeta}}_n^{\mathrm{noise},*,b}(x_0) - \overline{\boldsymbol{\zeta}}_n^{\mathrm{noise},*}(x_0)\right\} \leq \mathbf{u}  \mid \mathcal{D}_n\right) \right. \\
& \left.- \mathbb{P}\left( \sqrt{n(n\gamma)^{-1/\nu}} \overline{\boldsymbol{\zeta}}_n^{\mathrm{noise},*}(x_0) \leq \mathbf{u}  \right)\right| \leq C_2 \frac{\overline{\tau}^{6 - 3/(2\nu)}}{\underline{\tau}^{6 - 2/\nu}} \frac{(n\gamma)^{1/(2\nu)}(\log n)^3}{n^{1/2}}.
\end{align*}

\item[(2)] \textbf{Non-constant step-size} $(\gamma_i = i^{-\kappa})$. If $0<\kappa< 1/3$,  with probability at least $1 - C_1' n^{-1}$, we have
\begin{align*}
\sup_{\mathbf{u}\in\mathbb{R}^{p+1}} & \left|\mathbb{P}\left( \sqrt{n(n\gamma_n)^{-1/\nu}}\left\{\overline{\boldsymbol{\zeta}}_n^{\mathrm{noise},*,b}(x_0) - \overline{\boldsymbol{\zeta}}_n^{\mathrm{noise},*}(x_0)\right\} \leq \mathbf{u} \mid \mathcal{D}_n\right) \right. \\
& \left. -\mathbb{P}\left( \sqrt{n(n\gamma_n)^{-1/\nu}} \overline{\boldsymbol{\zeta}}_n^{\mathrm{noise},*}(x_0) \leq \mathbf{u} \right)\right| \leq C_2'\frac{\overline{\tau}^6}{\underline{\tau}^{(6-3\kappa)/(1-\kappa)}}\frac{(n\gamma_n)^{1/(2\nu)}(\log n)^3 }{n^{1/2}}.
\end{align*}
\end{enumerate}
\end{theorem}

\begin{remark}
Theorem \ref{TH3} shows that bootstrap differences $\overline{\boldsymbol{\zeta}}_n^{\mathrm{noise},*,b}(x_0)-\overline{\boldsymbol{\zeta}}_n^{\mathrm{noise},*}(x_0)$ capture the stochastic fluctuation of $\overline{\boldsymbol{\zeta}}_n^{\mathrm{noise},*}(x_0)$. The stricter condition $0<\kappa<1/3$ for non-constant step-sizes reflects the extra randomness from the bootstrap weights $w_{i,b}$. Controlling the higher-order moments of the bootstrapped weighted gradients requires faster step-size decay to suppress early observations, which is critical for bounding the bootstrap remainders and the Kolmogorov distance. 
\end{remark}

Let $q_{n,k}(\alpha, x_0)$ be the $\alpha$-th quantile of the distribution of $\sqrt{n(n\gamma_n)^{-1/\nu}}\{\overline{\beta}_{n,k}(x_0) - \beta_k^*(x_0) - \overline{\zeta}_{n,k}^{\mathrm{bias},*}(x_0)\}$, where $\overline{\zeta}_{n,k}^{\mathrm{bias},*}$ denotes the $k$-th component of the leading bias term $\overline{\boldsymbol{\zeta}}_n^{\mathrm{bias},*}$. Let $\widetilde{q}_{n,k}(\alpha, x_0)$ be the $\alpha$-th quantile of $\sqrt{n(n\gamma_n)^{-1/\nu}}\{\overline{\beta}_{n,k}^b(x_0) - \overline{\beta}_{n,k}(x_0)\}$ given $\mathcal{D}_n$. Theorem \ref{TH4} establishes the asymptotic validity of the online PCI.

\begin{theorem}[Asymptotically valid PCI for online inference]\label{TH4}
Under Lemma \ref{Kbound} and Assumptions \ref{A5}--\ref{A8}, for any fixed component $k=0,\ldots,p$ and fixed point $x_0\in\mathcal{X}$, as $n \to \infty$, the following statements hold.

\begin{enumerate}
    \item[(1)] \textbf{Constant step-size} $(\gamma_i \equiv \gamma=\Gamma(n))$. If $0<\gamma< 1/(2\mu_1)$, we have
    \begin{align*}
    \mathbb{P}\Big( &\widetilde{q}_{n,k}(\alpha/2, x_0) \leq \sqrt{n(n\gamma)^{-1/\nu}}\left\{\overline{\beta}_{n,k}(x_0) - \beta_k^*(x_0) - \overline{\zeta}_{n,k}^{\mathrm{bias},*}(x_0)\right\}\\
    & \leq \widetilde{q}_{n,k}(1 - \alpha/2, x_0) \Big) \to 1 - \alpha.
    \end{align*}

    \item[(2)]
    \textbf{Non-constant step-size} $(\gamma_i = i^{-\kappa})$. If $0<\kappa< 1/3$, we have
    \begin{align*}
    \mathbb{P}\Big( &\widetilde{q}_{n,k}(\alpha/2, x_0) \leq \sqrt{n(n\gamma_n)^{-1/\nu}}\left\{\overline{\beta}_{n,k}(x_0) - \beta_k^*(x_0) - \overline{\zeta}_{n,k}^{\mathrm{bias},*}(x_0)\right\} \\
    &\leq \widetilde{q}_{n,k}(1 - \alpha/2, x_0) \Big) \to 1 - \alpha.
    \end{align*}
\end{enumerate}
\end{theorem}

\begin{remark}[Undersmoothing for asymptotically valid online PCI]
As discussed in Remark \ref{rem:undersmooth}, choosing a slightly slower decaying step-size (undersmoothing) makes the leading bias term negligible relative to the standard deviation of the leading noise term.
In this regime we have $\sqrt{n(n\gamma_n)^{-1/\nu}} \overline{\zeta}_{n,k}^{\mathrm{bias},*}(x_0) = o\left\{\overline{\beta}_{n,k}(x_0) - \beta_k^*(x_0)\right\}$, so that Theorem \ref{TH4} implies
$$
\mathbb{P}\left( \widetilde{q}_{n,k}(\alpha/2, x_0) \leq \sqrt{n(n\gamma_n)^{-1/\nu}}\left\{\overline{\beta}_{n,k}(x_0) - \beta_k^*(x_0)\right\} \leq \widetilde{q}_{n,k}(1 - \alpha/2, x_0) \right) \to 1 - \alpha,
$$
which establishes an asymptotically valid online PCI for $\beta_k^*(x_0)$. 
\end{remark}

In practice, we use the percentile method to construct a nominal level $1-\alpha$ PCI, denoted by $CI_{n,k}(x_0; \alpha) = [U_{n,k}(x_0; \alpha), V_{n,k}(x_0; \alpha)]$. This requires only extracting the values at $x_0$ from the $B$ stored bootstrap trajectories, with no additional pass over the data. Specifically, let $\widehat{q}_{n,k}(\alpha/2, x_0)$ and $\widehat{q}_{n,k}(1 - \alpha/2, x_0)$ be the $\alpha/2$ and $1-\alpha/2$ sample quantiles of the bootstrap differences $\{\overline{\beta}_{n,k}^b(x_0) - \overline{\beta}_{n,k}(x_0)\}_{b=1}^B$. Then set $U_{n,k}(x_0; \alpha) = \overline{\beta}_{n,k}(x_0) + \widehat{q}_{n,k}(\alpha/2, x_0)$ and $V_{n,k}(x_0; \alpha) = \overline{\beta}_{n,k}(x_0) + \widehat{q}_{n,k}(1 - \alpha/2, x_0)$. Algorithm \ref{alg:pointwise} summarizes the construction procedure.

\begin{algorithm}[htbp]
\def\baselinestretch{1}
\caption{Online PCI Construction for $\beta_k^*(x)$}
\label{alg:pointwise}
\begin{algorithmic}
\REQUIRE Estimators $\overline{\beta}_{n,k}$ and $\overline{\beta}_{n,k}^b$, $b=1,\ldots,B$, any fixed point $x_0$, confidence level $\alpha$.
\ENSURE PCI for $\beta_k^*(x)$.
\STATE Compute $\widehat{q}_{n,k}(\alpha/2, x_0)$ and $\widehat{q}_{n,k}(1 - \alpha/2, x_0)$ as the $\alpha/2$ and $1-\alpha/2$ sample quantiles of $\left\{\overline{\beta}_{n,k}^b(x_0) - \overline{\beta}_{n,k}(x_0)\right\}_{b=1}^B$;
\STATE Obtain $U_{n,k}(x_0; \alpha) = \overline{\beta}_{n,k}(x_0) + \widehat{q}_{n,k}(\alpha/2, x_0)$ and $V_{n,k}(x_0; \alpha) = \overline{\beta}_{n,k}(x_0) + \widehat{q}_{n,k}(1 - \alpha/2, x_0)$.
\RETURN $CI_{n,k}(x_0; \alpha) = [U_{n,k}(x_0; \alpha), V_{n,k}(x_0; \alpha)]$.
\end{algorithmic}
\end{algorithm}

\begin{remark}[Normal-based online PCI]
A normal-based online PCI also can be constructed by replacing the bootstrap quantiles with a normal approximation. We compute the bootstrap variance $\Sigma_{n,k}^b(x_0) = (B-1)^{-1}\sum_{b=1}^B \left\{ \overline{\beta}_{n,k}^b(x_0) - \overline{\beta}_{n,k}(x_0) \right\}^2$  and set
$$
U_{n,k}(x_0; \alpha) = \overline{\beta}_{n,k}(x_0) + z_{\alpha/2} \sqrt{\Sigma_{n,k}^b(x_0)}, \quad
V_{n,k}(x_0; \alpha) = \overline{\beta}_{n,k}(x_0) + z_{1-\alpha/2} \sqrt{\Sigma_{n,k}^b(x_0)},
$$
where $z_{\alpha/2}$ is the $\alpha/2$ quantile of the standard normal distribution.
\end{remark}

\section{Simulation Study}\label{Sec4}

In this section, we evaluate the finite-sample performance of the proposed online method. We generate data from model \eqref{model} with $p=2$ on the design domain $\mathcal{X}=[0,1]$. The covariates are independently drawn as $X\sim \mathrm{Unif}(0,1)$, $Z_1\sim \mathrm{Unif}(-\sqrt{3},\sqrt{3})$, and $Z_2\sim \mathrm{N}(0,1)$ truncated to the interval $[-4,4]$. The true coefficient functions are specified as
\begin{align*}
\beta_0^*(x) = 2\sin(3\pi x/2), \quad \beta_1^*(x) = \frac{1}{2}\exp(x/2)\sin(3\pi x), \quad \beta_2^*(x) = \frac{\cos(4\pi x)}{1+\exp(-10x+2)}.
\end{align*}
At each fixed expectile level $\tau$, the random error $\varepsilon$ is drawn from four distributions: (1) the normal distribution $\mathrm{N}(0,2)$; (2) the contaminated normal mixture $0.95\,\mathrm{N}(0,1)+0.05\,\mathrm{N}(0,20)$; (3) the standard Laplace distribution $\mathrm{Laplace}(0,1)$; (4) the Student-$t$ distribution $\mathrm{t}(4)$. And each distribution is recentered by subtracting its $\tau$-expectile. The response is then generated as $Y=\beta_0^*(X)+Z_1\beta_1^*(X)+Z_2\beta_2^*(X)+\varepsilon$.

We consider expectile levels $\tau\in\{0.25,0.50,0.75\}$, corresponding to different tail regions. For each Monte Carlo replication ($M=100$), we generate a training stream of total $N_{\text{train}}=5000$ observations and a test set of size $N_{\text{test}}=5000$. We evaluate the performance at sample sizes $n=500,1000,\ldots,5000$ as the data stream arrives.

For the proposed online FSGD method, we use the exponential kernel $K(x,x')=\exp(-|x-x'|/h)$ on $\mathcal{X}$, with the bandwidth $h$ determined by the median heuristic based on the first $n_0=500$ observations. We consider two step-size schemes: a constant step-size $\gamma\equiv N^{-c}$ (FSGD-C) and a non-constant step-size $\gamma_i=i^{-\kappa}$ (FSGD-NC). To mitigate finite-sample variance inflation, we truncate the non-constant step-size as $\gamma_i=\min\{i^{-\kappa}, n_0^{-\kappa}\}$. This preserves the online decaying structure while reducing the variance contribution of early large step-sizes to the bootstrap trajectories. Similar strategies have been adopted in the literature. For example, \cite{xie2026scalable} and \cite{liu2026scalable} adopt a delayed update, in which the first $n_0$ observations are excluded from the Polyak--Ruppert average to obtain a stable initial estimate, whereas \cite{fang2018online} and \cite{chen2022onlinefda} control the magnitude of early updates by choosing an initial step-size satisfying $0<\gamma_0<1$. Since the exponential kernel has eigenvalue decay rate $\nu=2$, the optimal step-size exponent is $1/(\nu+1)=1/3$. To ensure valid PCIs, we adopt an undersmoothing strategy that restricts the step-size exponent to $(0,1/3)$, so that the leading bias is asymptotically negligible. Accordingly, the step-size parameters $c$ and $\kappa$ are selected by 5-fold cross-validation on the first $n_0$ observations, using a grid of 10 equally spaced values in $(0,1/3)$. For multiplier bootstrap, we draw i.i.d. weights $w_{i,b}\sim \mathrm{Unif}(1-\sqrt{3},1+\sqrt{3})$ for $b=1,\ldots,B$ with $B=200$, and construct nominal $95\%$ PCIs using both the percentile and normal methods.

We compare the proposed online method with three offline methods: the iteratively weighted local least squares (IWLLS) estimator of \cite{xie2014varying}, the kernel expectile regression via majorization--minimization (KERMM) estimator of \cite{yang2018flexible}, and the iterative polynomial-based local expectile regression (IPLER) estimator of \cite{adam2022local}. For IWLLS and IPLER, we construct plug-in normal PCIs based on the variance estimators of \cite{xie2014varying} and \cite{adam2022local}, respectively. In addition, for all offline methods, we construct percentile PCIs based on the wild bootstrap (\cite{wu1986jackknife} and \cite{mammen1993wild}), where bootstrap responses are generated as $Y_i^*=\widehat{Y}_i+w_{i,b}\widehat{\varepsilon}_i$ with $w_{i,b}\sim\mathrm{Unif}(1-\sqrt{3},1+\sqrt{3})$. For KERMM, we additionally construct a normal PCI based on the wild bootstrap.

To compare computational efficiency, we record estimation and inference time separately. For the online FSGD method, both times are accumulated during a one-pass over the data stream. For offline competitors, the model is refit on the first $n$ observations, and both times are recorded. All experiments were conducted on a workstation equipped with an Intel(R) Xeon(R) Gold 6348 28-core CPU @ 2.60 GHz and 256 GB of RAM.

\begin{figure}[htbp]
\def\baselinestretch{1}
\centering
\begin{subfigure}{0.23\textwidth}
  \includegraphics[width=\textwidth]{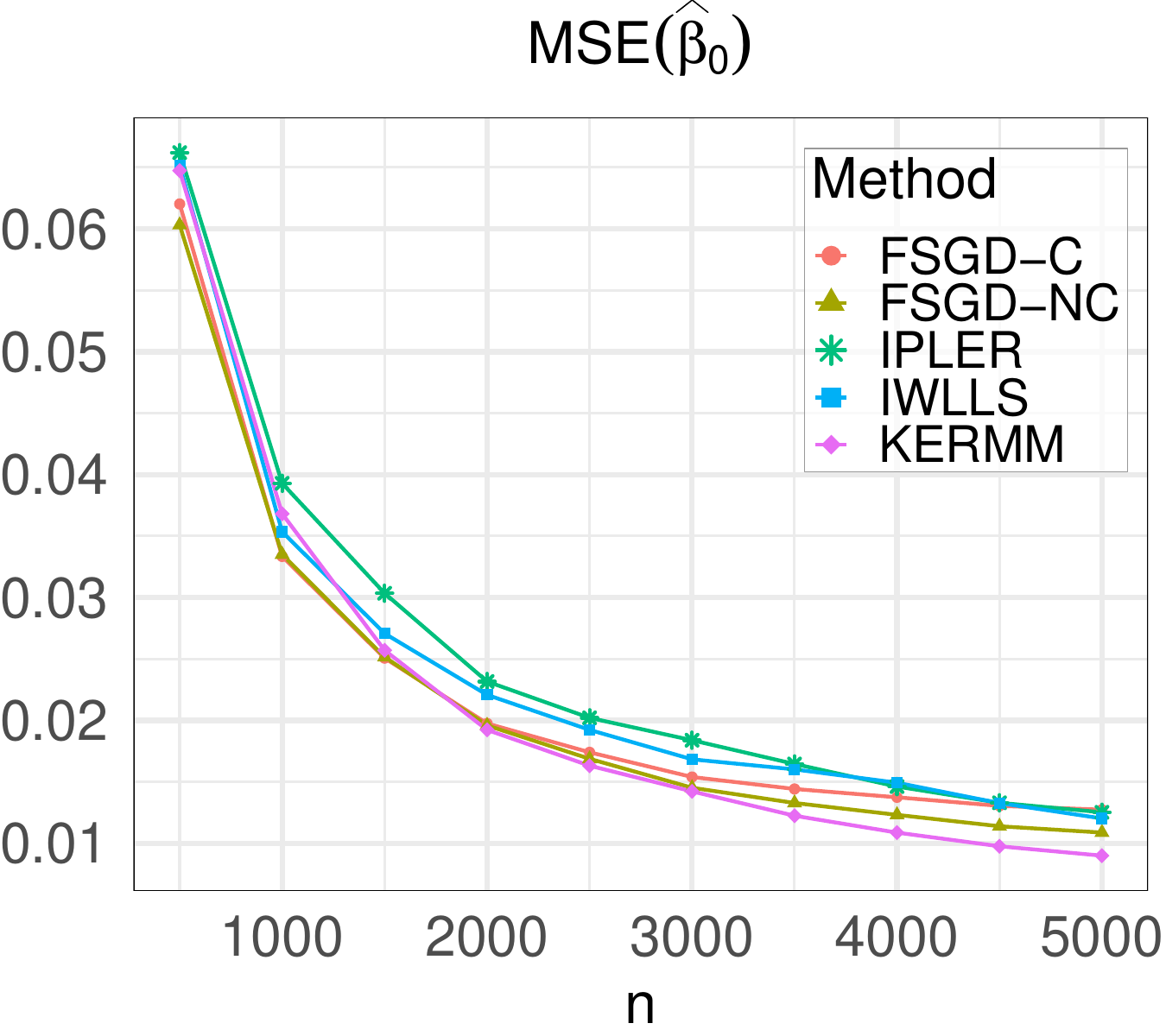}
\end{subfigure}
\begin{subfigure}{0.23\textwidth}
  \includegraphics[width=\textwidth]{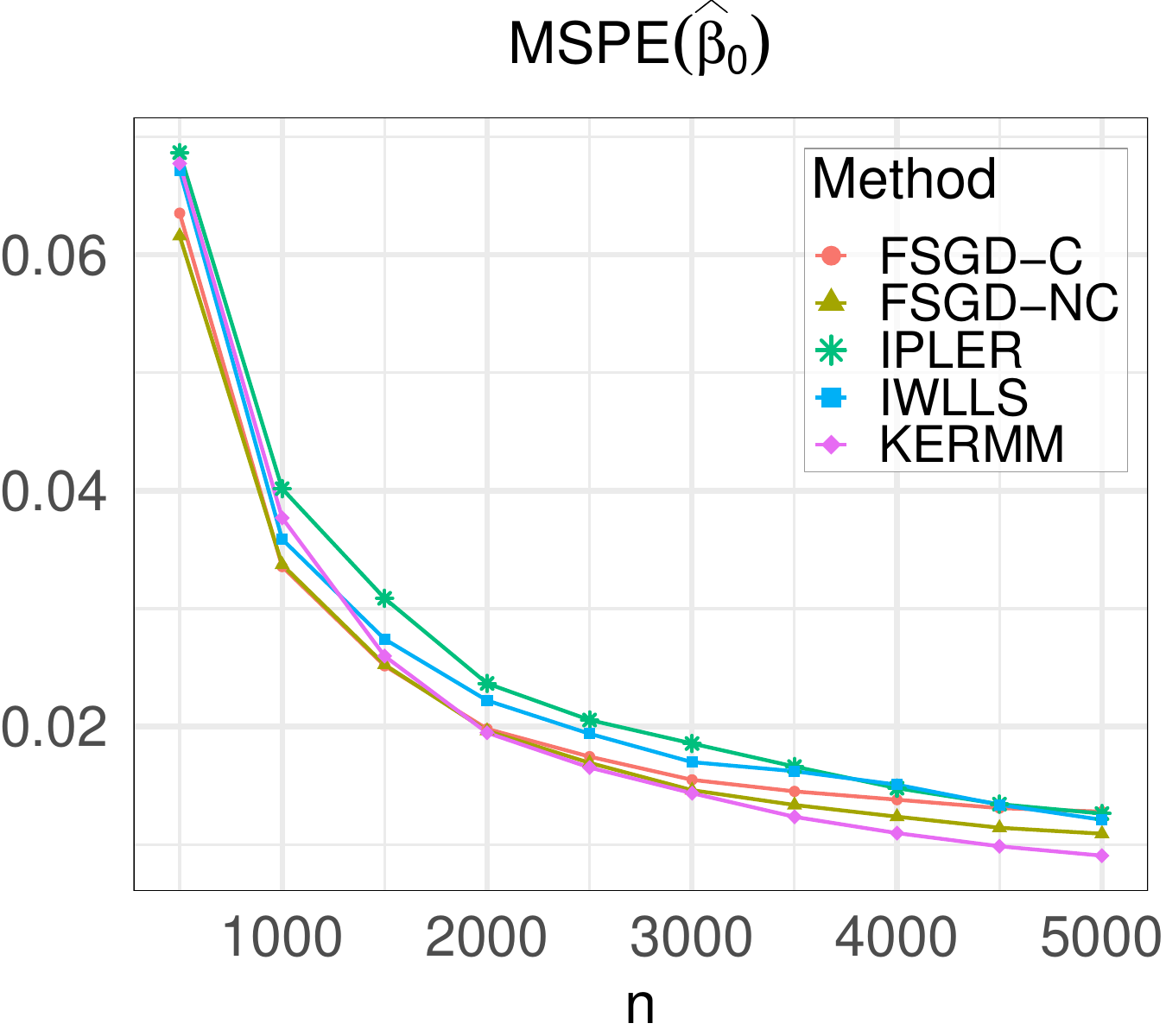}
\end{subfigure}
\begin{subfigure}{0.23\textwidth}
  \includegraphics[width=\textwidth]{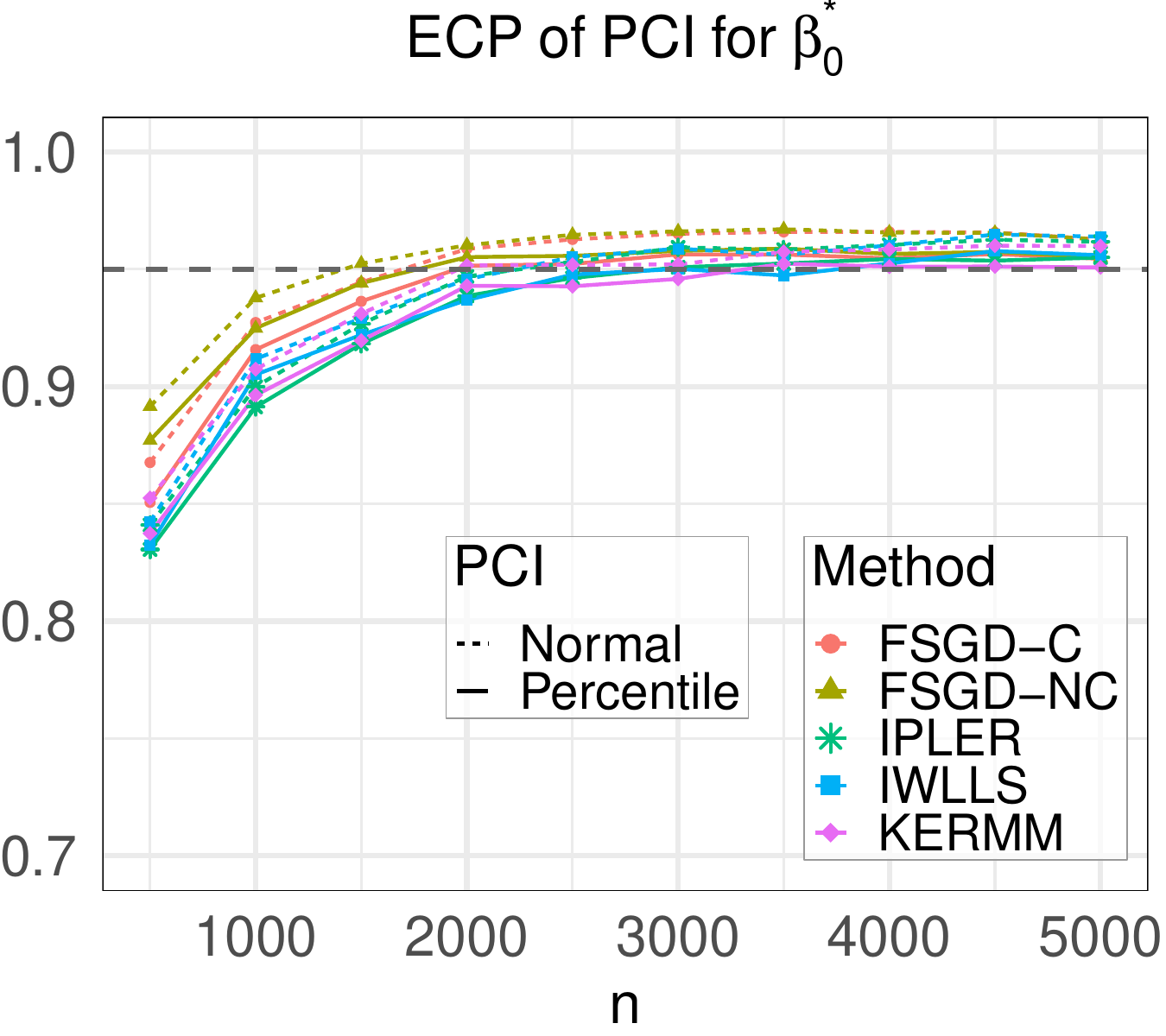}
\end{subfigure}
\begin{subfigure}{0.23\textwidth}
  \includegraphics[width=\textwidth]{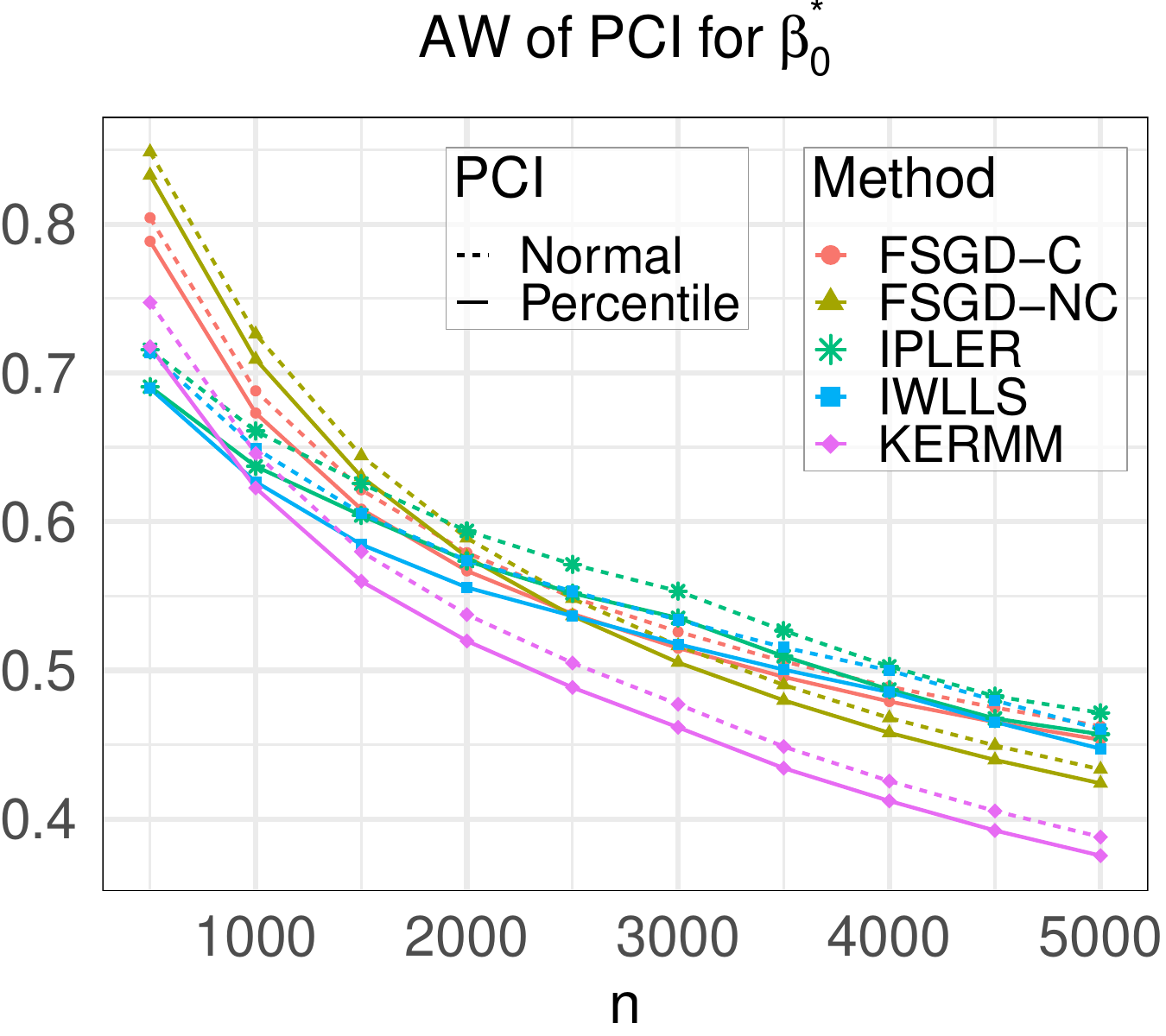}
\end{subfigure}

\begin{subfigure}{0.23\textwidth}
  \includegraphics[width=\textwidth]{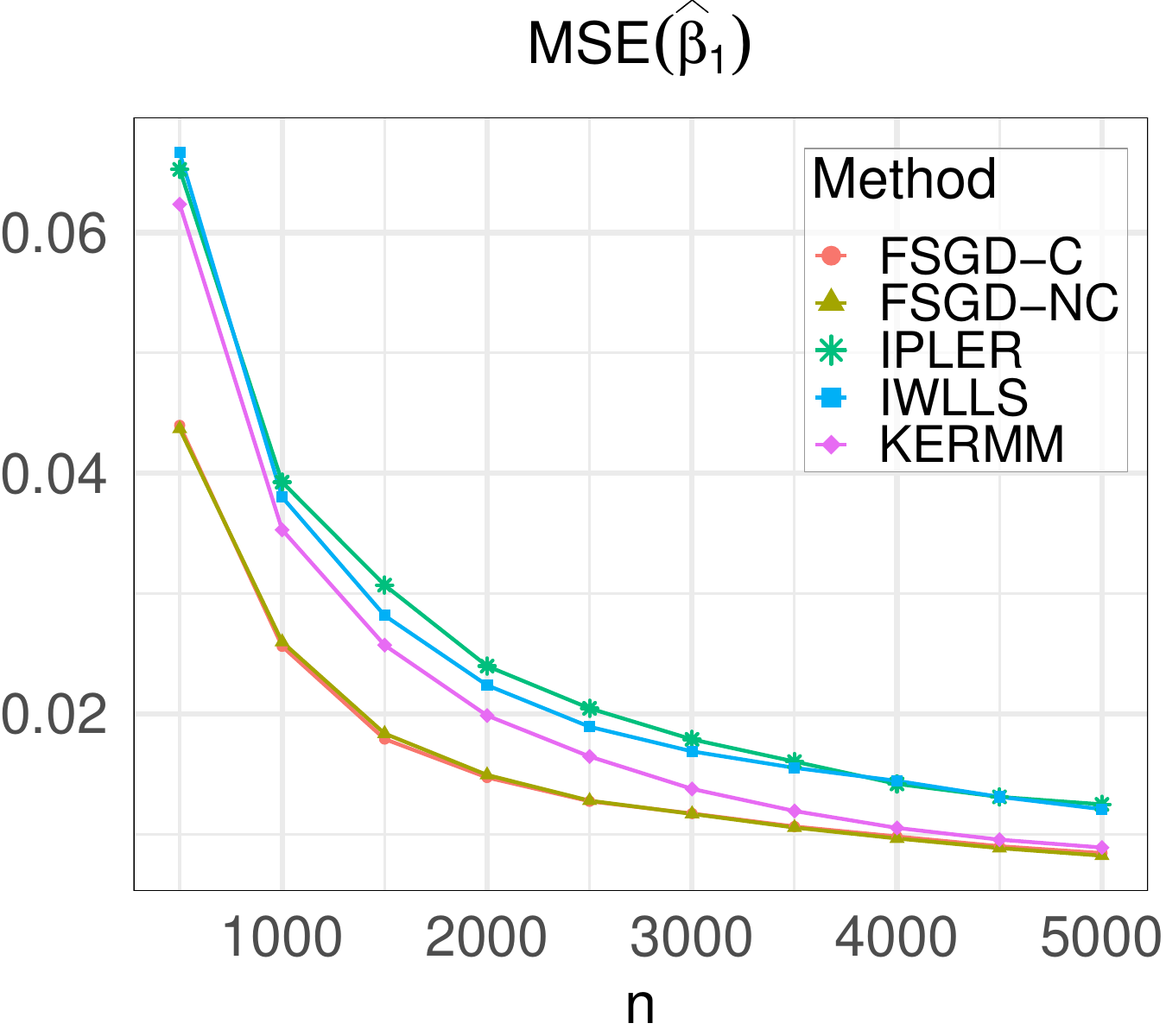}
\end{subfigure}
\begin{subfigure}{0.23\textwidth}
  \includegraphics[width=\textwidth]{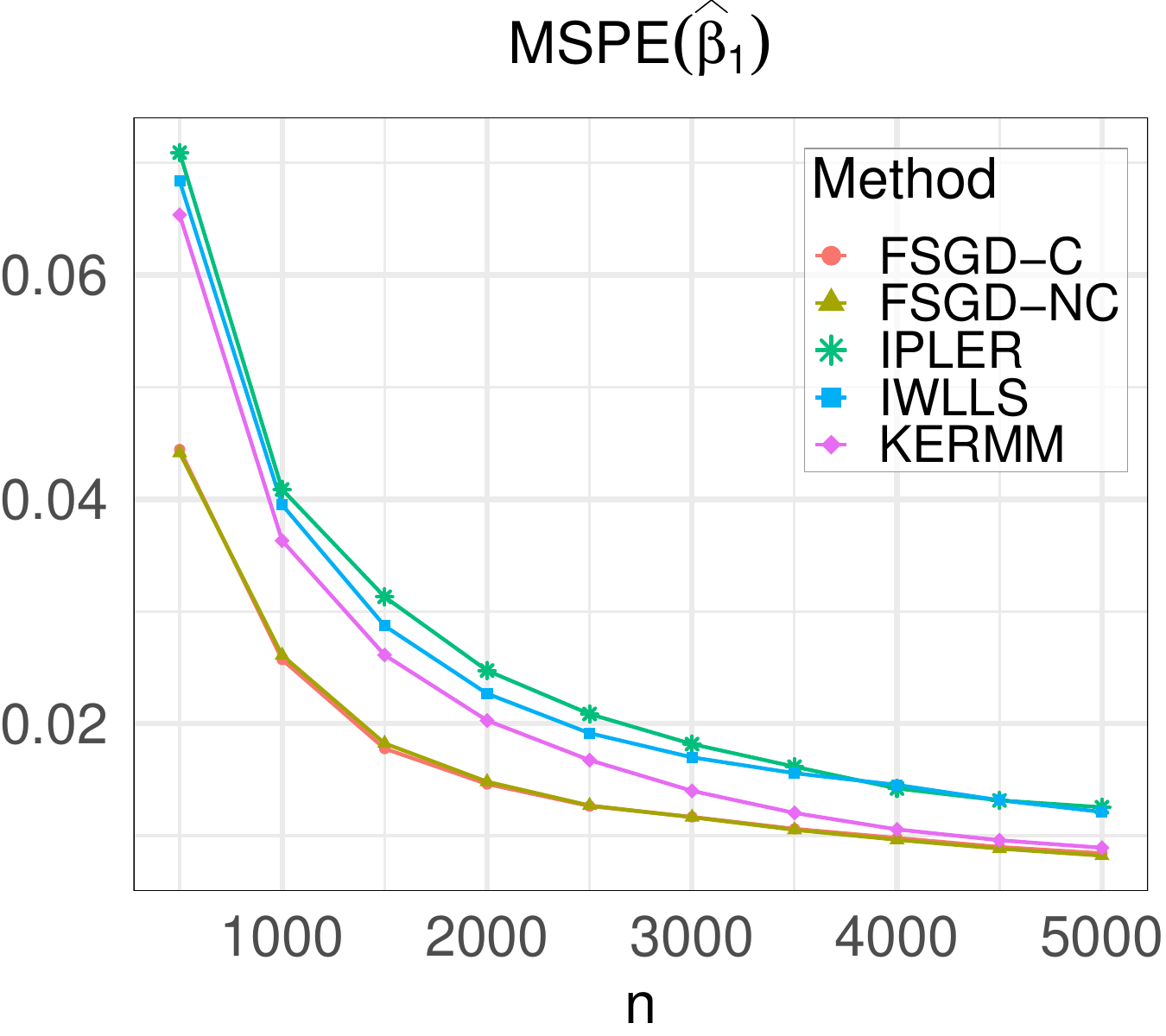}
\end{subfigure}
\begin{subfigure}{0.23\textwidth}
  \includegraphics[width=\textwidth]{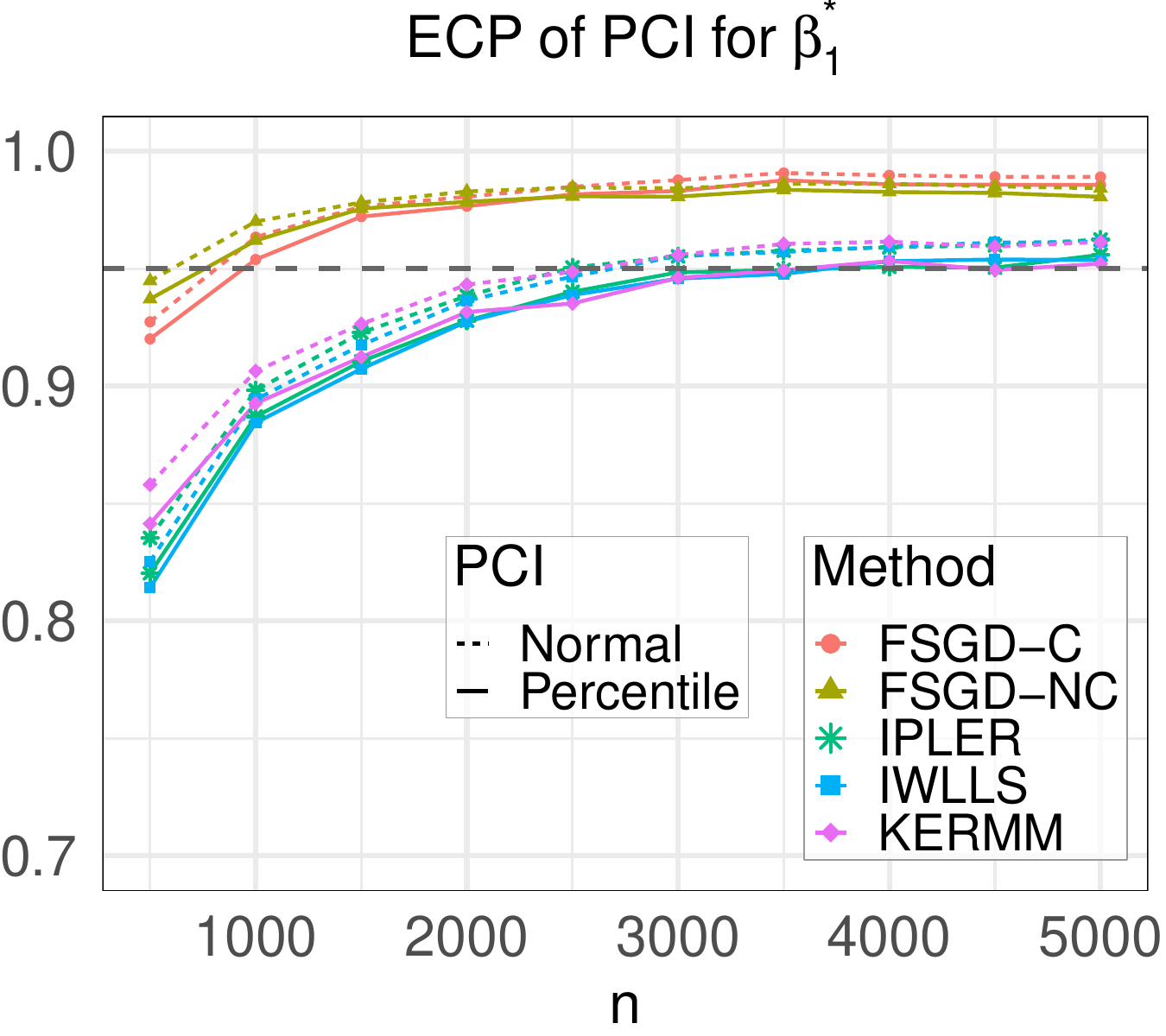}
\end{subfigure}
\begin{subfigure}{0.23\textwidth}
  \includegraphics[width=\textwidth]{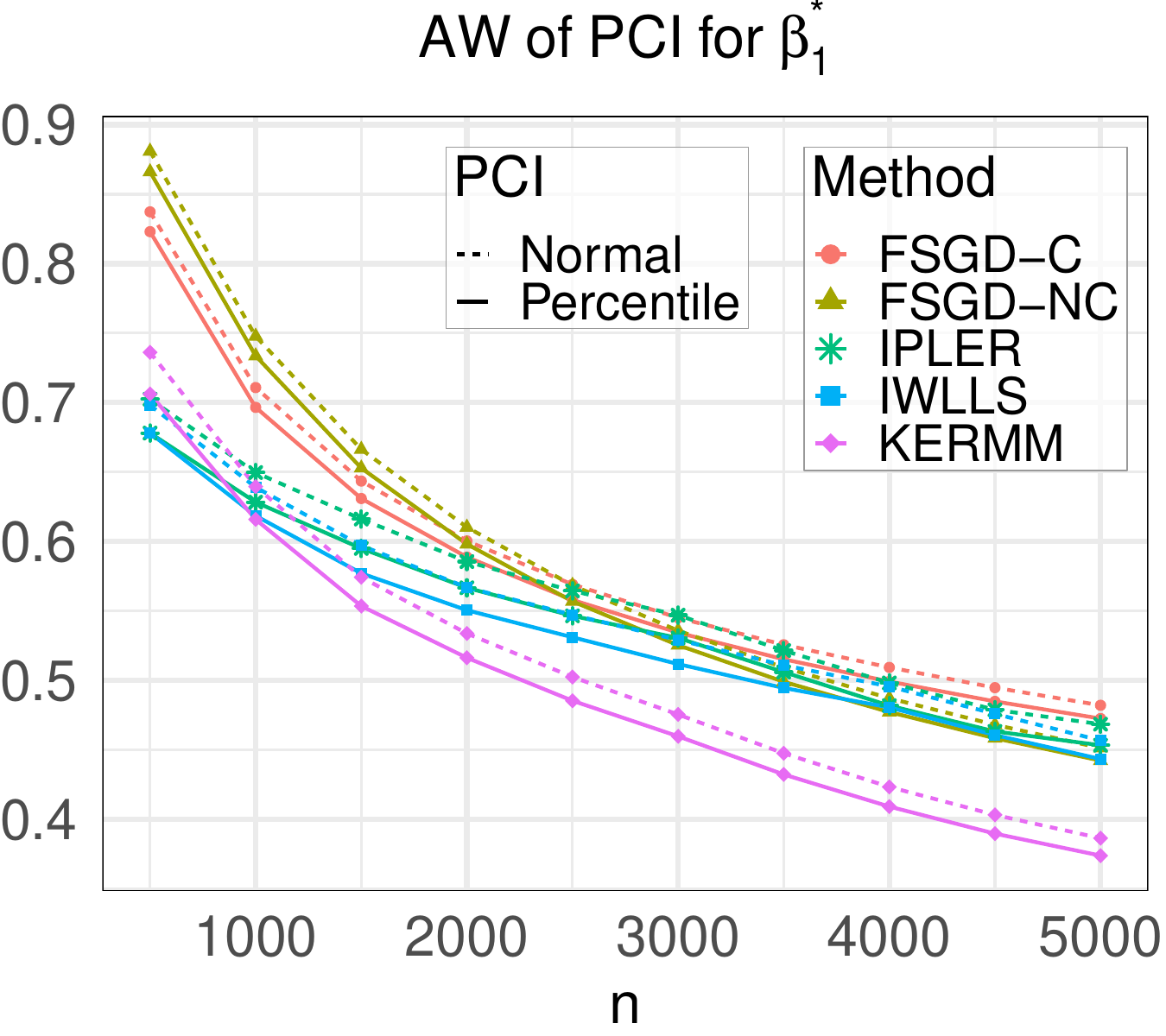}
\end{subfigure}

\begin{subfigure}{0.23\textwidth}
  \includegraphics[width=\textwidth]{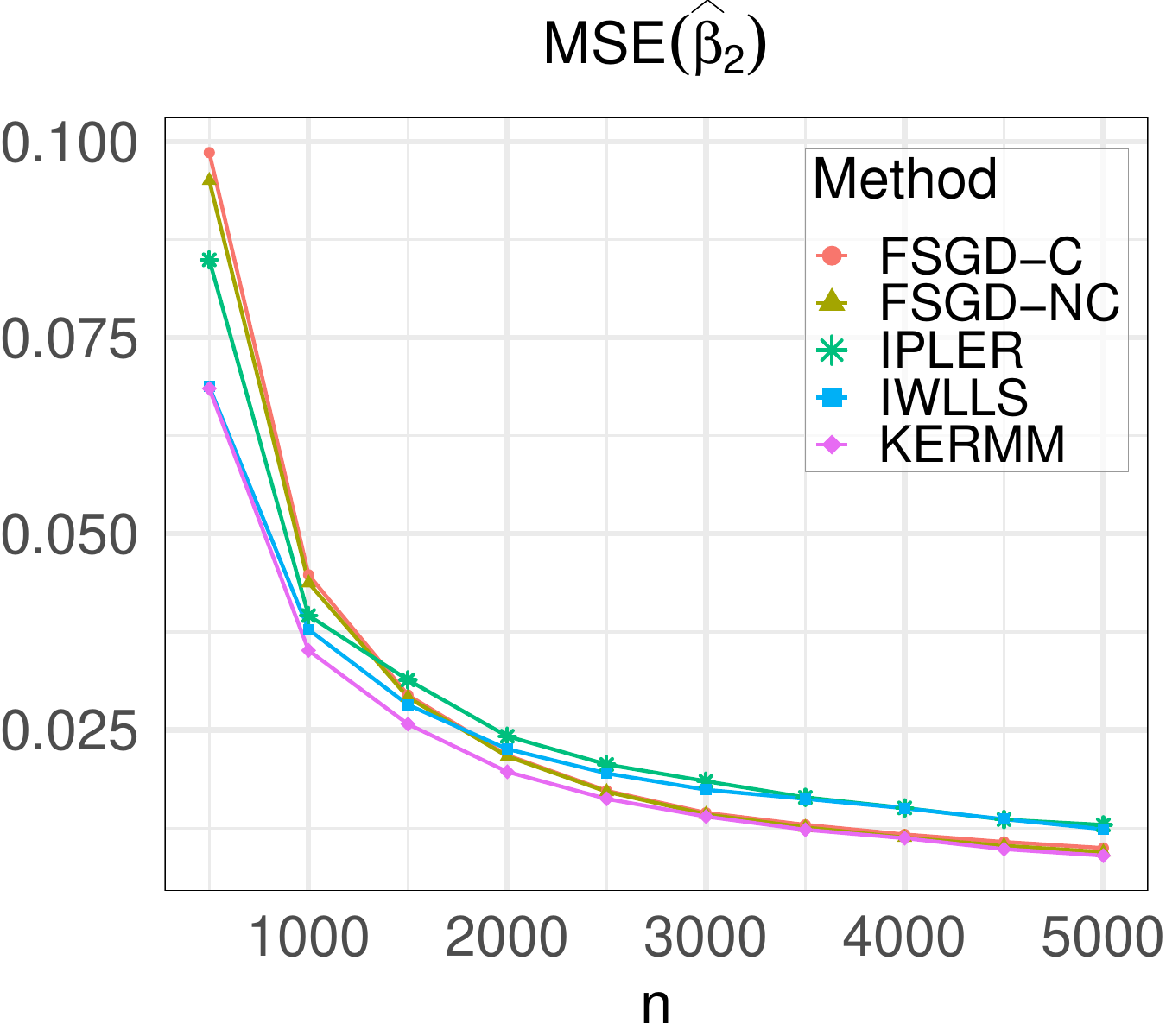}
\end{subfigure}
\begin{subfigure}{0.23\textwidth}
  \includegraphics[width=\textwidth]{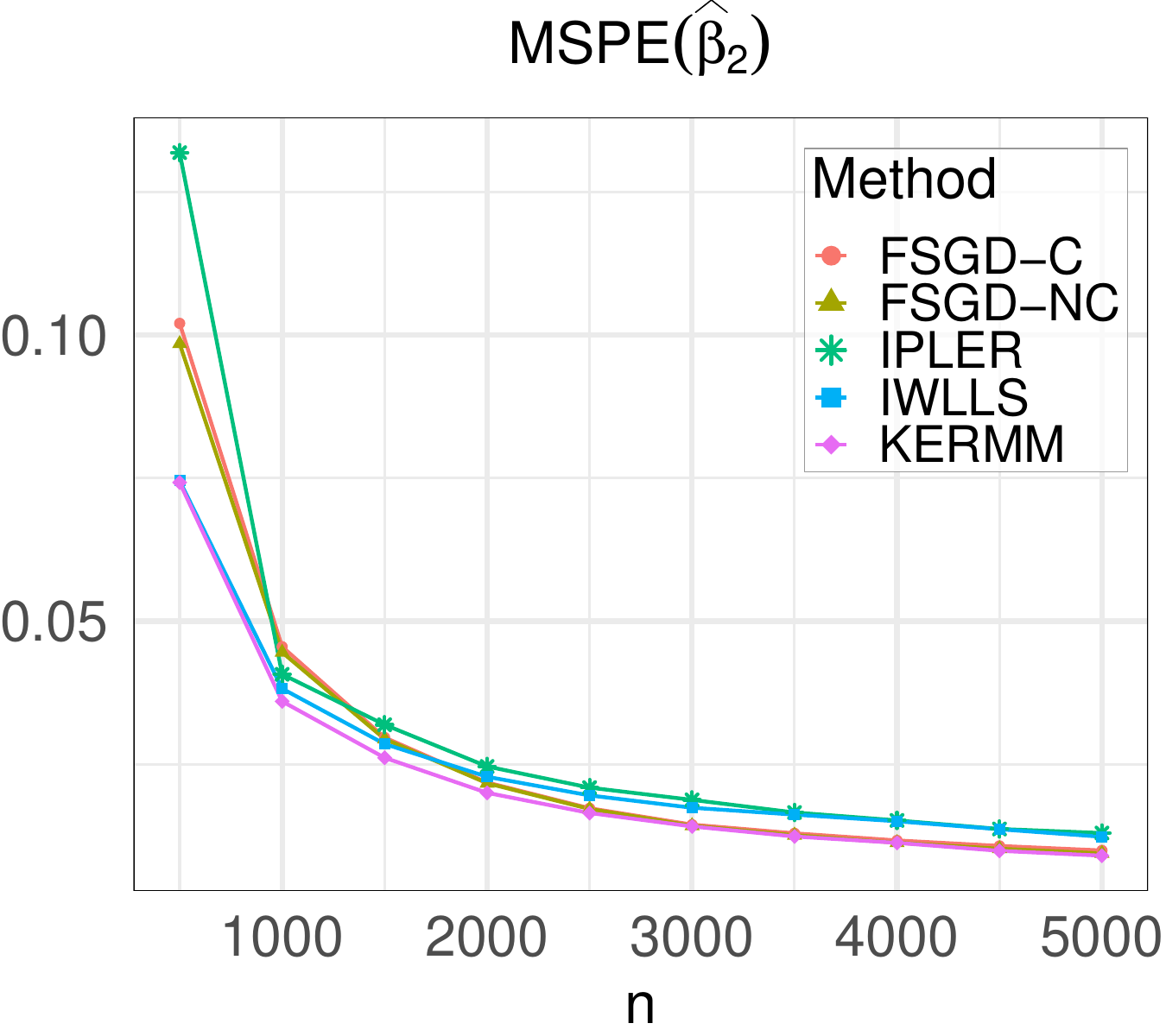}
\end{subfigure}
\begin{subfigure}{0.23\textwidth}
  \includegraphics[width=\textwidth]{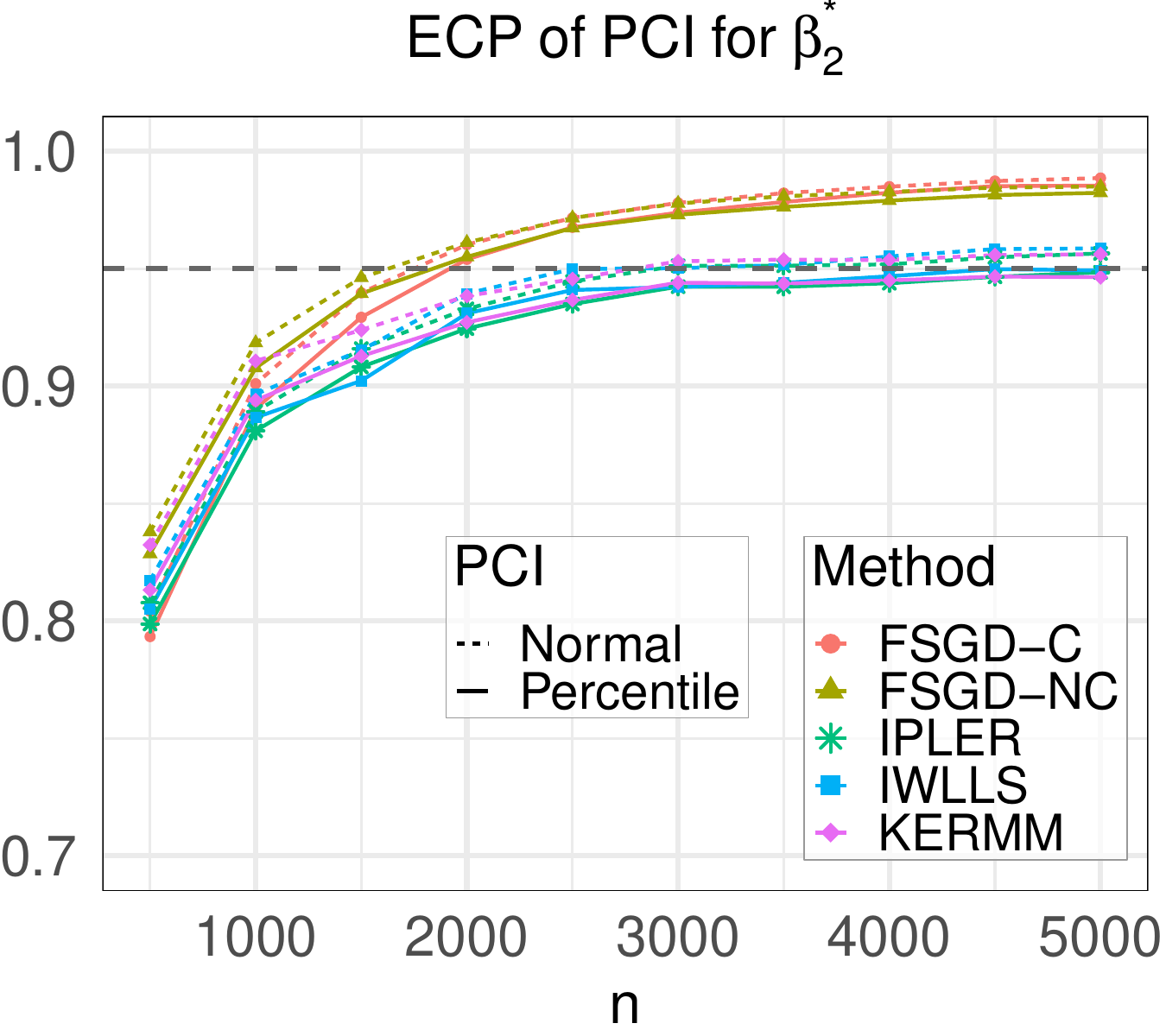}
\end{subfigure}
\begin{subfigure}{0.23\textwidth}
  \includegraphics[width=\textwidth]{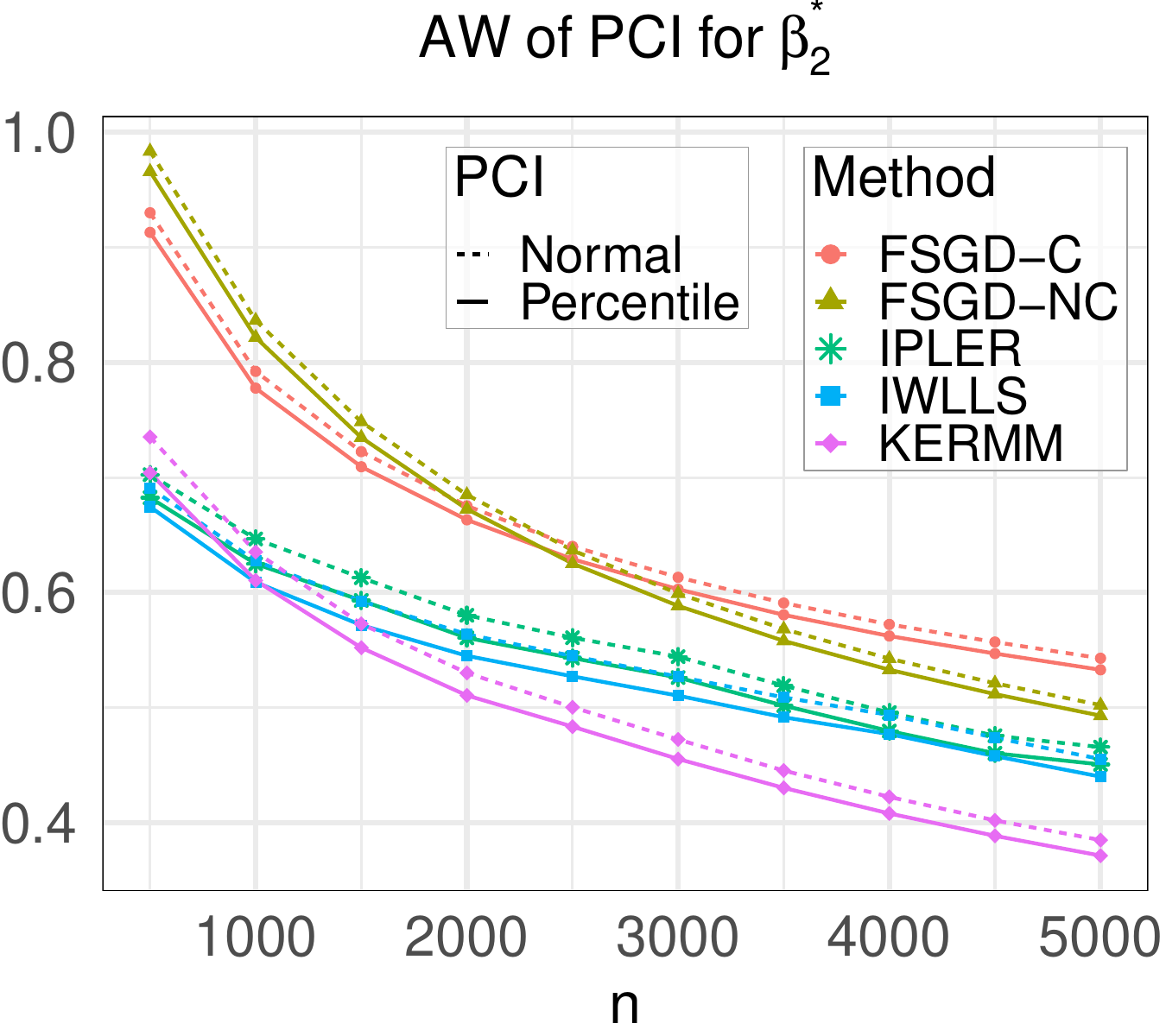}
\end{subfigure}
\caption{Simulation results under normal errors at $\tau=0.25$.}
\label{fig:normal_0.25}
\end{figure}

\begin{figure}[htbp]
\def\baselinestretch{1}
\centering
\begin{subfigure}{0.23\textwidth}
  \includegraphics[width=\textwidth]{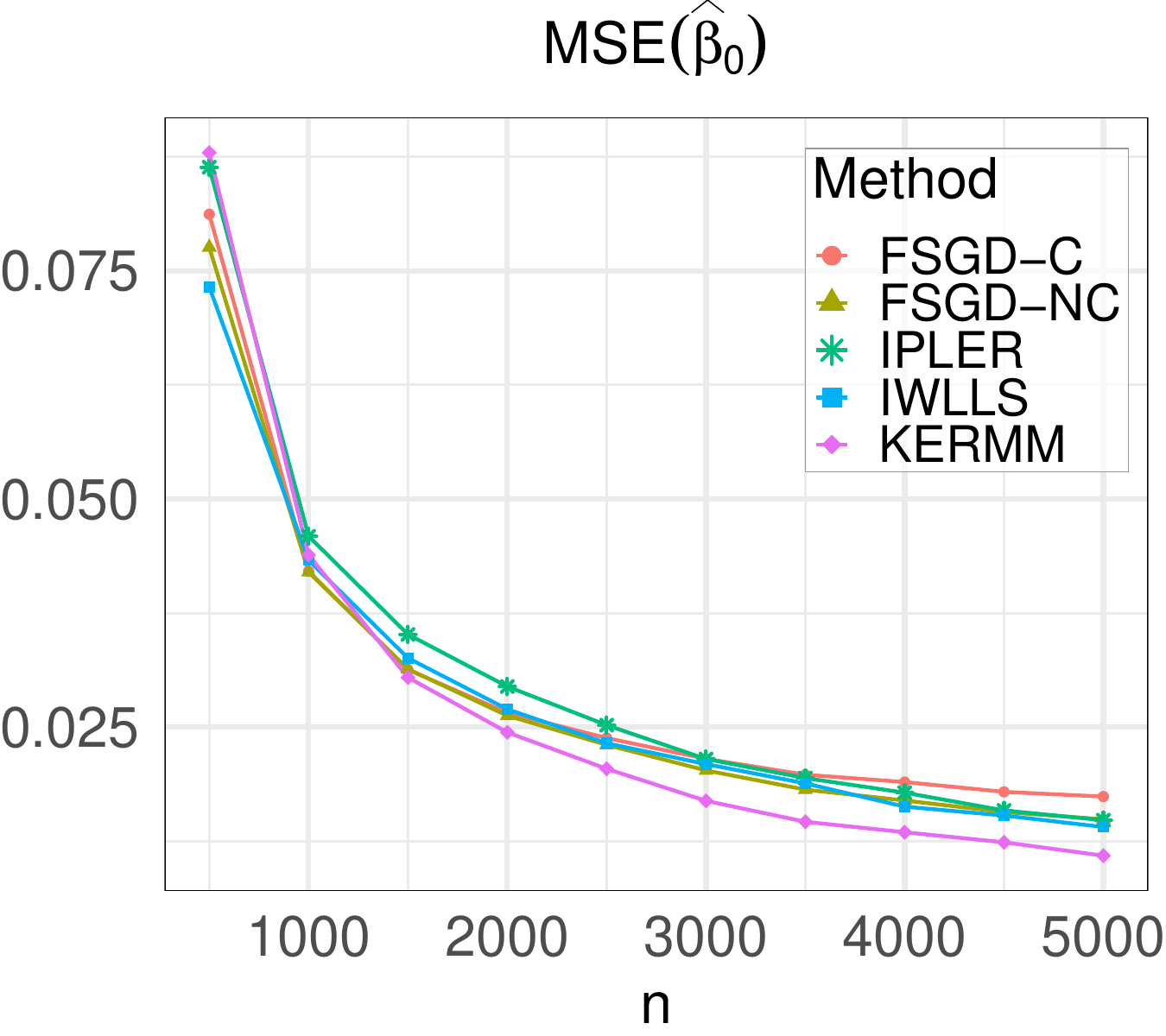}
\end{subfigure}
\begin{subfigure}{0.23\textwidth}
  \includegraphics[width=\textwidth]{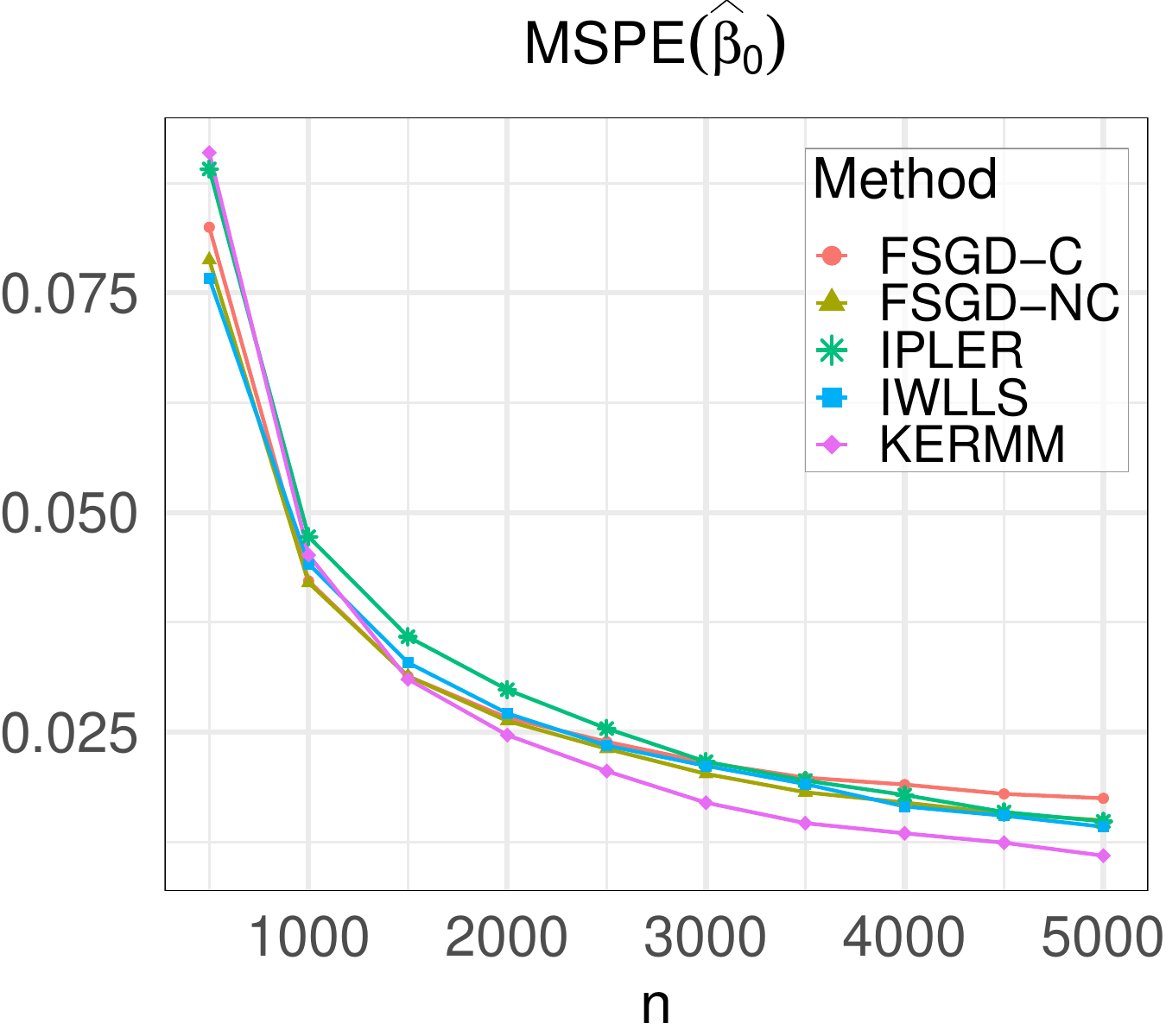}
\end{subfigure}
\begin{subfigure}{0.23\textwidth}
  \includegraphics[width=\textwidth]{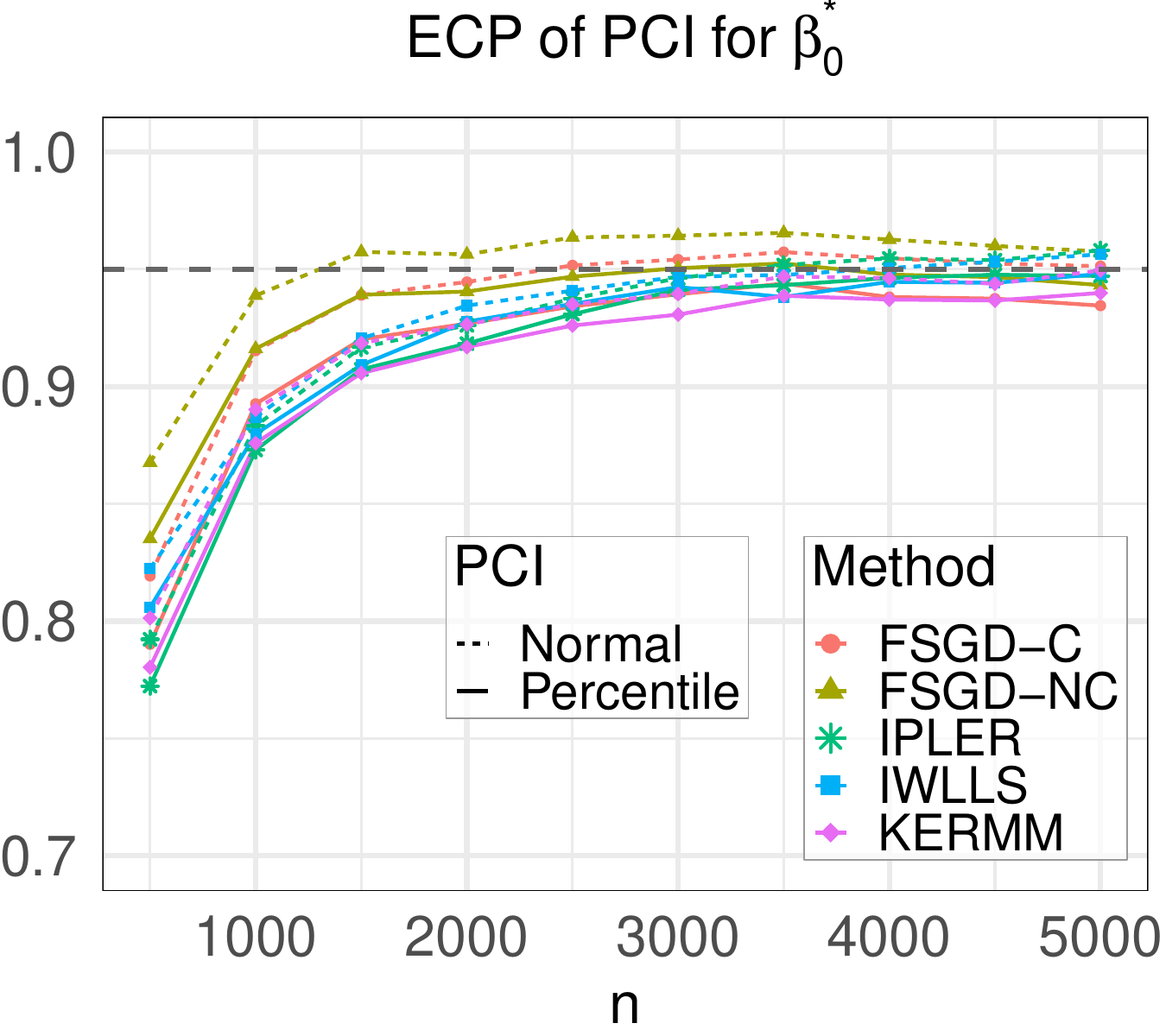}
\end{subfigure}
\begin{subfigure}{0.23\textwidth}
  \includegraphics[width=\textwidth]{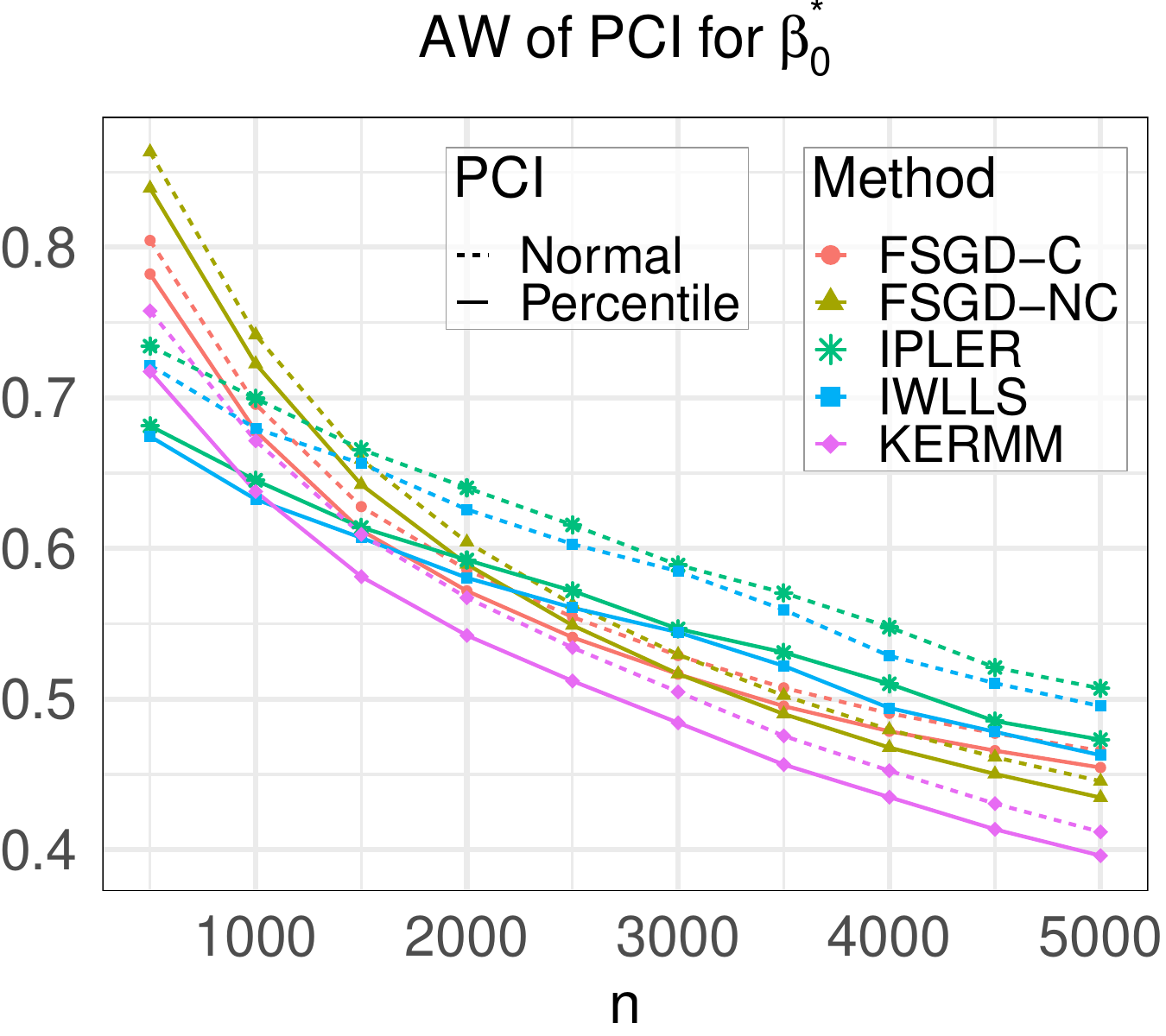}
\end{subfigure}

\begin{subfigure}{0.23\textwidth}
  \includegraphics[width=\textwidth]{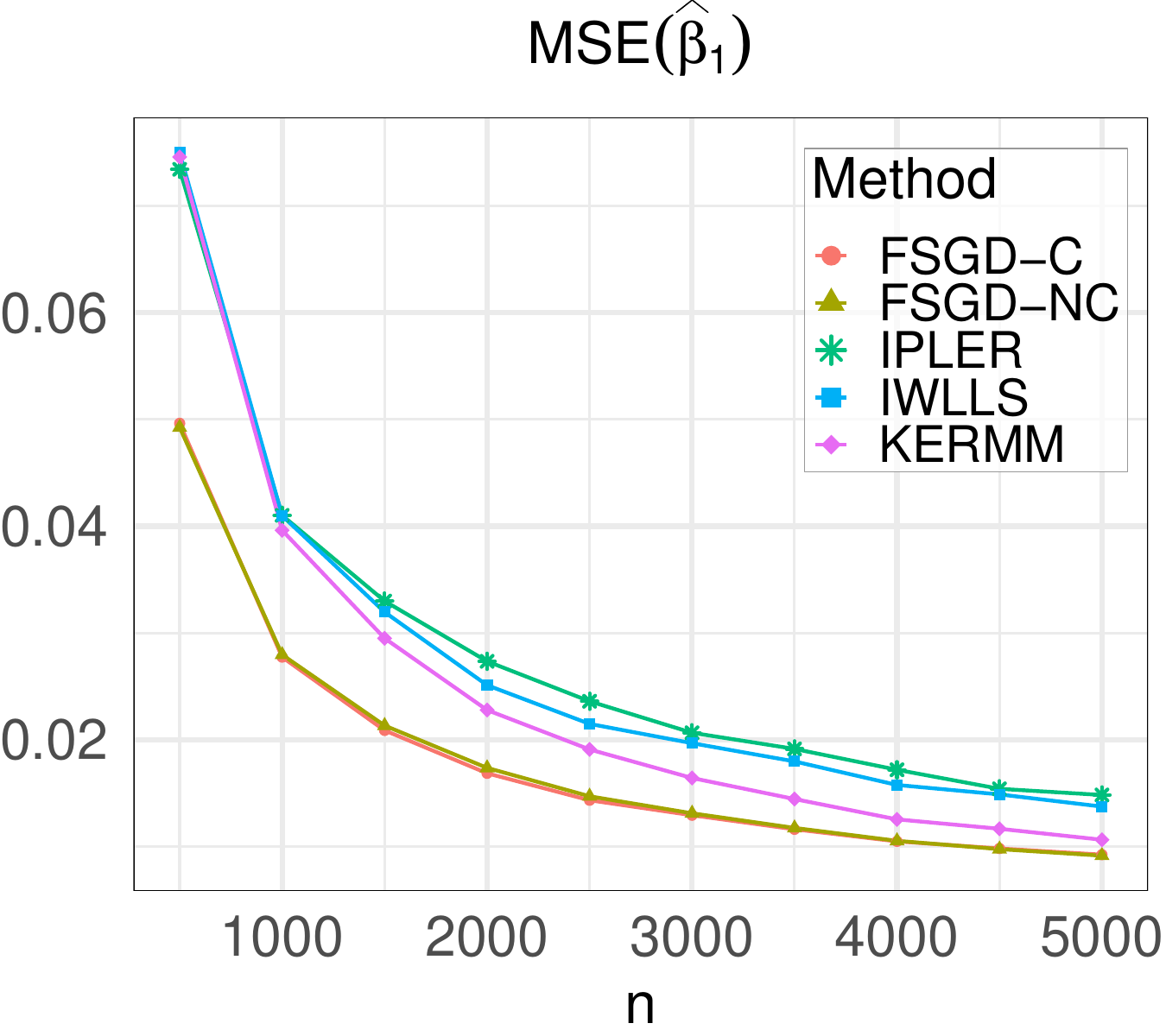}
\end{subfigure}
\begin{subfigure}{0.23\textwidth}
  \includegraphics[width=\textwidth]{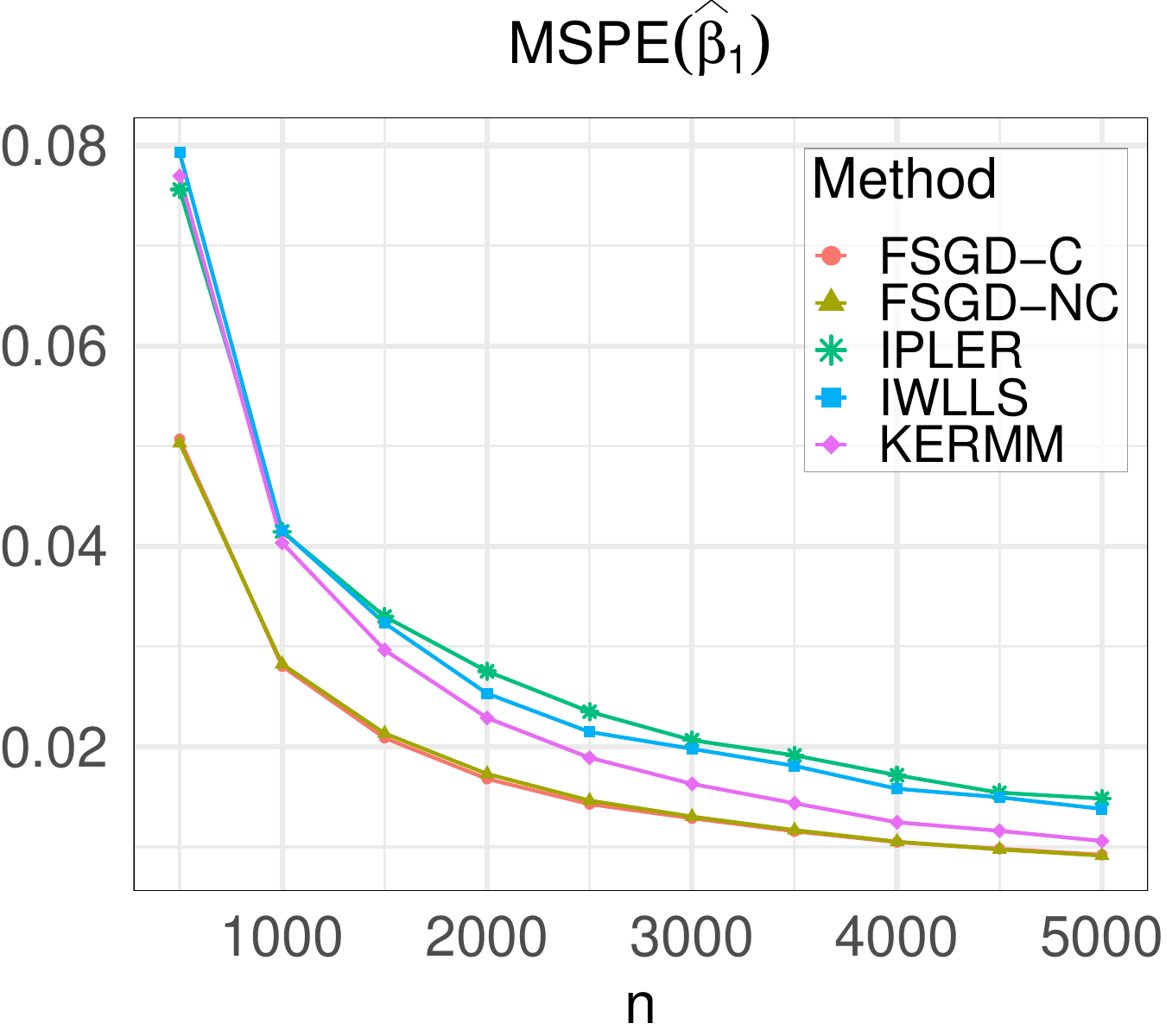}
\end{subfigure}
\begin{subfigure}{0.23\textwidth}
  \includegraphics[width=\textwidth]{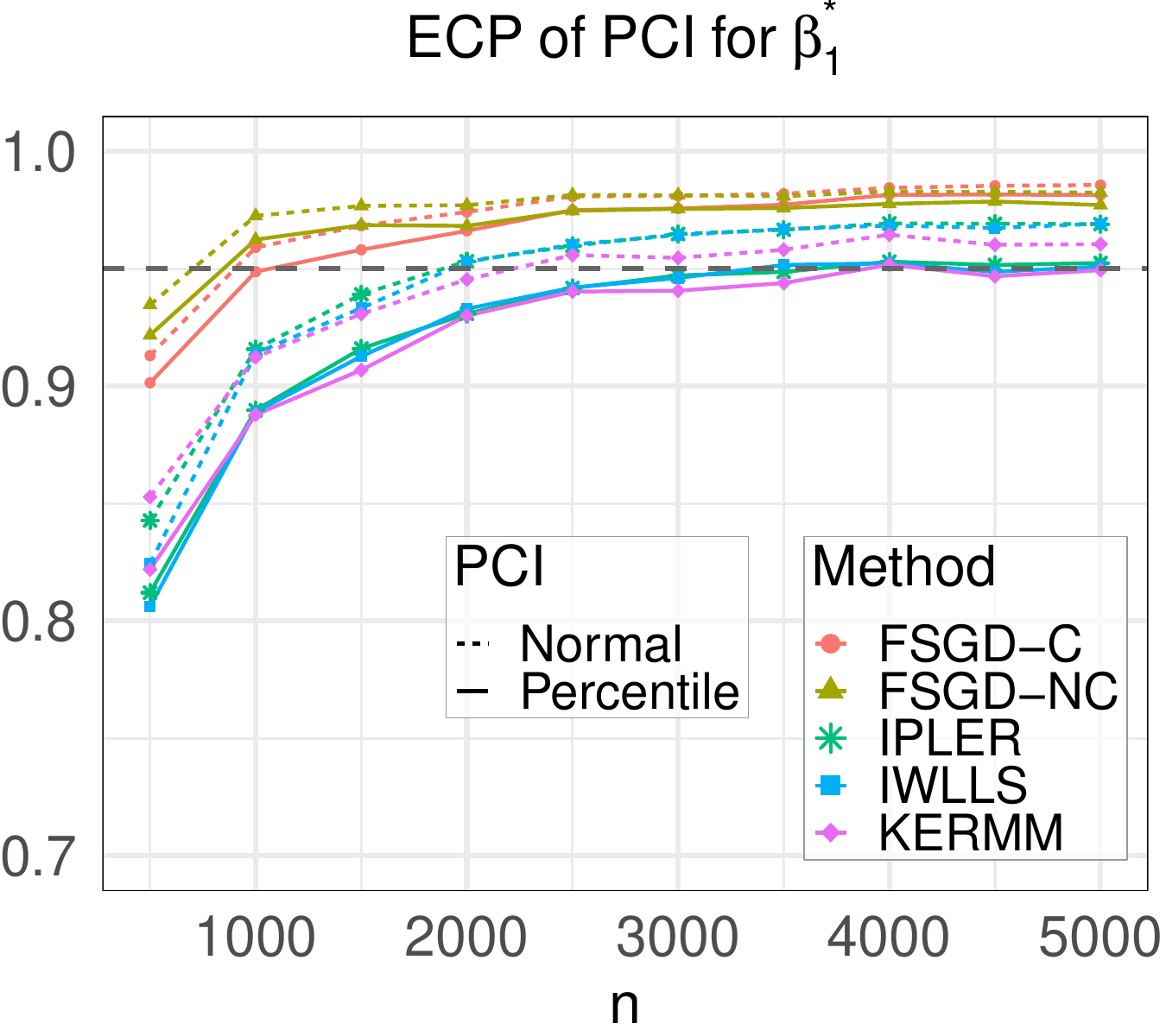}
\end{subfigure}
\begin{subfigure}{0.23\textwidth}
  \includegraphics[width=\textwidth]{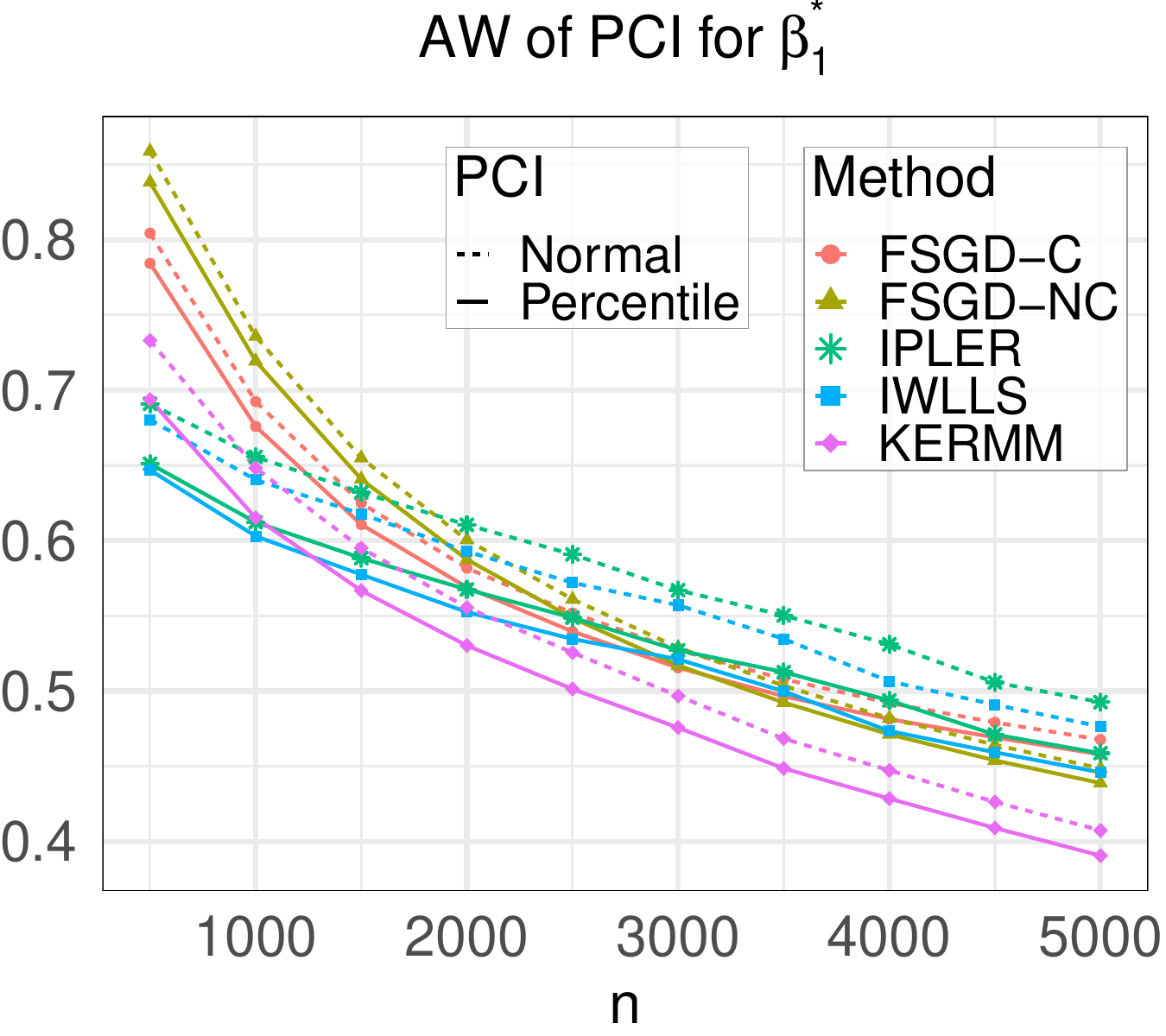}
\end{subfigure}

\begin{subfigure}{0.23\textwidth}
  \includegraphics[width=\textwidth]{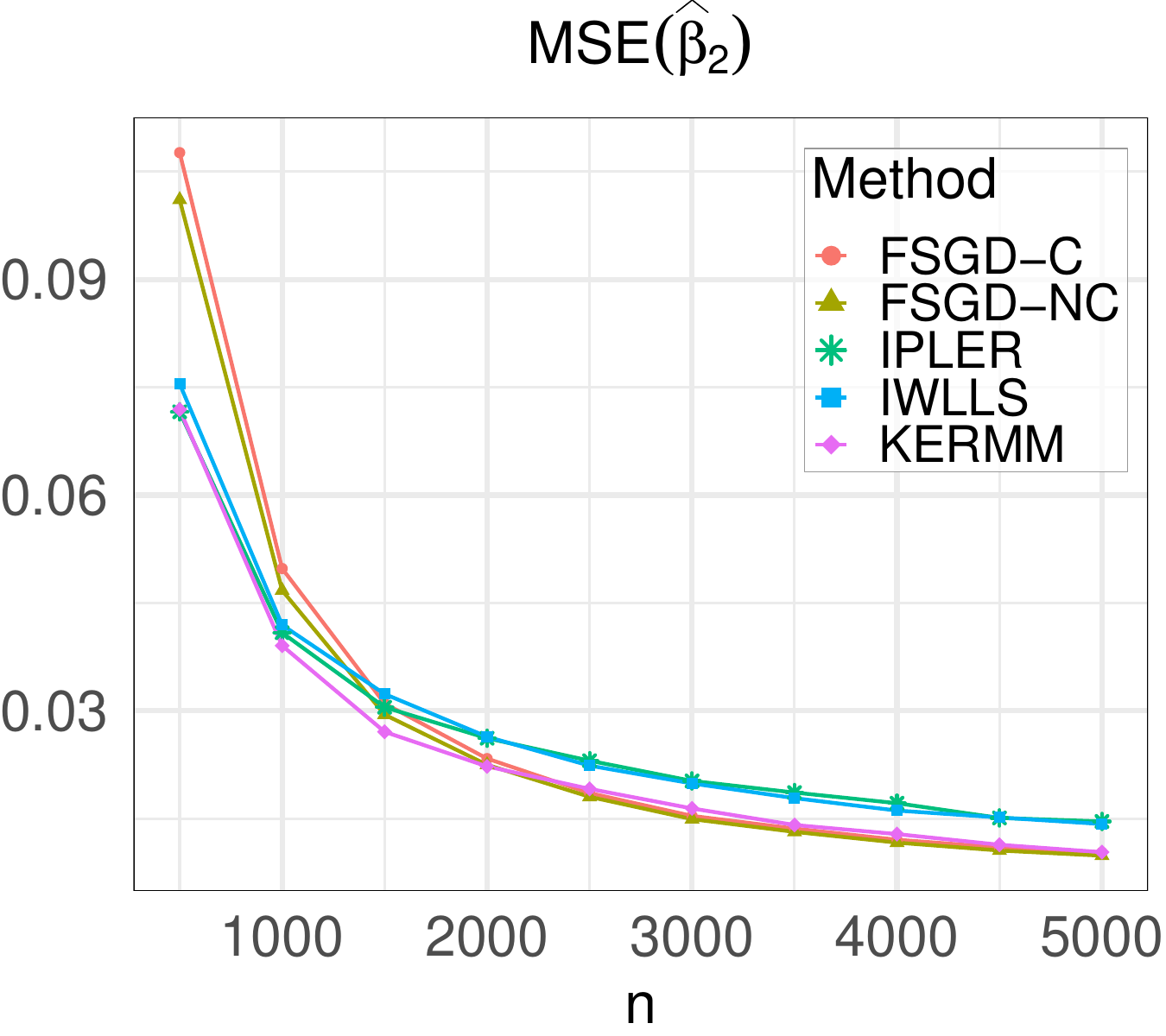}
\end{subfigure}
\begin{subfigure}{0.23\textwidth}
  \includegraphics[width=\textwidth]{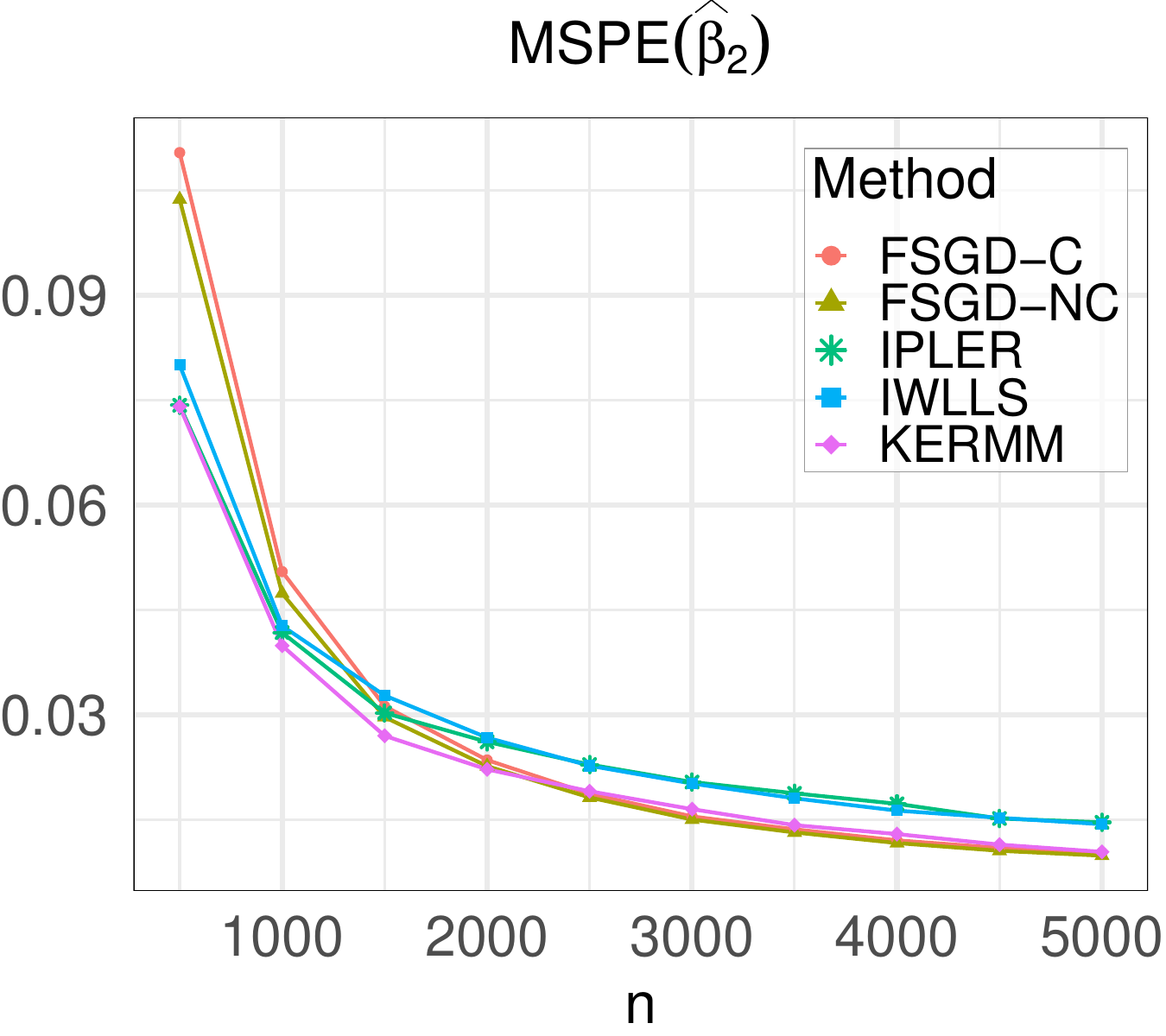}
\end{subfigure}
\begin{subfigure}{0.23\textwidth}
  \includegraphics[width=\textwidth]{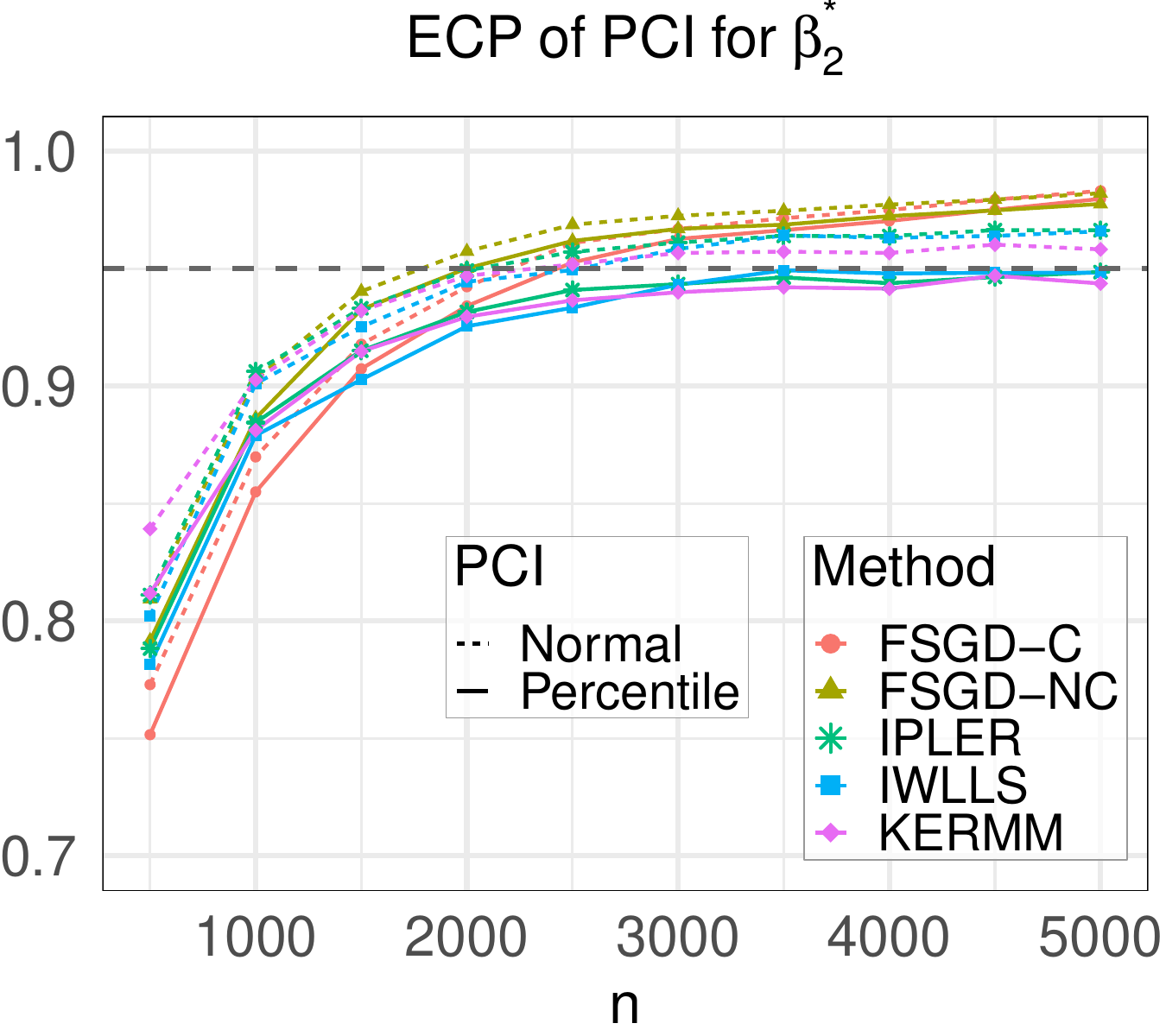}
\end{subfigure}
\begin{subfigure}{0.23\textwidth}
  \includegraphics[width=\textwidth]{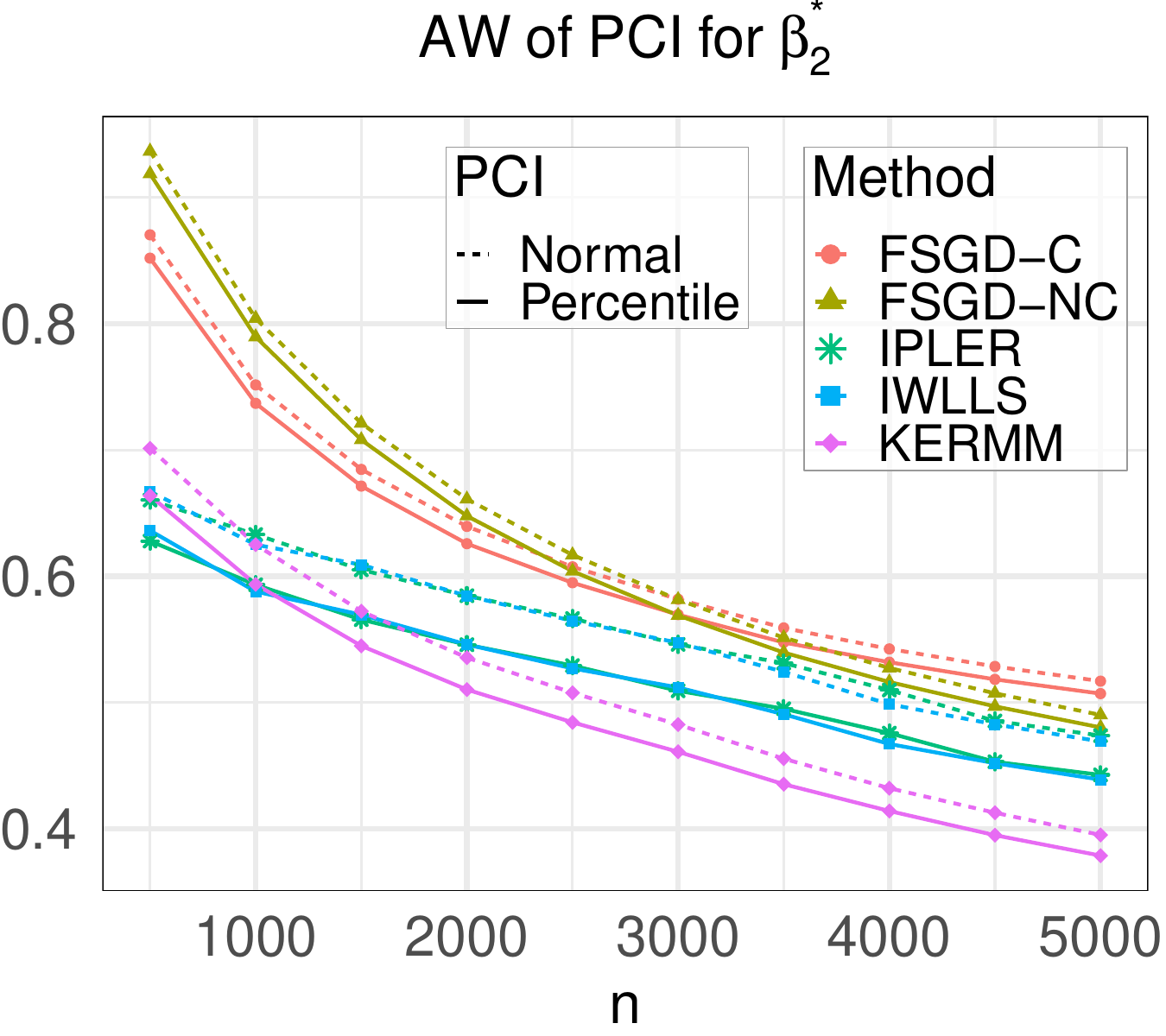}
\end{subfigure}
\caption{Simulation results under mixture errors at $\tau=0.25$.}
\label{fig:mixture_0.25}
\end{figure}

\begin{figure}[htbp]
\def\baselinestretch{1}
\centering
\begin{subfigure}{0.23\textwidth}
  \includegraphics[width=\textwidth]{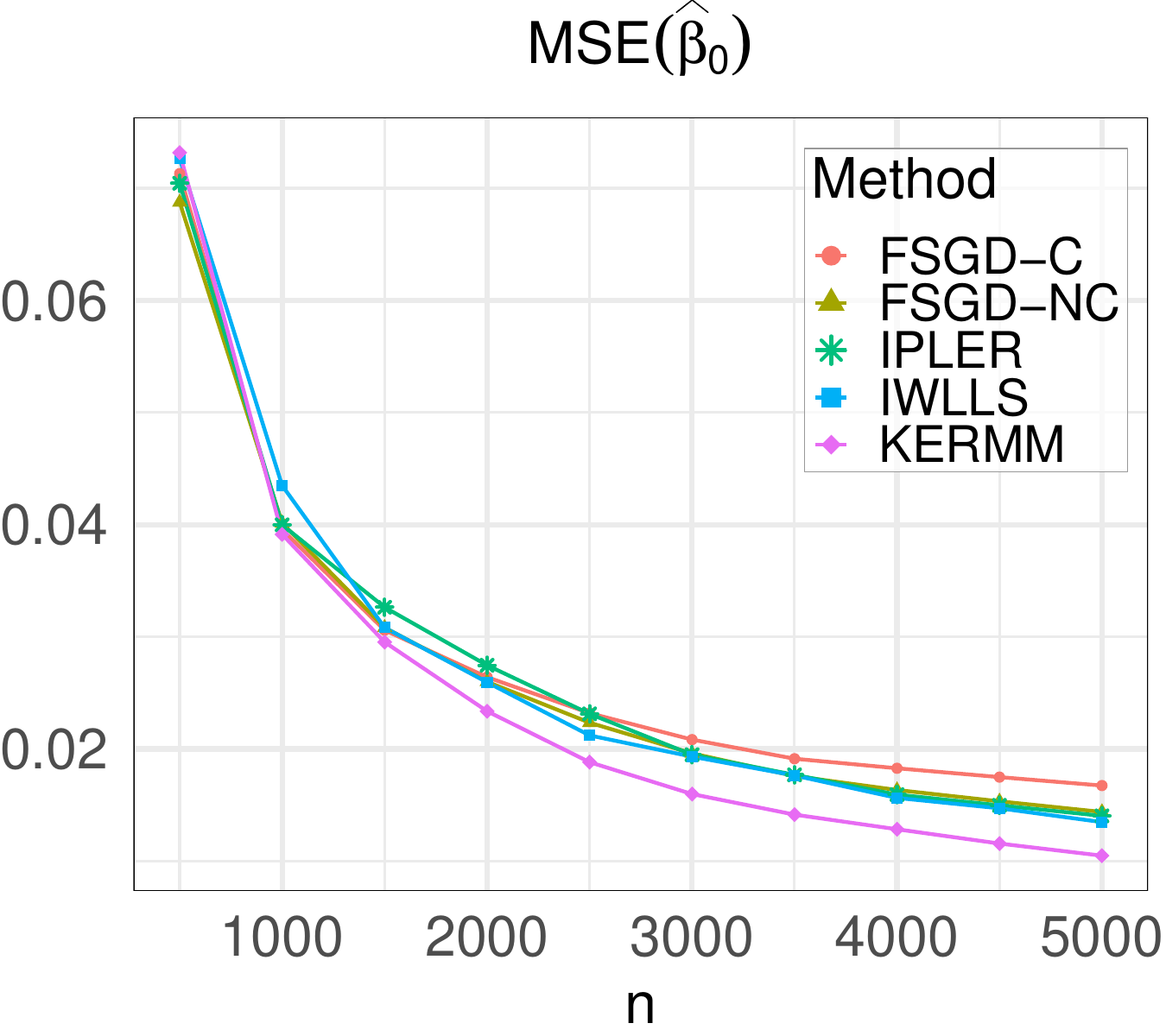}
\end{subfigure}
\begin{subfigure}{0.23\textwidth}
  \includegraphics[width=\textwidth]{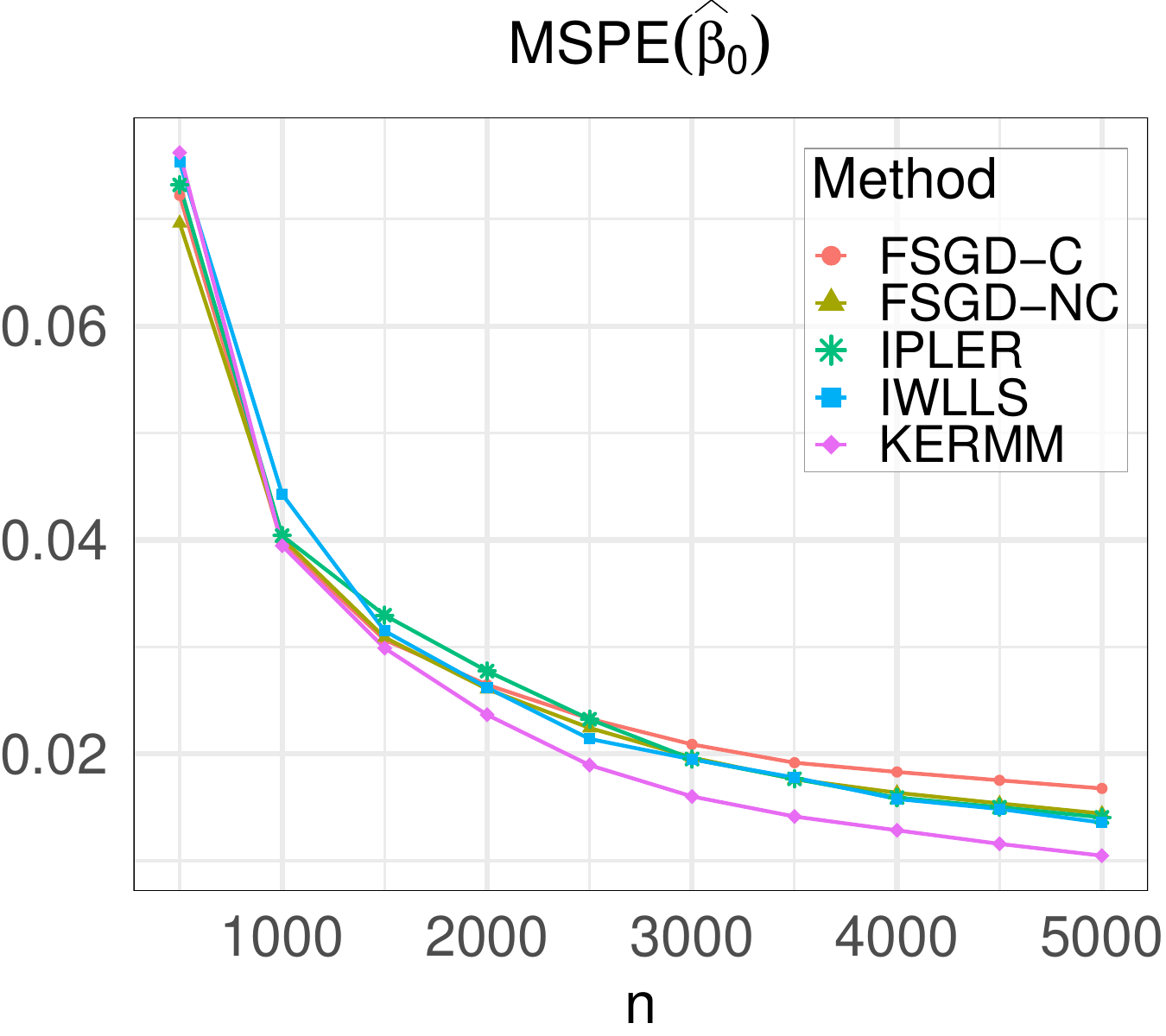}
\end{subfigure}
\begin{subfigure}{0.23\textwidth}
  \includegraphics[width=\textwidth]{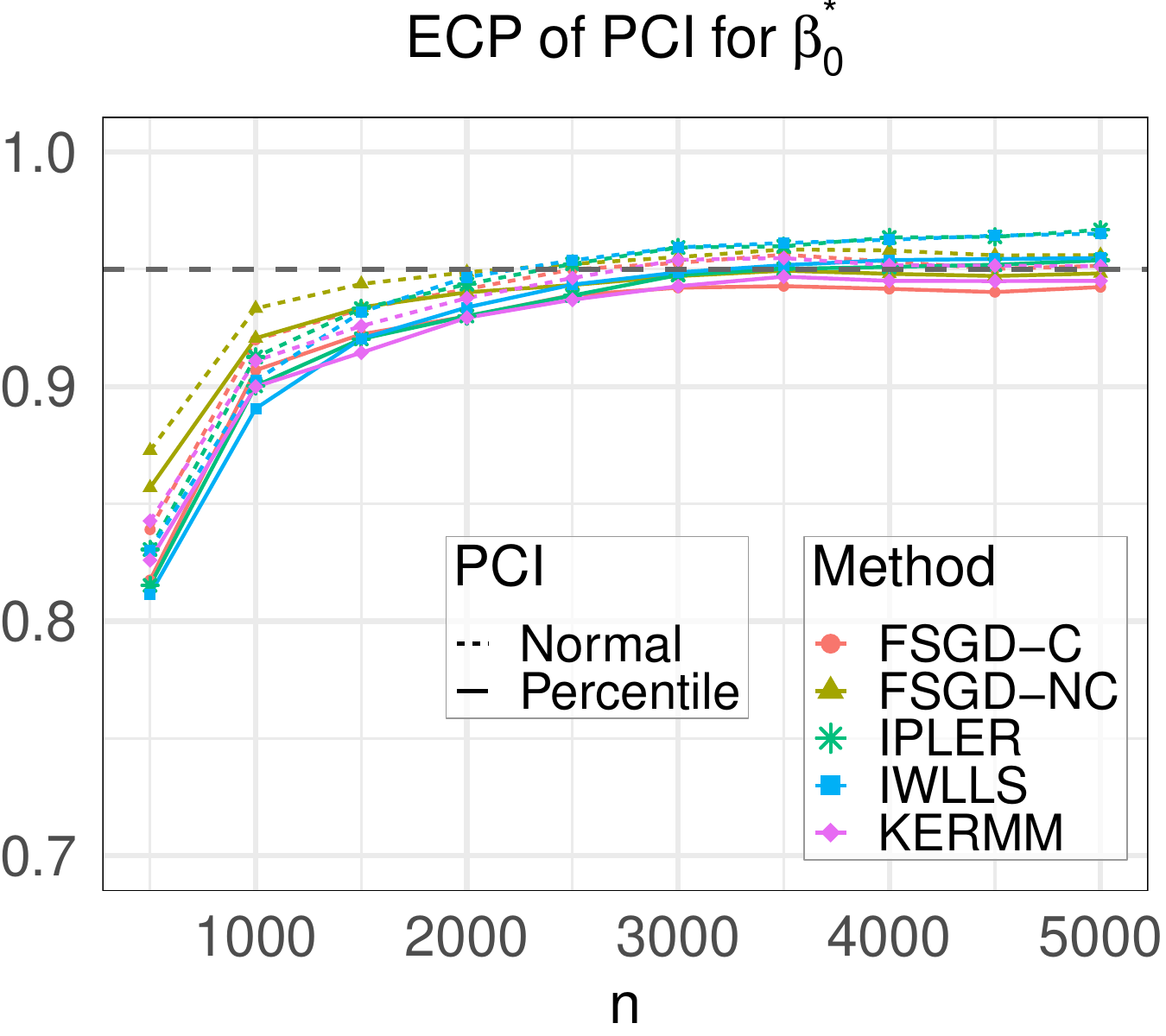}
\end{subfigure}
\begin{subfigure}{0.23\textwidth}
  \includegraphics[width=\textwidth]{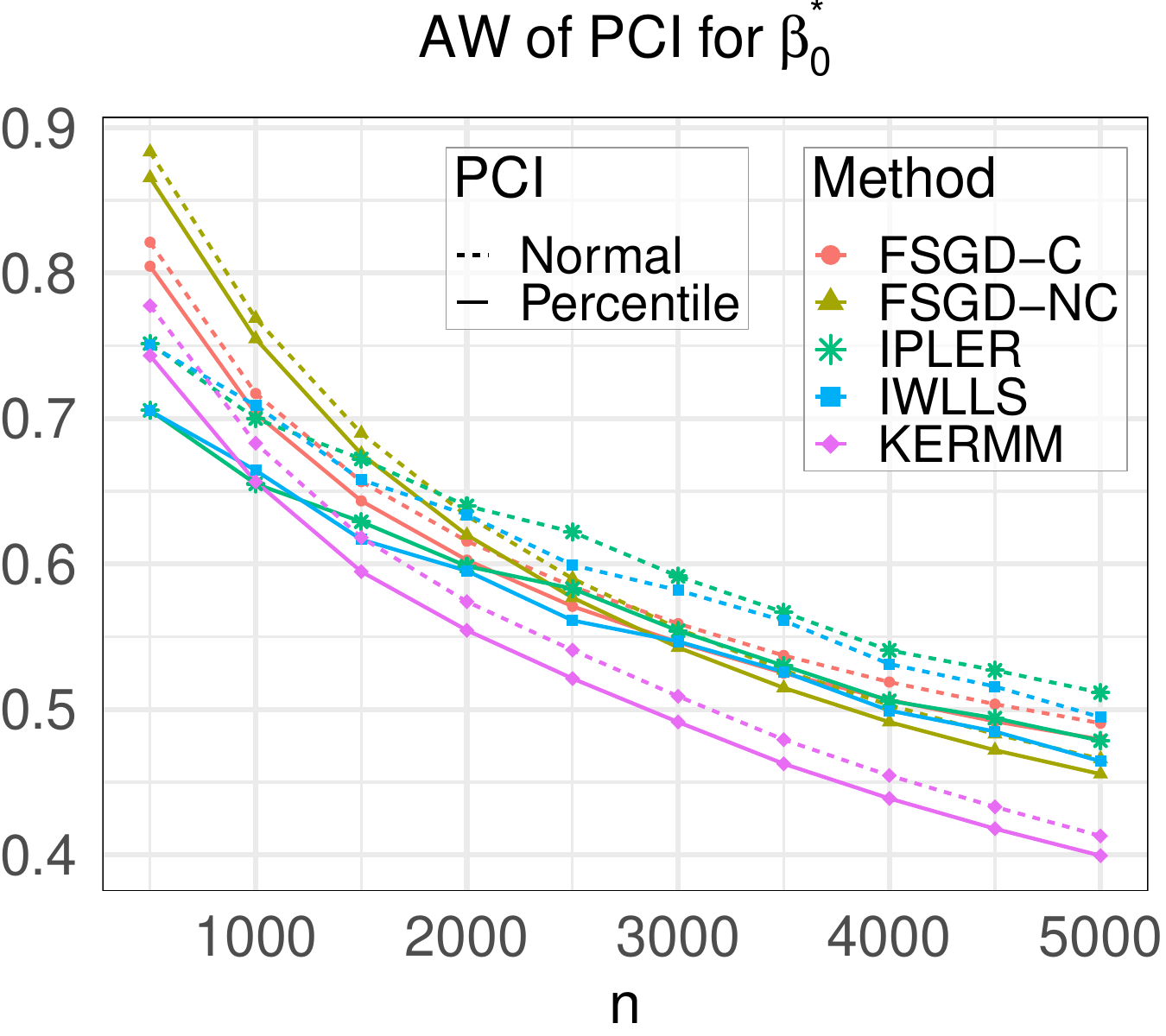}
\end{subfigure}

\begin{subfigure}{0.23\textwidth}
  \includegraphics[width=\textwidth]{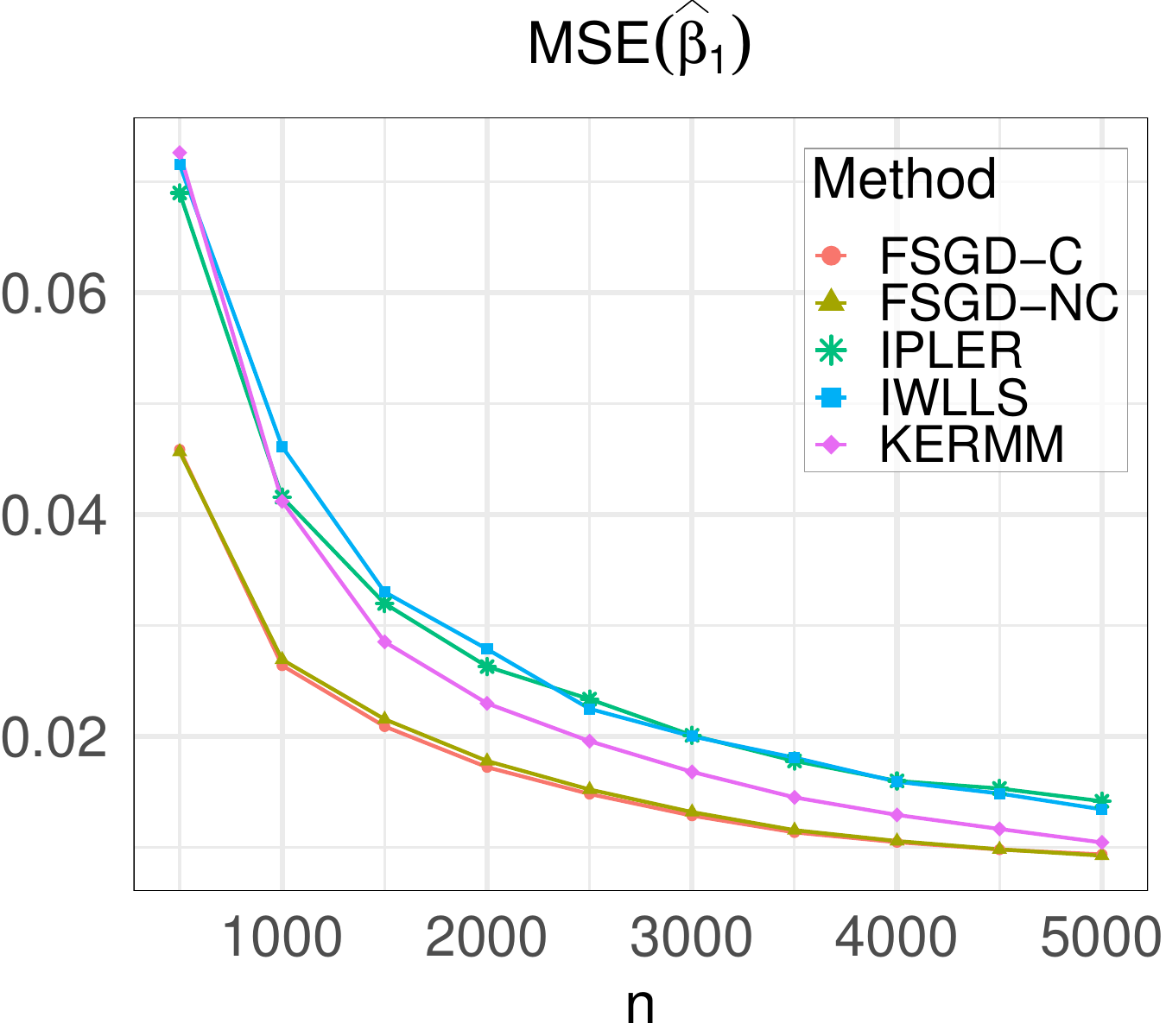}
\end{subfigure}
\begin{subfigure}{0.23\textwidth}
  \includegraphics[width=\textwidth]{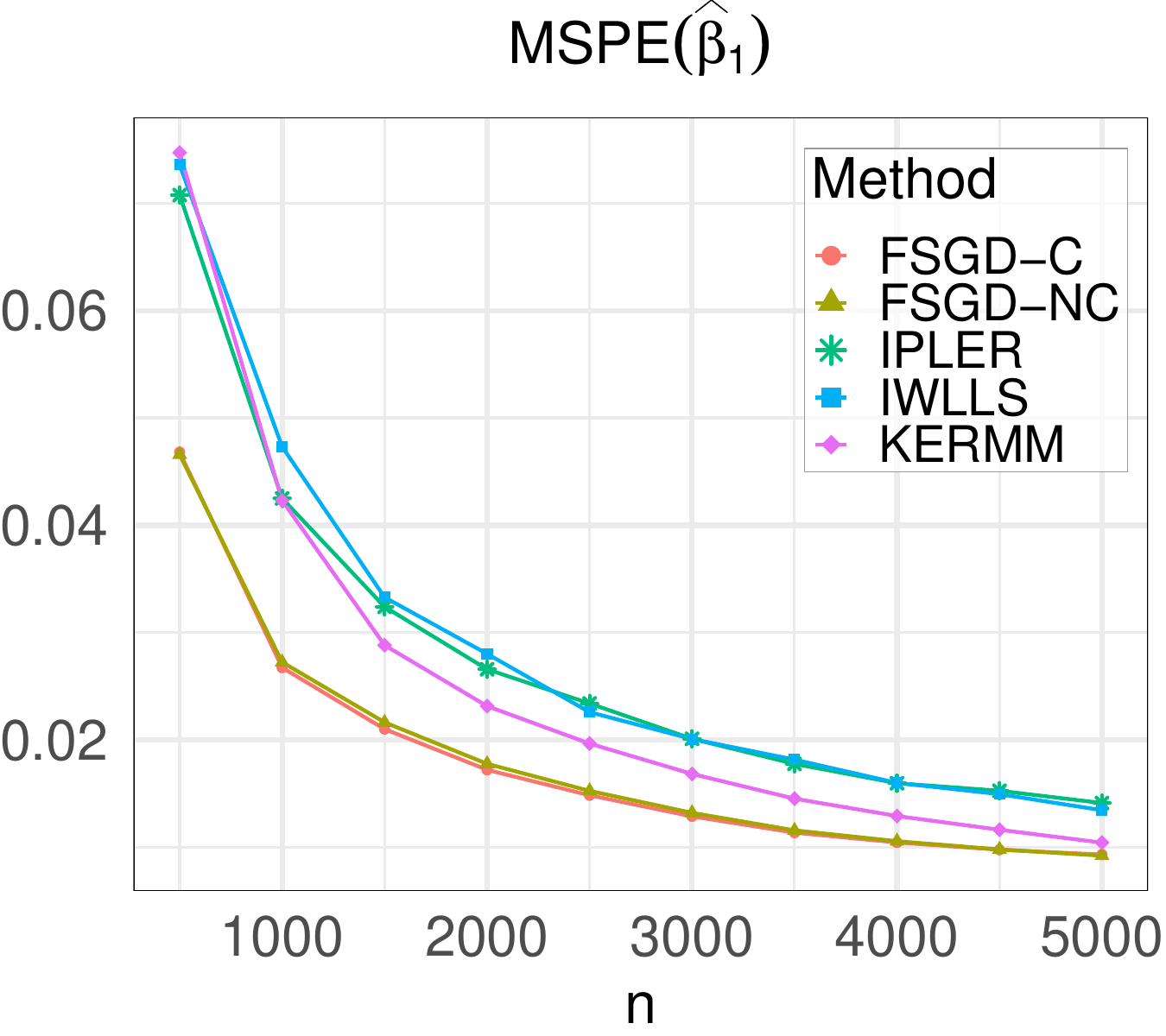}
\end{subfigure}
\begin{subfigure}{0.23\textwidth}
  \includegraphics[width=\textwidth]{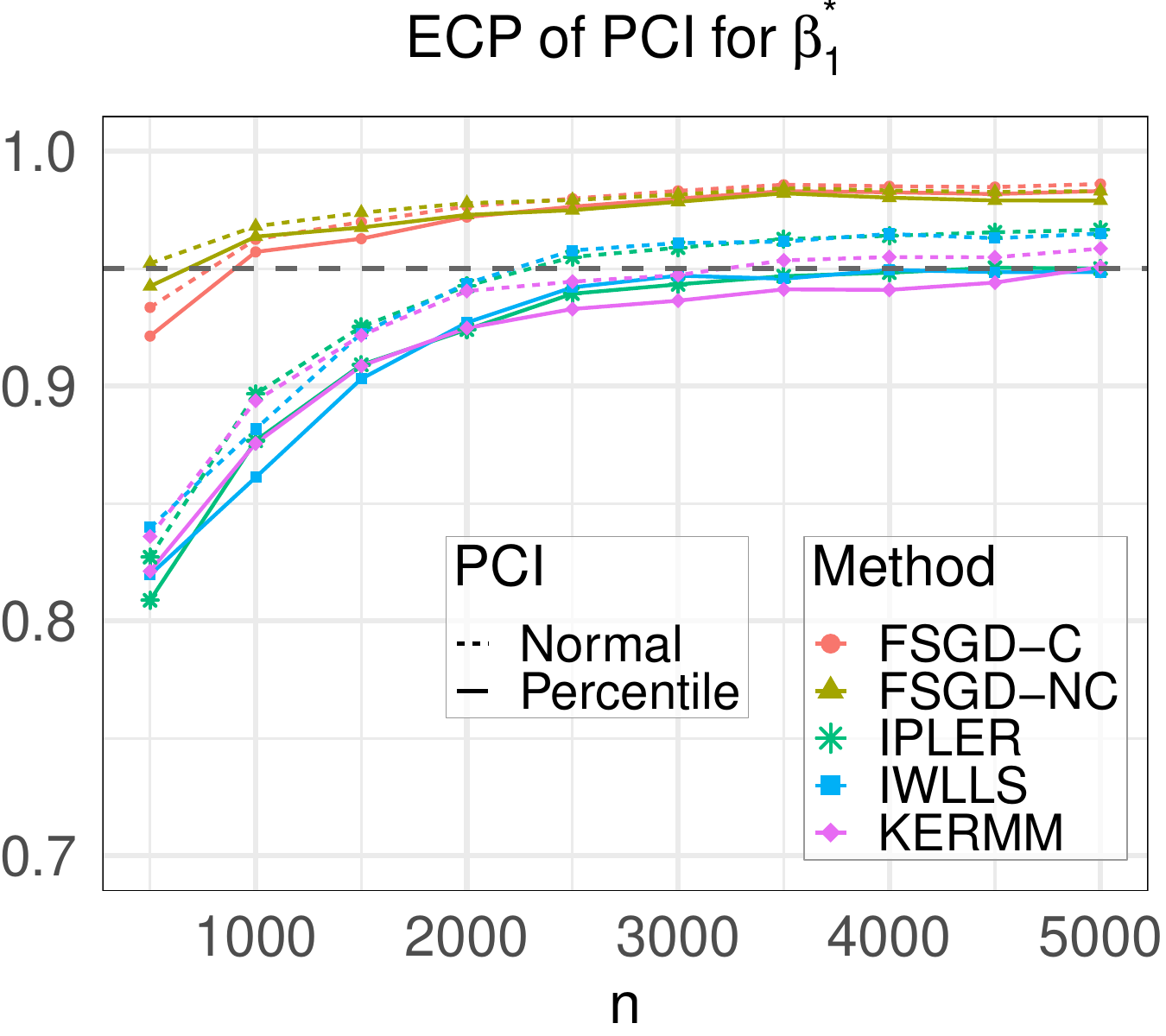}
\end{subfigure}
\begin{subfigure}{0.23\textwidth}
  \includegraphics[width=\textwidth]{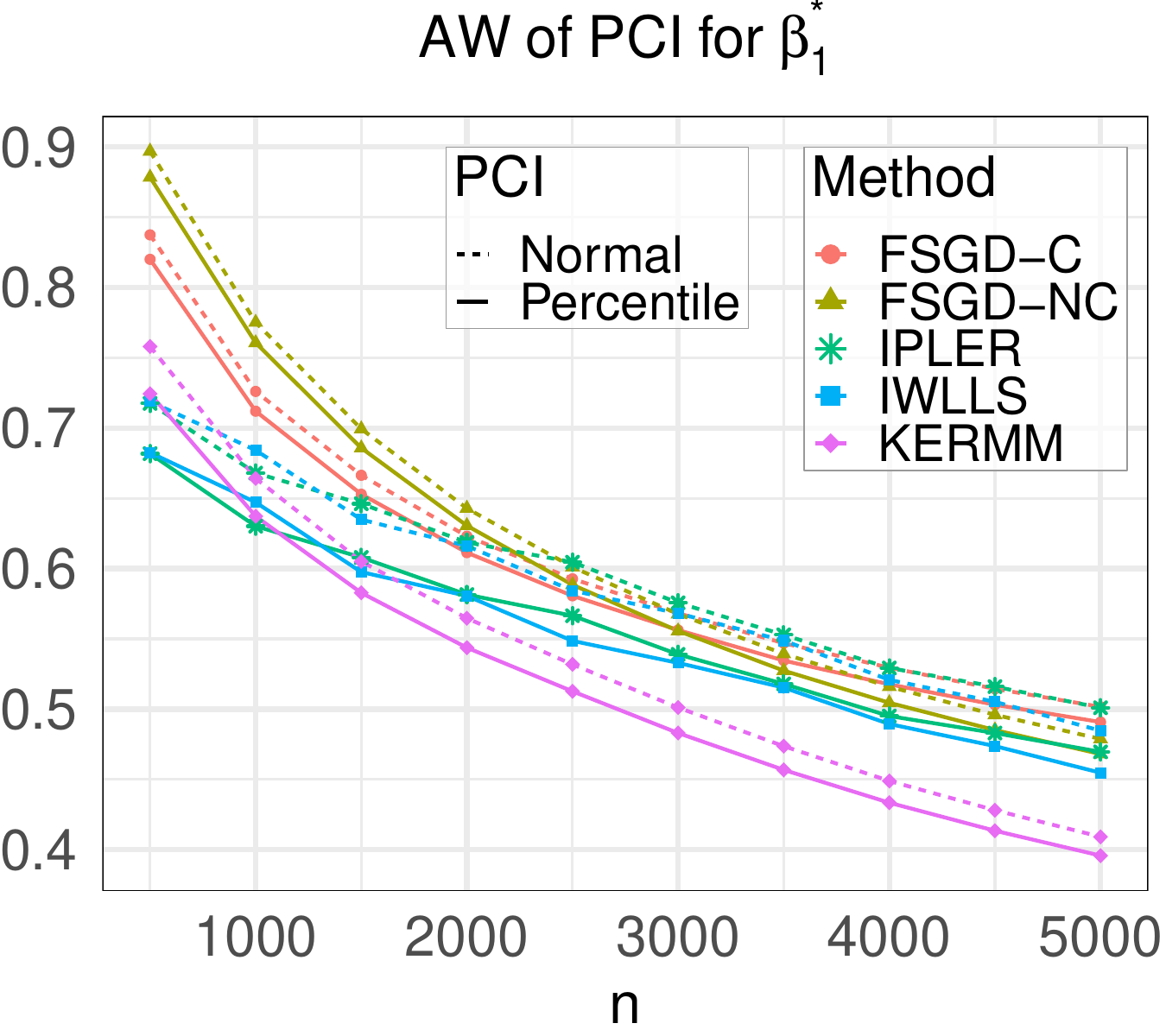}
\end{subfigure}

\begin{subfigure}{0.23\textwidth}
  \includegraphics[width=\textwidth]{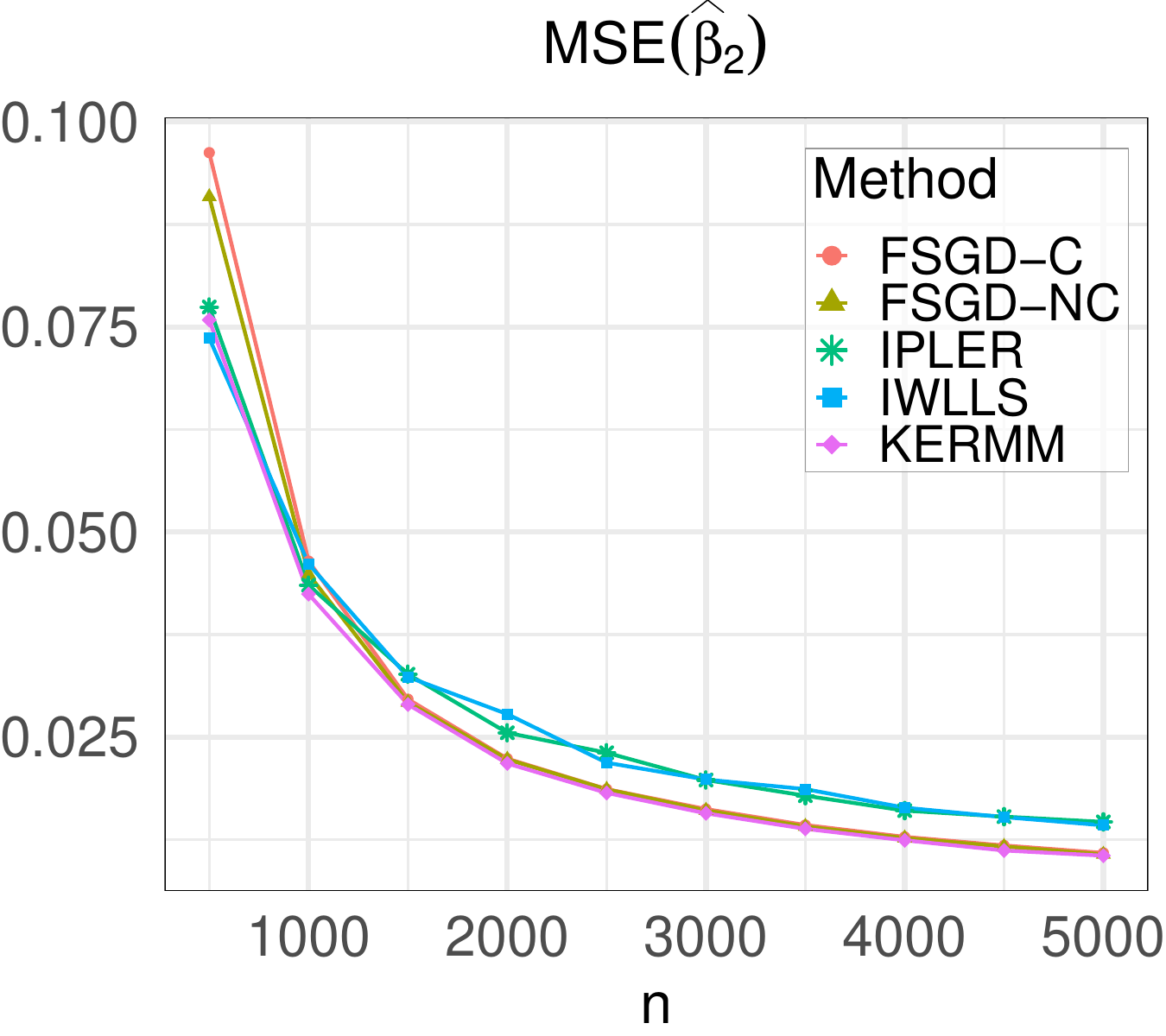}
\end{subfigure}
\begin{subfigure}{0.23\textwidth}
  \includegraphics[width=\textwidth]{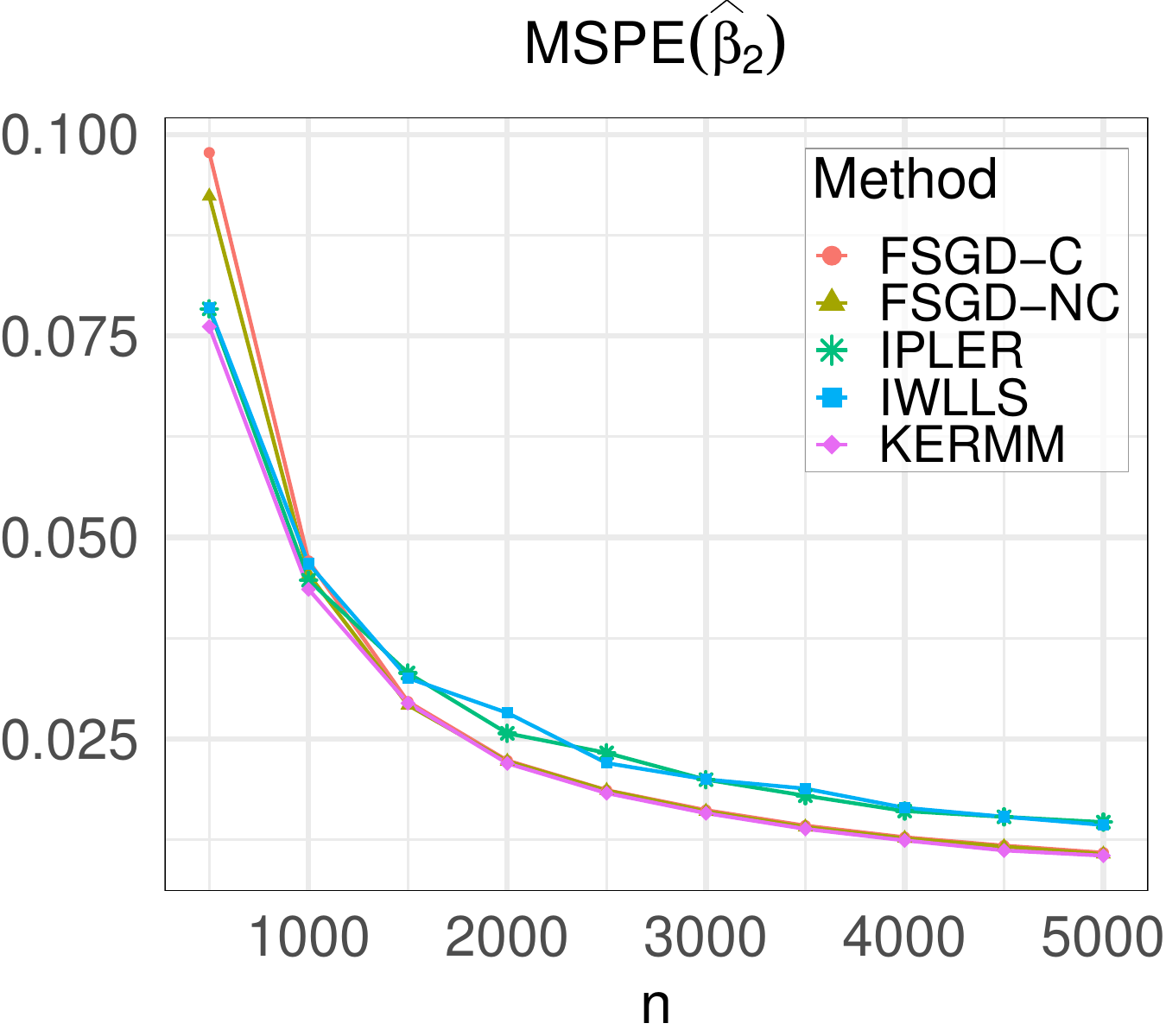}
\end{subfigure}
\begin{subfigure}{0.23\textwidth}
  \includegraphics[width=\textwidth]{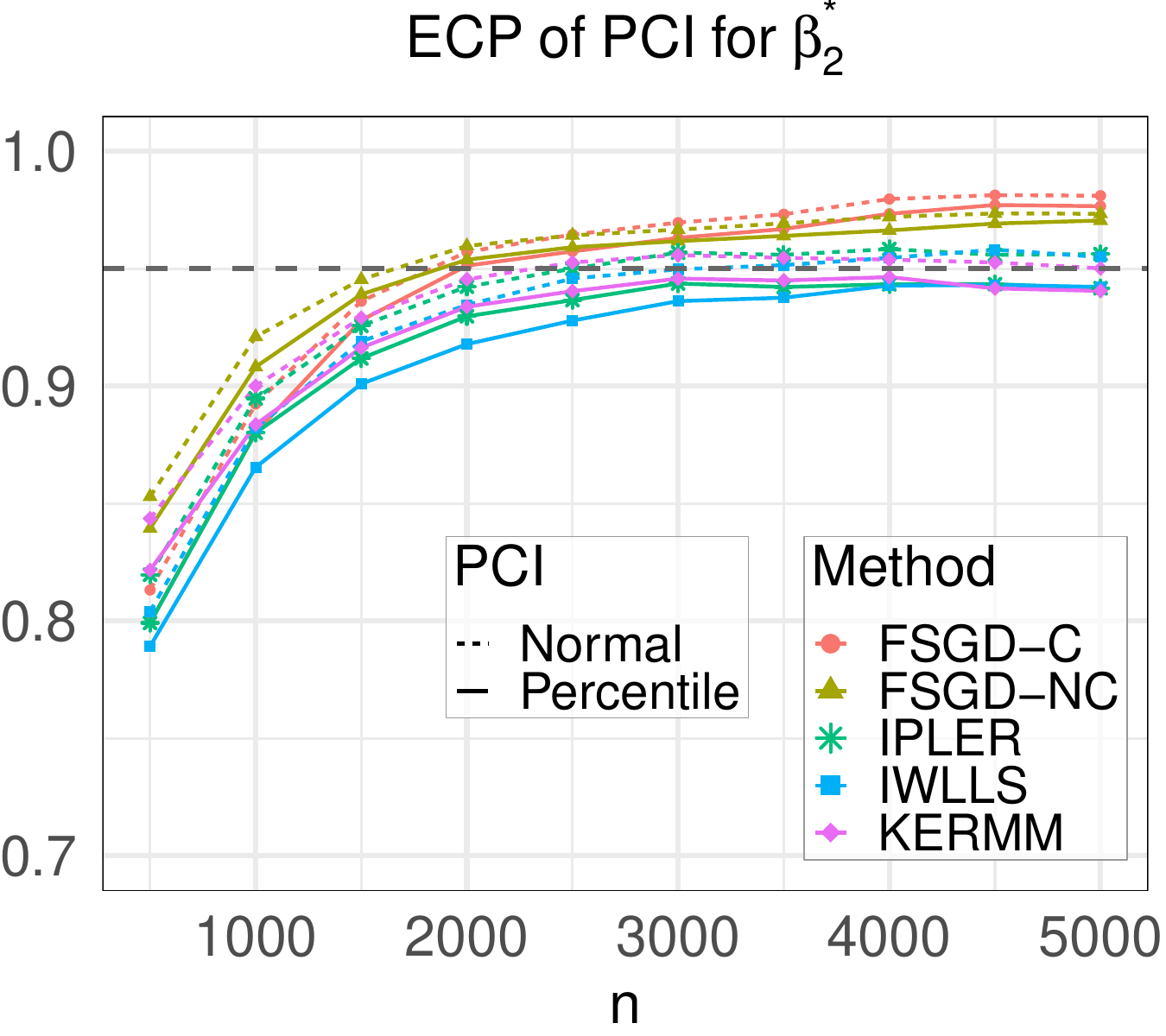}
\end{subfigure}
\begin{subfigure}{0.23\textwidth}
  \includegraphics[width=\textwidth]{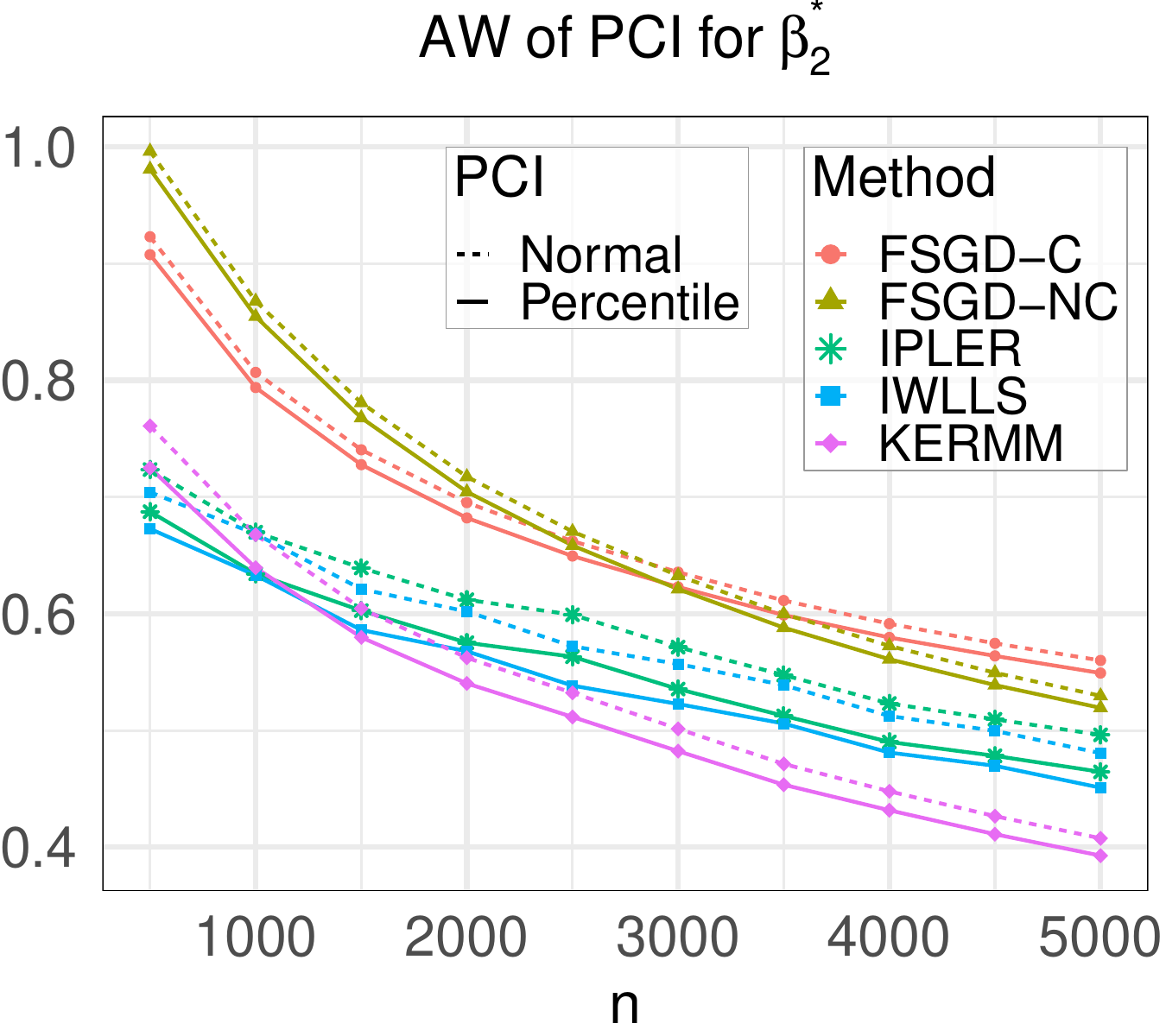}
\end{subfigure}
\caption{Simulation results under Laplace errors at $\tau=0.25$.}
\label{fig:laplace_0.25}
\end{figure}

\begin{figure}[htbp]
\def\baselinestretch{1}
\centering
\begin{subfigure}{0.23\textwidth}
  \includegraphics[width=\textwidth]{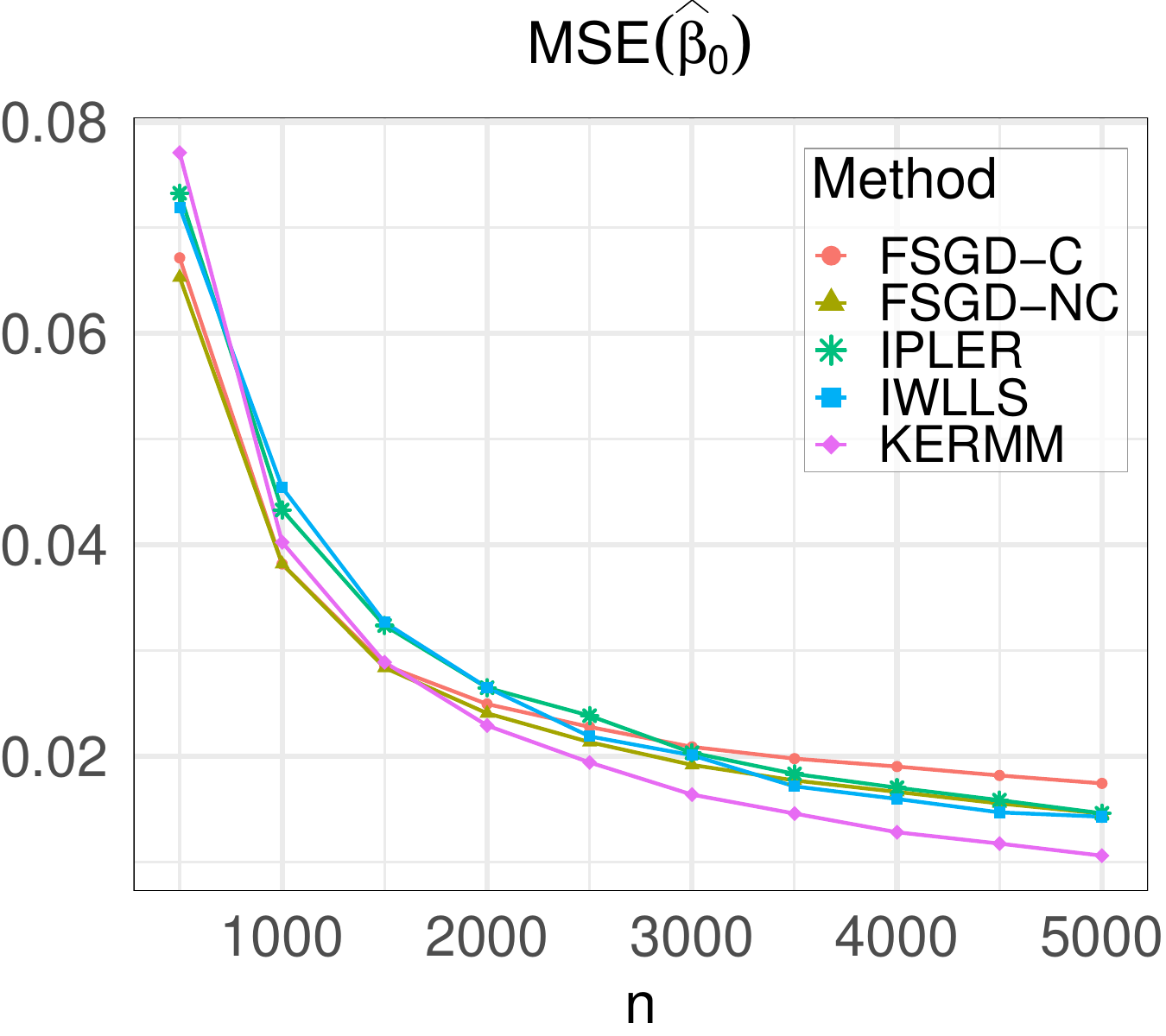}
\end{subfigure}
\begin{subfigure}{0.23\textwidth}
  \includegraphics[width=\textwidth]{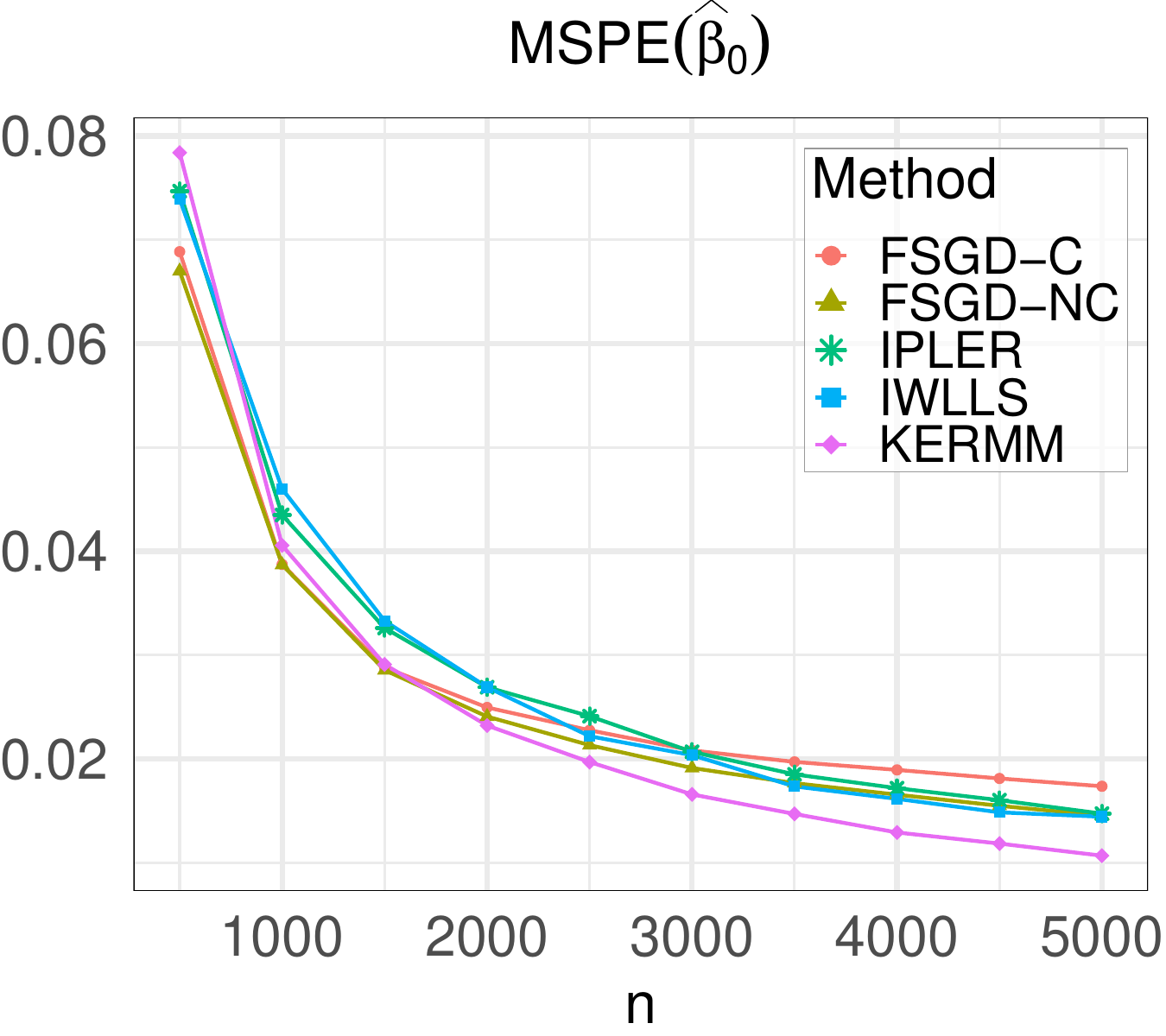}
\end{subfigure}
\begin{subfigure}{0.23\textwidth}
  \includegraphics[width=\textwidth]{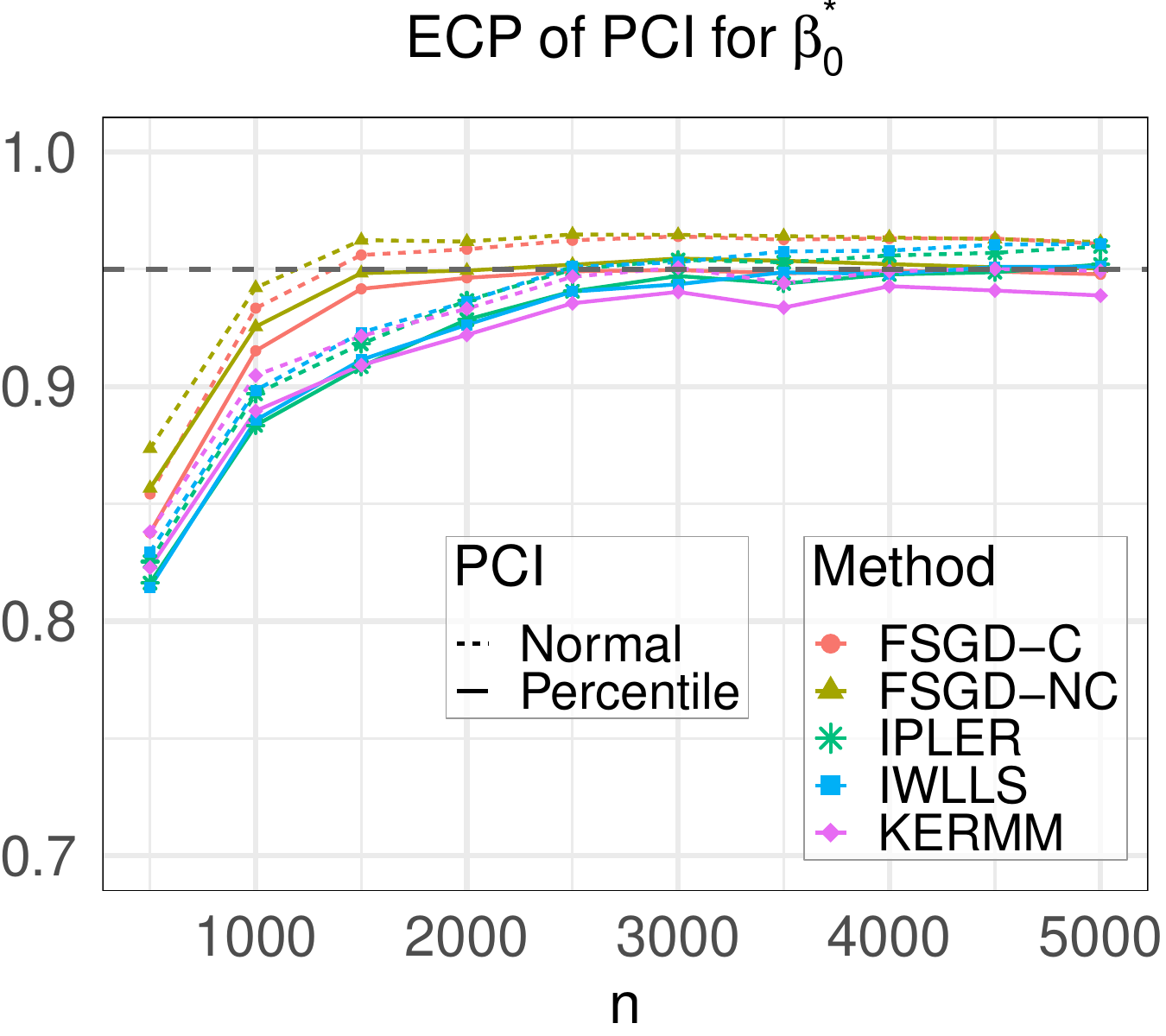}
\end{subfigure}
\begin{subfigure}{0.23\textwidth}
  \includegraphics[width=\textwidth]{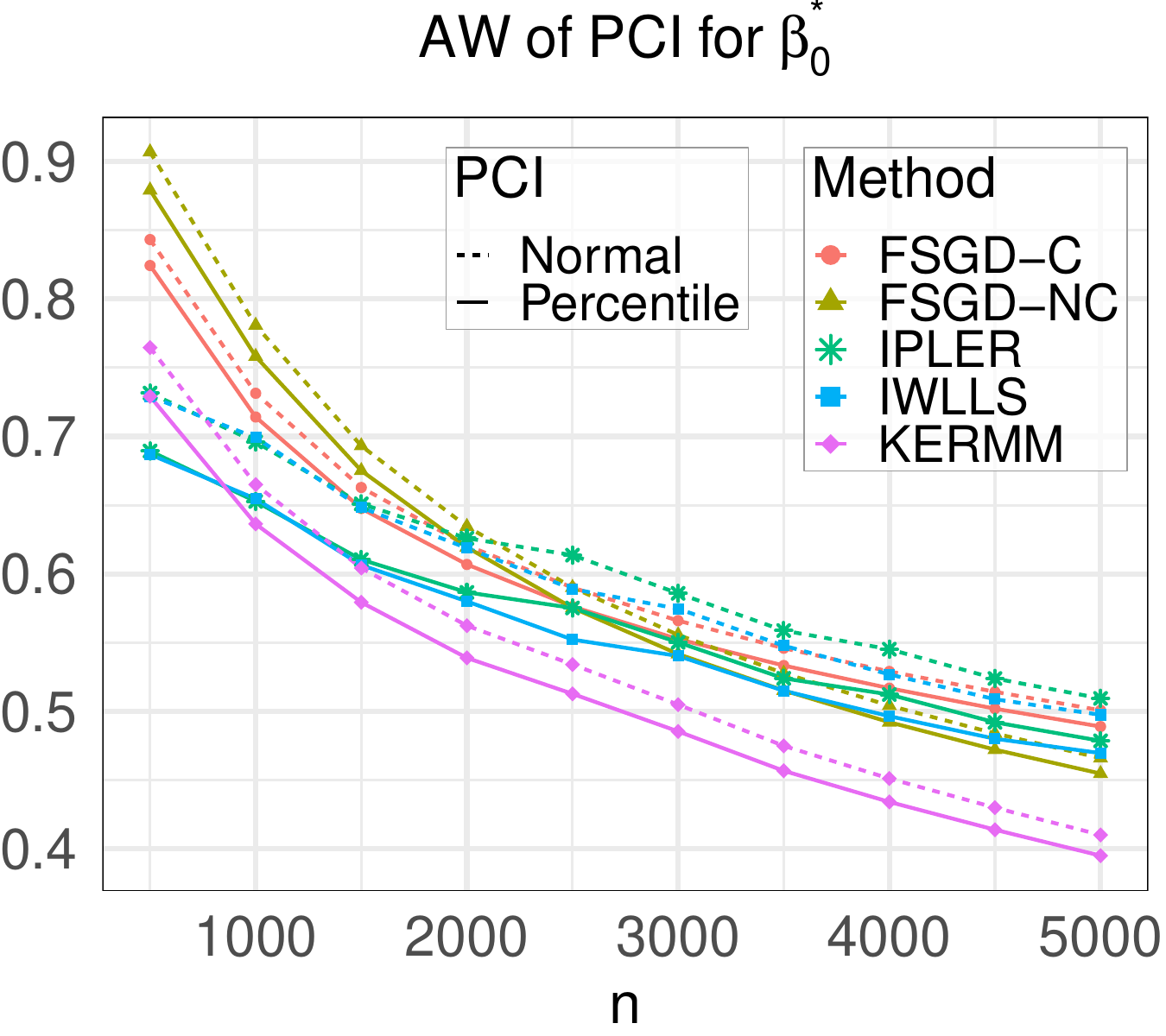}
\end{subfigure}

\begin{subfigure}{0.23\textwidth}
  \includegraphics[width=\textwidth]{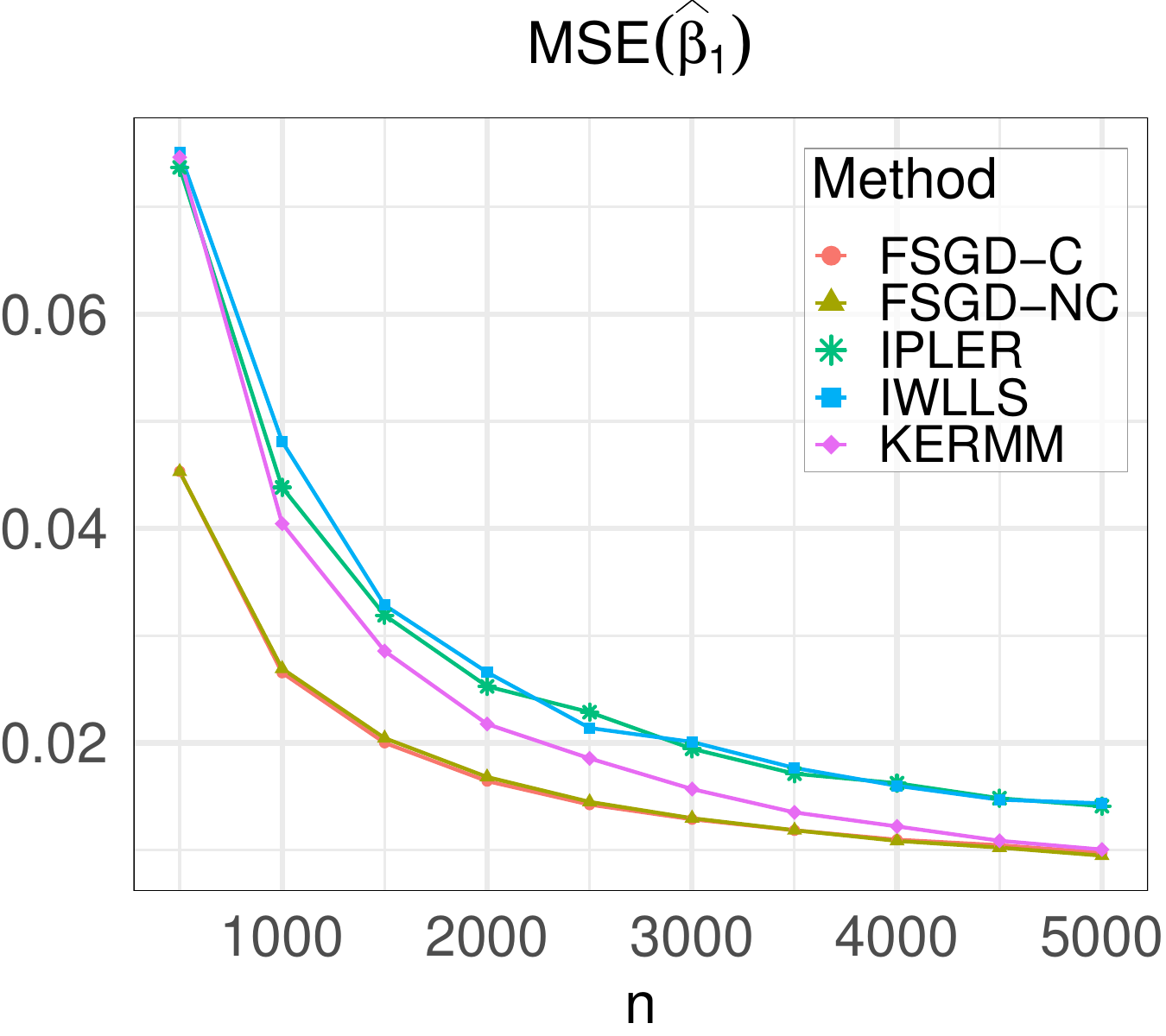}
\end{subfigure}
\begin{subfigure}{0.23\textwidth}
  \includegraphics[width=\textwidth]{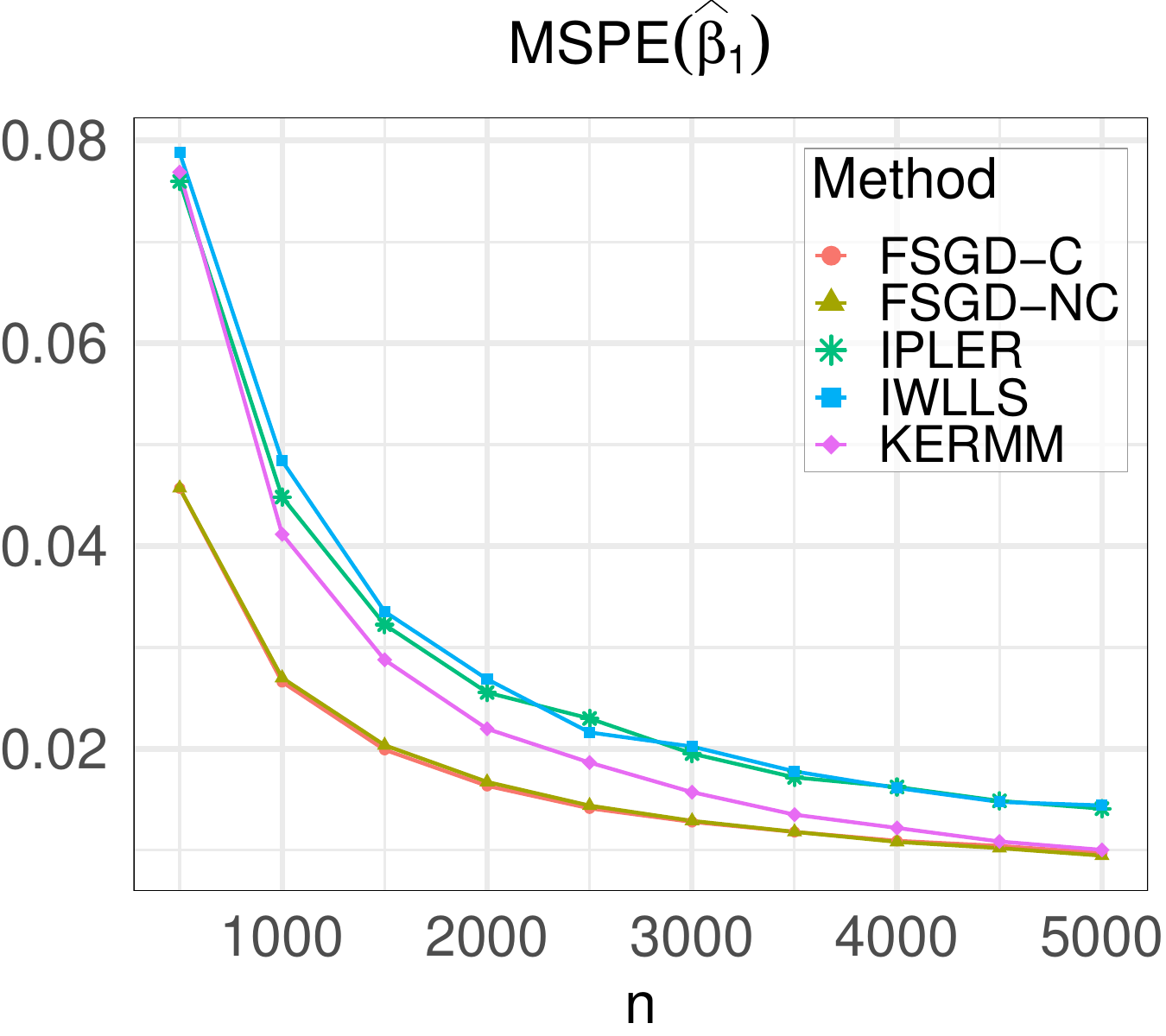}
\end{subfigure}
\begin{subfigure}{0.23\textwidth}
  \includegraphics[width=\textwidth]{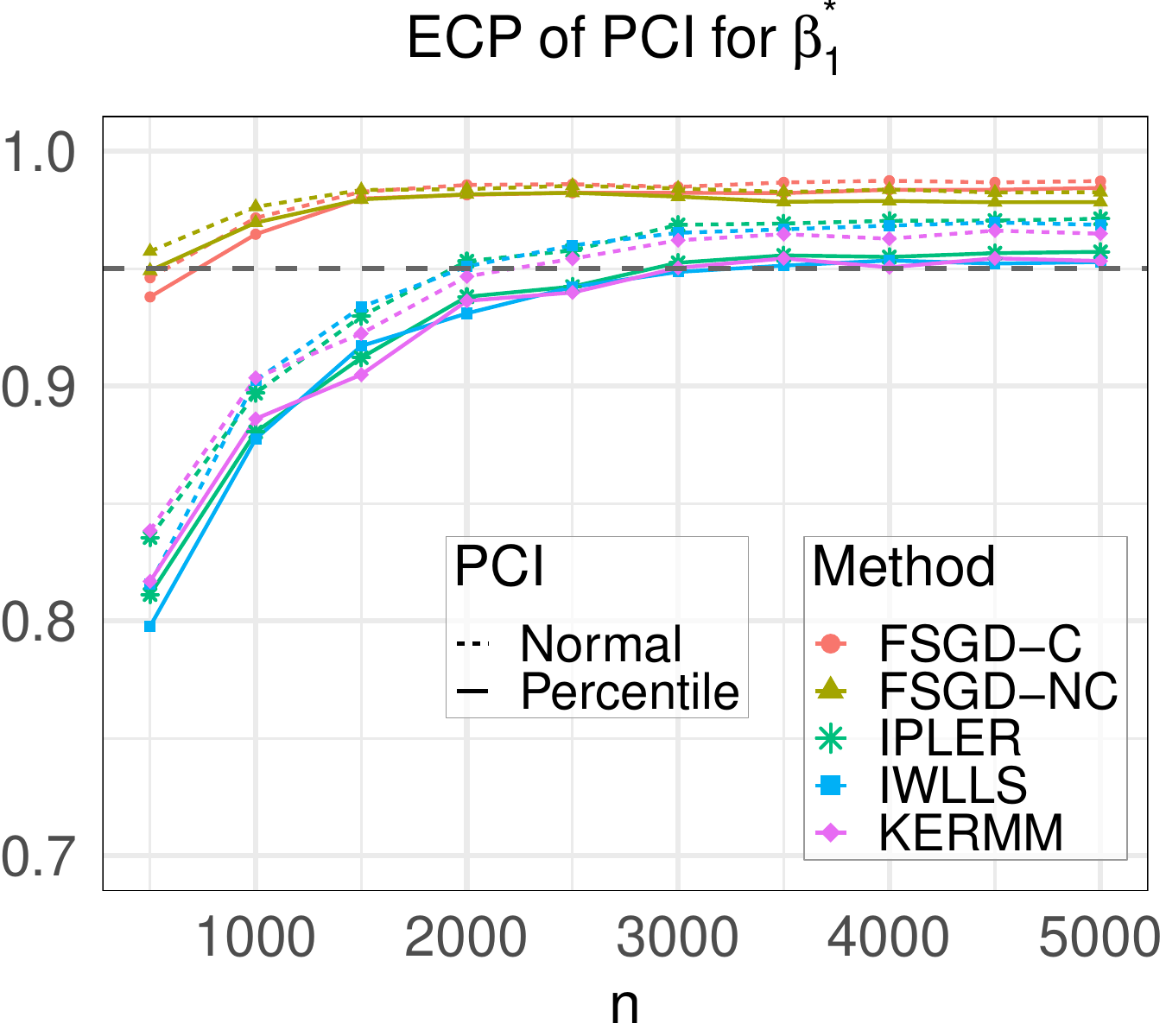}
\end{subfigure}
\begin{subfigure}{0.23\textwidth}
  \includegraphics[width=\textwidth]{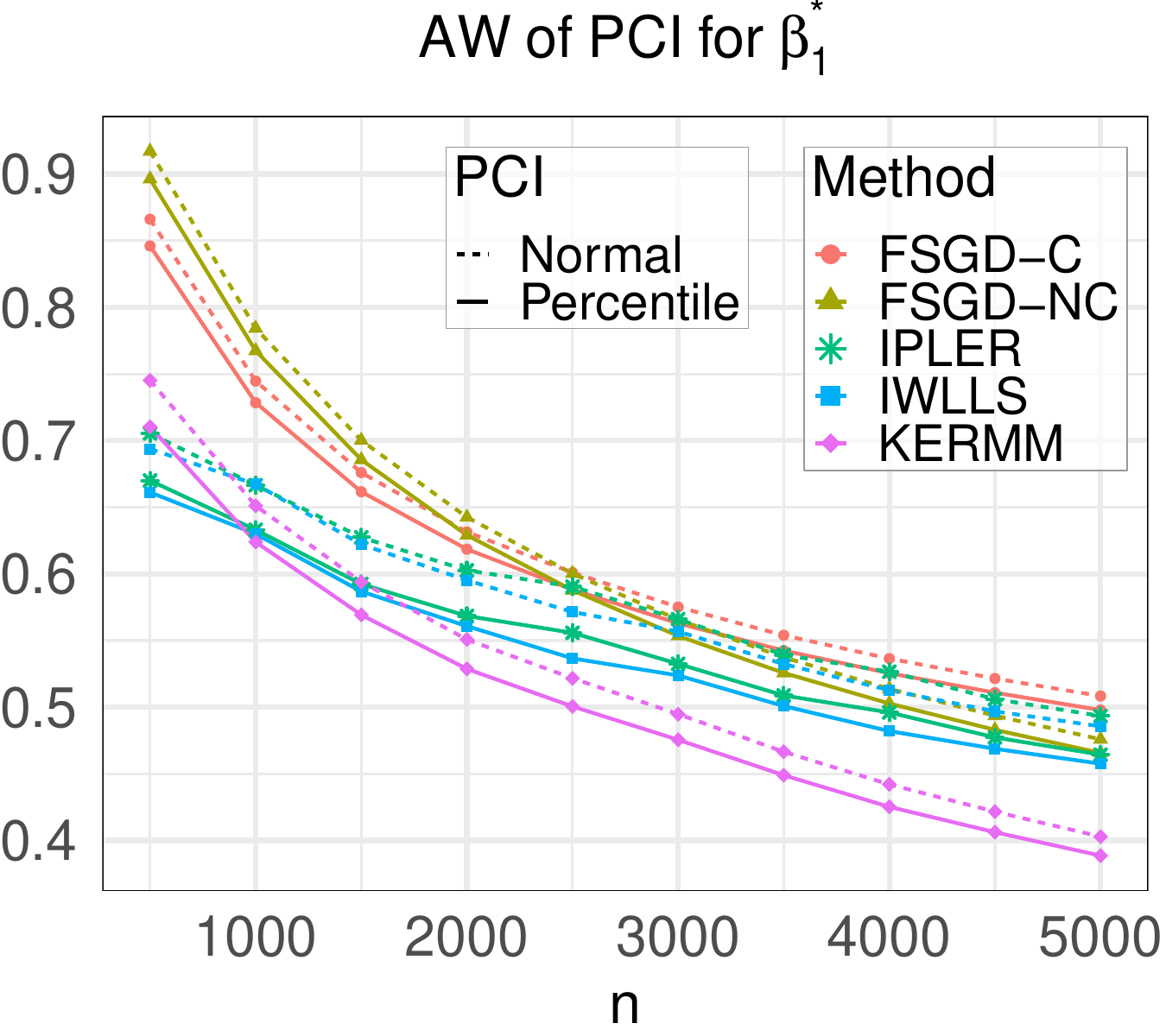}
\end{subfigure}

\begin{subfigure}{0.23\textwidth}
  \includegraphics[width=\textwidth]{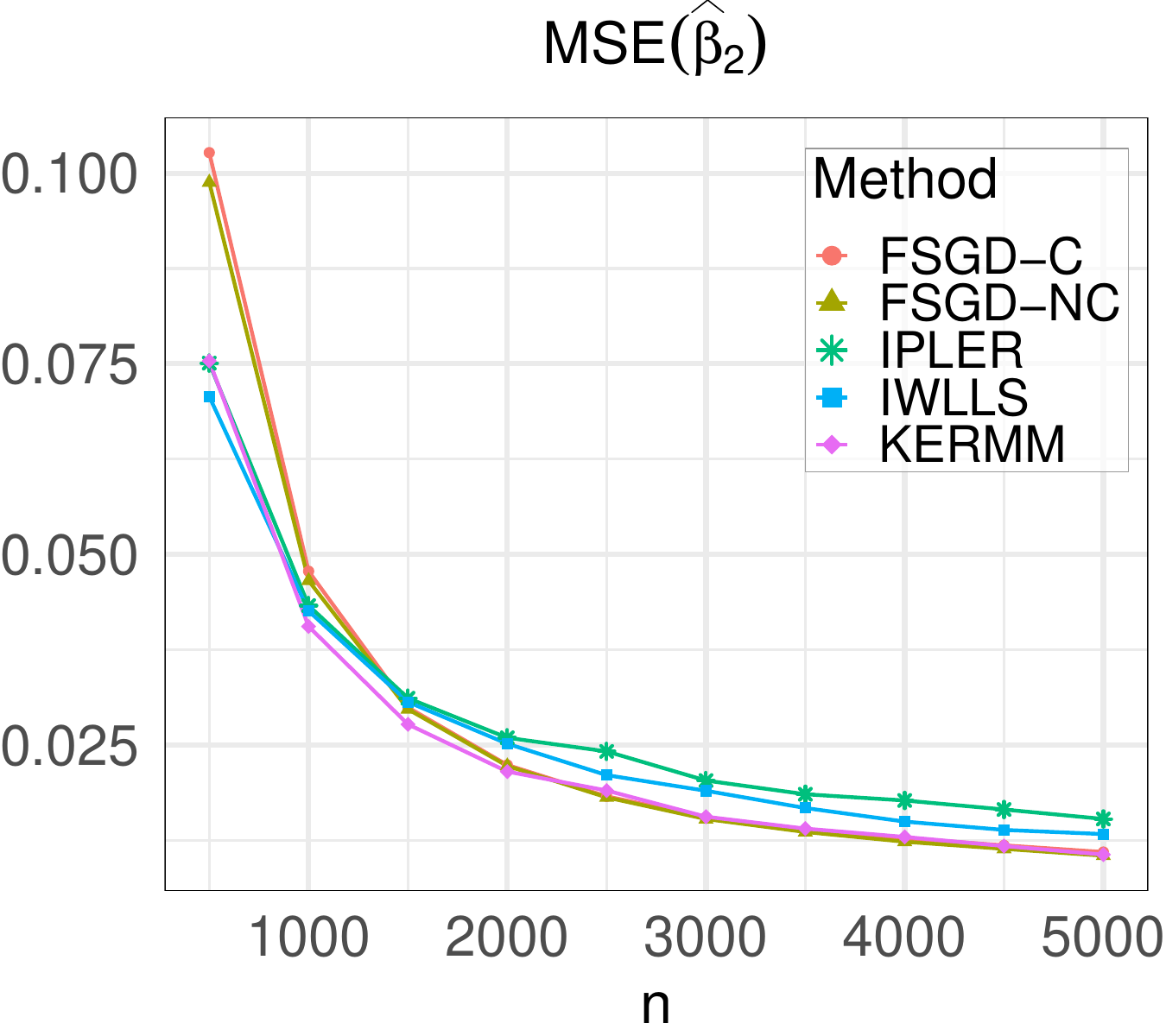}
\end{subfigure}
\begin{subfigure}{0.23\textwidth}
  \includegraphics[width=\textwidth]{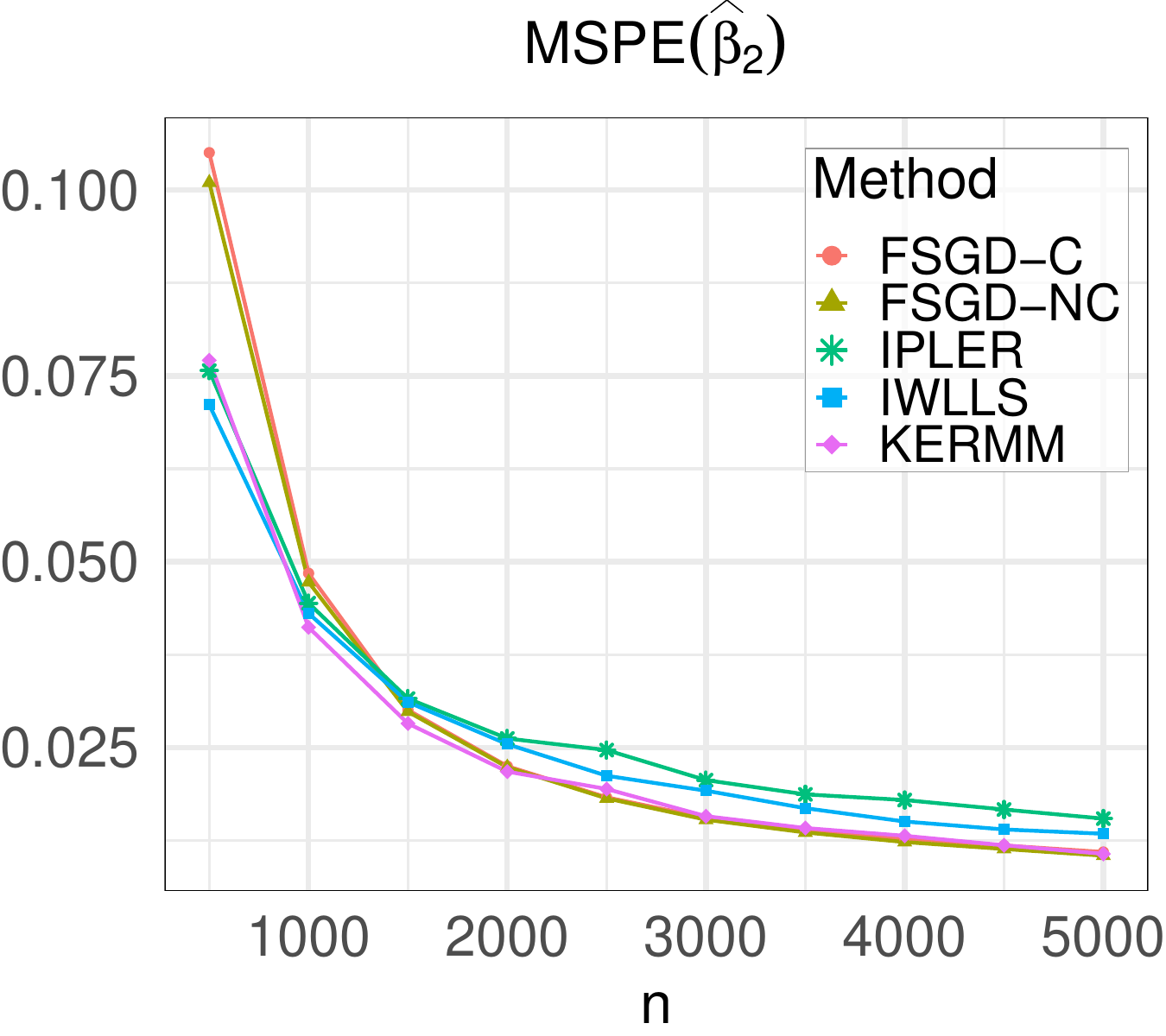}
\end{subfigure}
\begin{subfigure}{0.23\textwidth}
  \includegraphics[width=\textwidth]{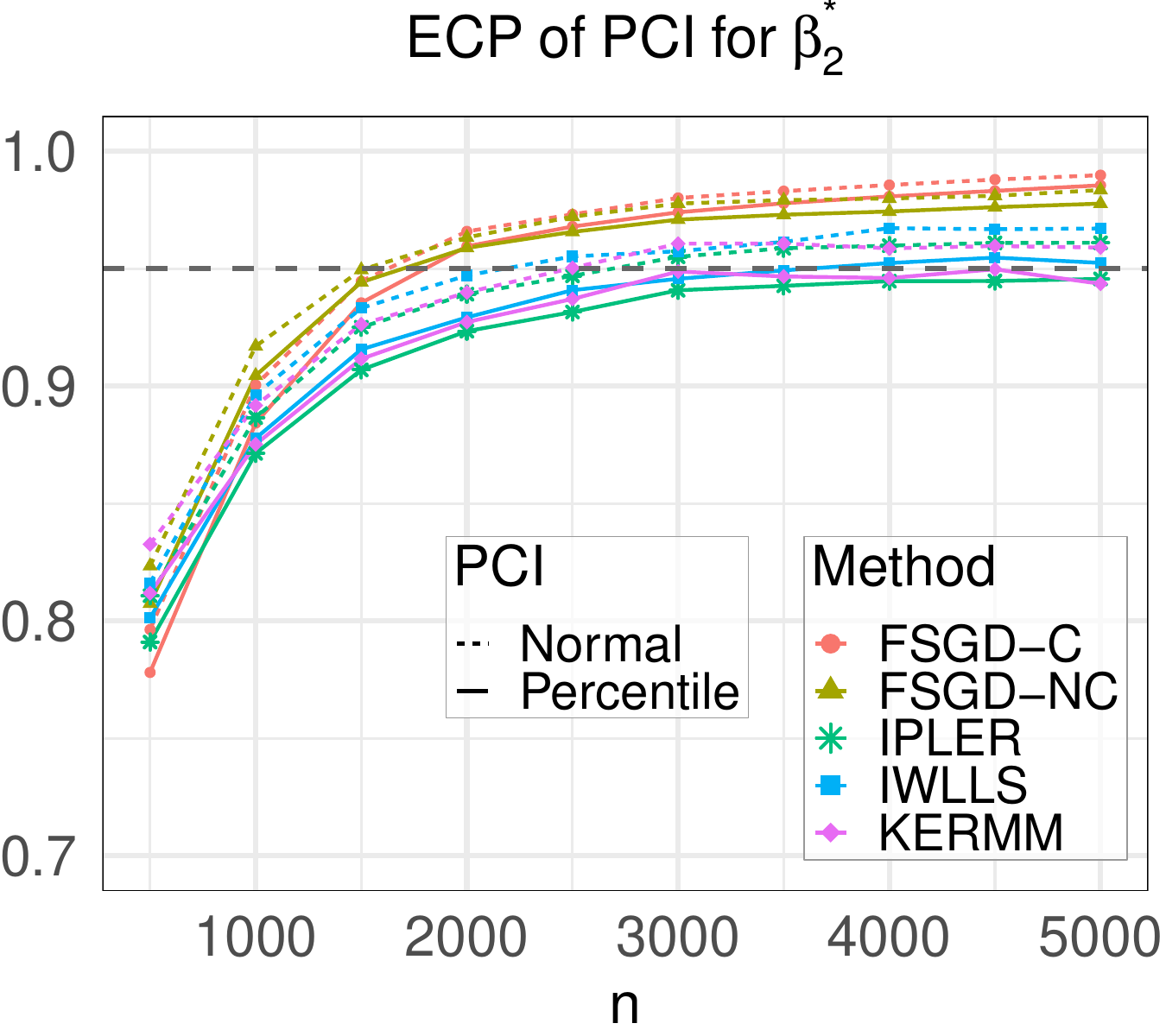}
\end{subfigure}
\begin{subfigure}{0.23\textwidth}
  \includegraphics[width=\textwidth]{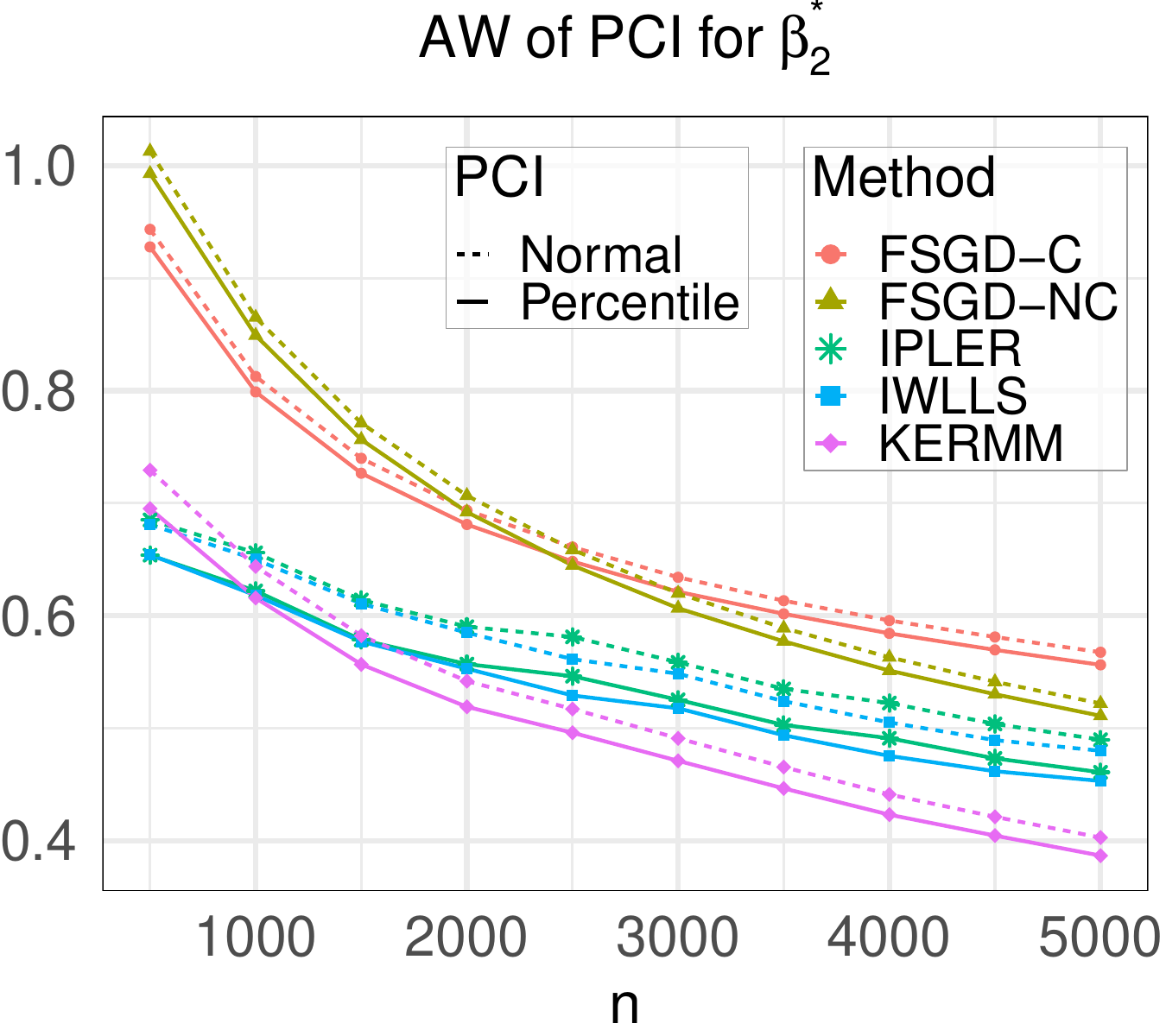}
\end{subfigure}
\caption{Simulation results under Student-$t$ errors at $\tau=0.25$.}
\label{fig:t_0.25}
\end{figure}

We first assess the estimation accuracy of the online and offline methods for the coefficient functions. Figures \ref{fig:normal_0.25}--\ref{fig:t_0.25} report the mean squared error (MSE) on the training set and the mean squared prediction error (MSPE) on the test set of the estimated coefficient functions at $\tau=0.25$. For every method and error distribution, the MSE and MSPE decrease with sample size, with MSPE closely tracking MSE. Across settings, the online estimator is competitive with the best offline method (KERMM) and more accurate than the local offline methods (IWLLS and IPLER). KERMM attains the smallest aggregate coefficient error, followed closely by FSGD-NC and FSGD-C, while IWLLS and IPLER are slightly less accurate. The online method attains this accuracy in one pass over the stream, because its regularization is driven by the step-size path and the Polyak--Ruppert average stabilizes the final estimate; this path-averaged structure enables competition with the best offline method without repeated refitting. Compared with normal errors, the estimation errors of all methods increase moderately under contaminated normal, Laplace, and Student-$t$ errors, but the relative ordering remains stable.

We next assess inferential accuracy. Figures \ref{fig:normal_0.25}--\ref{fig:t_0.25} also display the empirical coverage probability (ECP) and average width (AW) of the $95\%$ PCIs at $\tau=0.25$. As $n$ grows from 500 to 5000, every method's ECP stabilizes near the nominal level, and the AW narrows. Overall, the online multiplier bootstrap delivers valid, slightly conservative inference: its ECP settles slightly above nominal, whereas the offline methods settle near it. At comparable or slightly higher coverage, the online PCIs are comparable in width to local offline methods and somewhat wider than KERMM, which yields the narrowest PCIs but sits at or slightly below nominal. The Student-$t$ case stress-tests the error tail condition underlying the multiplier bootstrap consistency. Although its polynomial tails violate Assumption \ref{A8}, the online PCIs remain near nominal or slightly conservative. Thus, online inference remains reliable even when the formal tail condition is violated.

The two online schemes, FSGD-C and FSGD-NC, have similar estimation accuracy across four error distributions, with FSGD-NC consistently slightly more accurate at every sample size, though the differences remain small. This close agreement is expected asymptotically: both schemes achieve the same nonparametric convergence rate under their respective step-size choices, differing mainly in finite-sample trajectories. In online inference, because the multiplier bootstrap trajectories share the estimate recursion, FSGD-C uses a fixed step size, so early updates are forgotten only geometrically; FSGD-NC adapts to the current stream length and truncates the early learning rate, keeping the bootstrap trajectories more stable throughout the stream. Consequently, FSGD-NC yields somewhat narrower PCIs than FSGD-C at comparable coverage.

\begin{figure}[htbp]
\def\baselinestretch{1}
\centering
\begin{subfigure}{0.25\textwidth}
  \includegraphics[width=\textwidth]{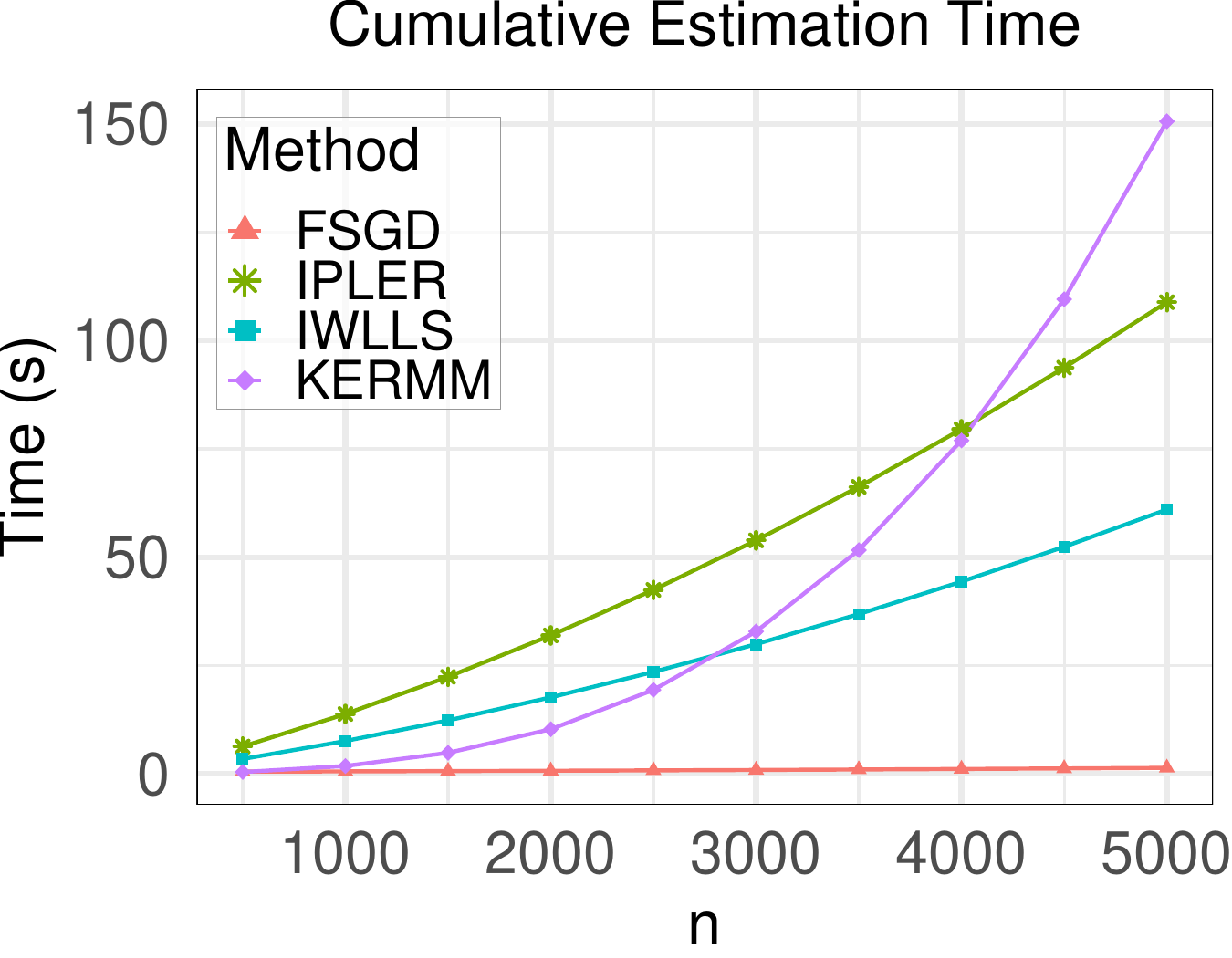}
\end{subfigure}
\hspace{0.01\textwidth}
\begin{subfigure}{0.25\textwidth}
  \includegraphics[width=\textwidth]{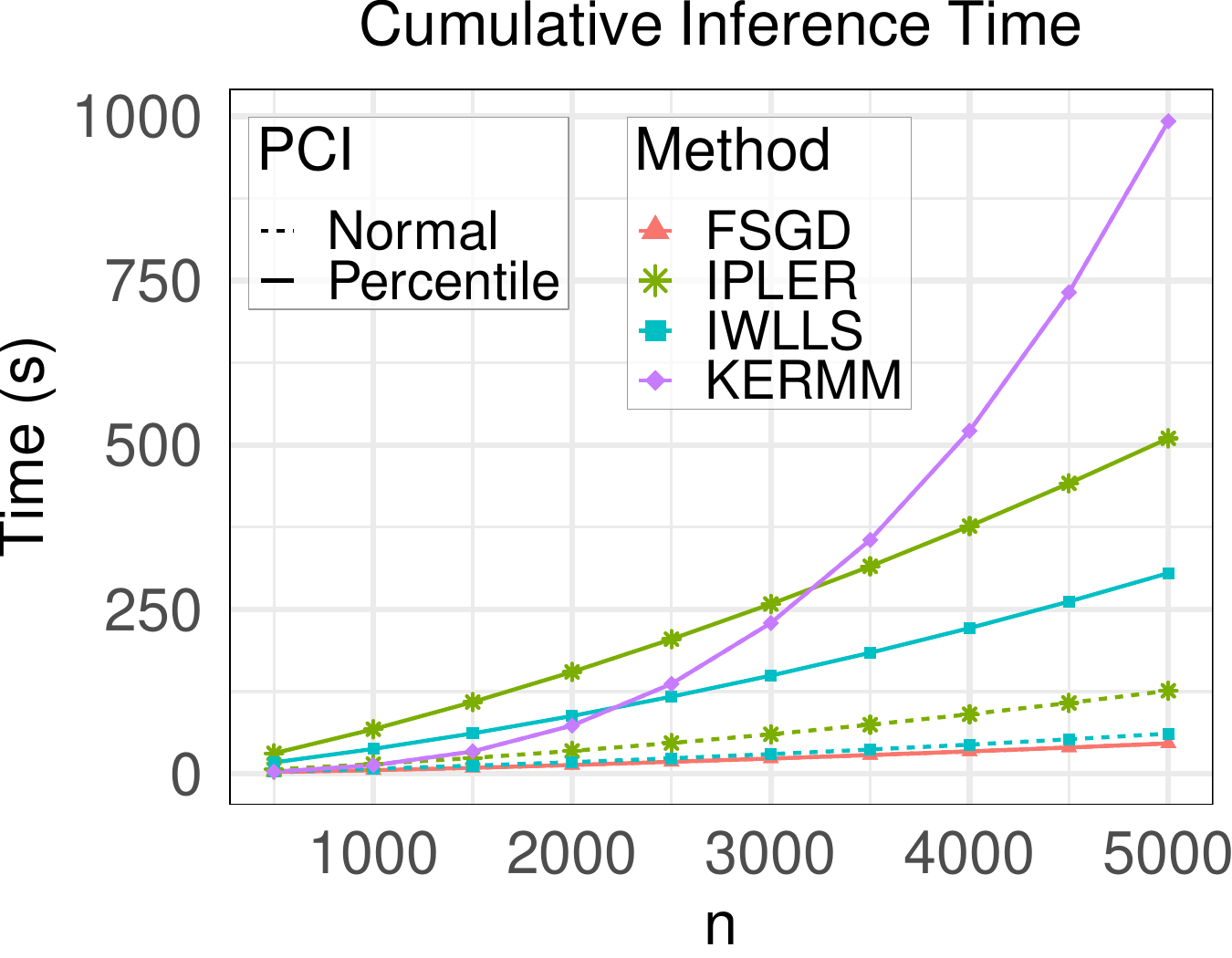}
\end{subfigure}
\caption{Comparison of cumulative computation time (seconds) in the simulation.}
\label{fig:time}
\end{figure}

Figure \ref{fig:time} compares estimation and inference times across stream lengths. For estimation, the online procedure is substantially faster than the offline alternatives, with the gap widening as $n$ grows. For inference, the online multiplier bootstrap is substantially faster than the offline wild bootstrap. For IWLLS and IPLER, plug-in PCIs are a cheaper offline option since they avoid the $B$ repeated fits, yet they remain slower than the online multiplier bootstrap. Thus, the online procedure retains a computational advantage over both plug-in and wild bootstrap offline inference. Figure \ref{fig:time} reports computation time only for FSGD-NC, as the two online schemes have essentially identical costs.

The conclusions remain essentially unchanged across expectile levels.

\section{Application}\label{Sec5}

To demonstrate the proposed online method, we use Binance's real-time public database. Binance, a centralized cryptocurrency exchange operating since 2017, continuously updates this database with transaction-level executions, quote updates, order-book depth, and candlestick bars aggregated over fixed intervals. A candlestick summarizes one trading pair's activity over a predefined interval (e.g., 30 minutes or an hour), reporting prices, trading volume, transaction count, and quote asset volume. Unfinished candlesticks update as new transactions arrive and finalize at the interval end. Completed records are archived daily and monthly at \url{https://data.binance.vision}, organized by trading pair, sampling interval, and calendar period.

From Binance's continuously updated database, we use the completed 30-minute spot market records of 100 USDT-quoted trading pairs (Tether USD, a stablecoin pegged to the U.S. dollar) from July 1 to July 15, 2026, and process them sequentially in chronological order. After removing the missing records, the analysis sample consists of 71,181 observations. Among them, July 1 to July 10 are used as the training stream with $N_{\text{train}}=47,421$, while July 11 to July 15 are used as a fixed test set with $N_{\text{test}}=23,760$.

\begin{figure}[htbp]
\def\baselinestretch{1}
  \centering
  \includegraphics[width=0.4\linewidth]{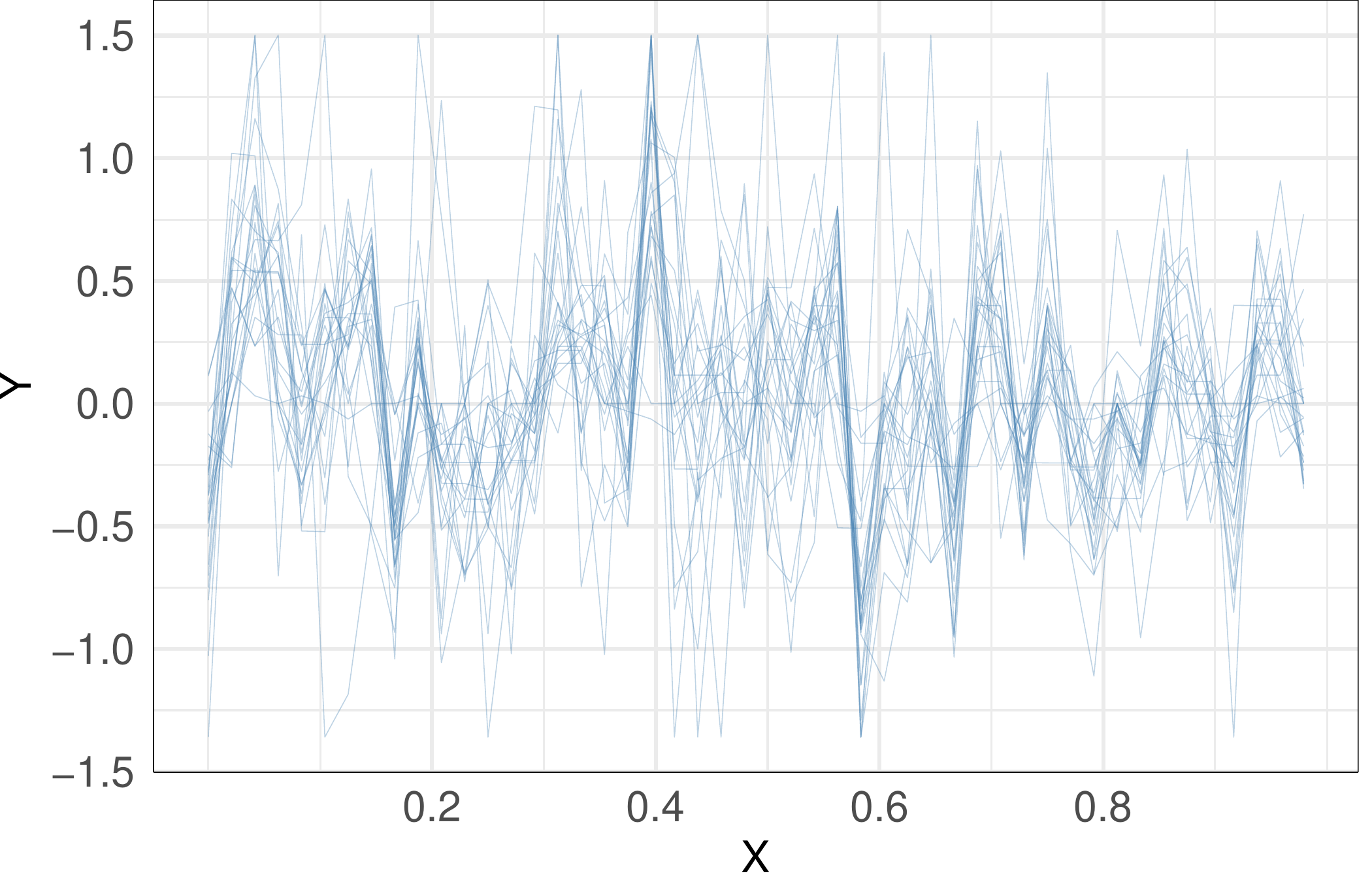}
  \caption{Intraday log-return curves of 25 randomly selected cryptocurrencies.}
  \label{fig:daydata}
\end{figure}

For each observation index $i$, the response is the within-pair scaled log return between consecutive close prices, $Y_i = 100\log(P_i/P_{i-1})$, where $P_i$ and $P_{i-1}$ are the close prices of the same trading pair in the current and preceding 30-minute intervals. The design variable $X_i$ is the time of day at which the interval opens, normalized to $[0,1)$. Figure \ref{fig:daydata} displays the intraday 30-minute log-return curves of 25 randomly selected cryptocurrencies, which show pronounced and distinct time-varying patterns, supporting nonparametric modeling. We consider three covariates. The market return $Z_{i,1}$ is the equal-weighted average of the scaled log returns of the other available trading pairs over the same interval, excluding the pair's own return to avoid self-dependence between the response and the market covariate. The order flow imbalance $Z_{i,2}$ is the difference between taker-buy and taker-sell volume divided by total volume, lying in $[-1,1]$, with positive values indicating buyer dominance and negative values indicating seller dominance. The scaled log quote volume is $Z_{i,3}=\log(1+\text{quote volume})/10$, where quote volume is the interval's trading volume denominated in USDT. At each fixed expectile level $\tau$, we fit the model
$$
e_{Y_i\mid (X_i,\mathbf{Z}_i)}(\tau)
=\beta_0^*(X_i,\tau)+Z_{i,1}\beta_1^*(X_i,\tau)
+Z_{i,2}\beta_2^*(X_i,\tau)+Z_{i,3}\beta_3^*(X_i,\tau),
$$
where $\mathbf{Z}_i=(Z_{i,1},Z_{i,2},Z_{i,3})^{\top}$. The coefficient functions $\beta_k^*(\cdot,\tau)$ describe how each covariate affects the conditional expectile of the return over time. We consider $\tau=0.25,0.5,0.75$, corresponding to the different regions of the conditional return distribution.

The training stream is processed one observation at a time, with a checkpoint at the end of each training day corresponding to cumulative sample sizes $n=4{,}653$, $9{,}405$, $\ldots$, $47{,}421$. For online FSGD, we select the bandwidth and step-size using only the first training day ($n_0=4{,}653$ observations). At each checkpoint, online methods and offline methods are evaluated on the fixed test set via the empirical expectile loss
$$
\mathrm{EL}(\widehat{Y}, \tau)
=\frac{1}{N_{\text{test}}}\sum_{i=1}^{N_{\text{test}}}
\left|\tau-\mathbb{I}\left(Y_i-\widehat{Y}_i<0\right)\right|
\left(Y_i-\widehat{Y}_i\right)^2.
$$
Computational cost is summarized by cumulative estimation time, recorded as in the simulation study. In addition, online coefficient estimates and their PCIs are recorded at each checkpoint to trace their evolution as the data stream grows.

\begin{figure}[htbp]
\def\baselinestretch{1}
\centering
\begin{subfigure}{0.23\textwidth}
  \includegraphics[width=\textwidth]{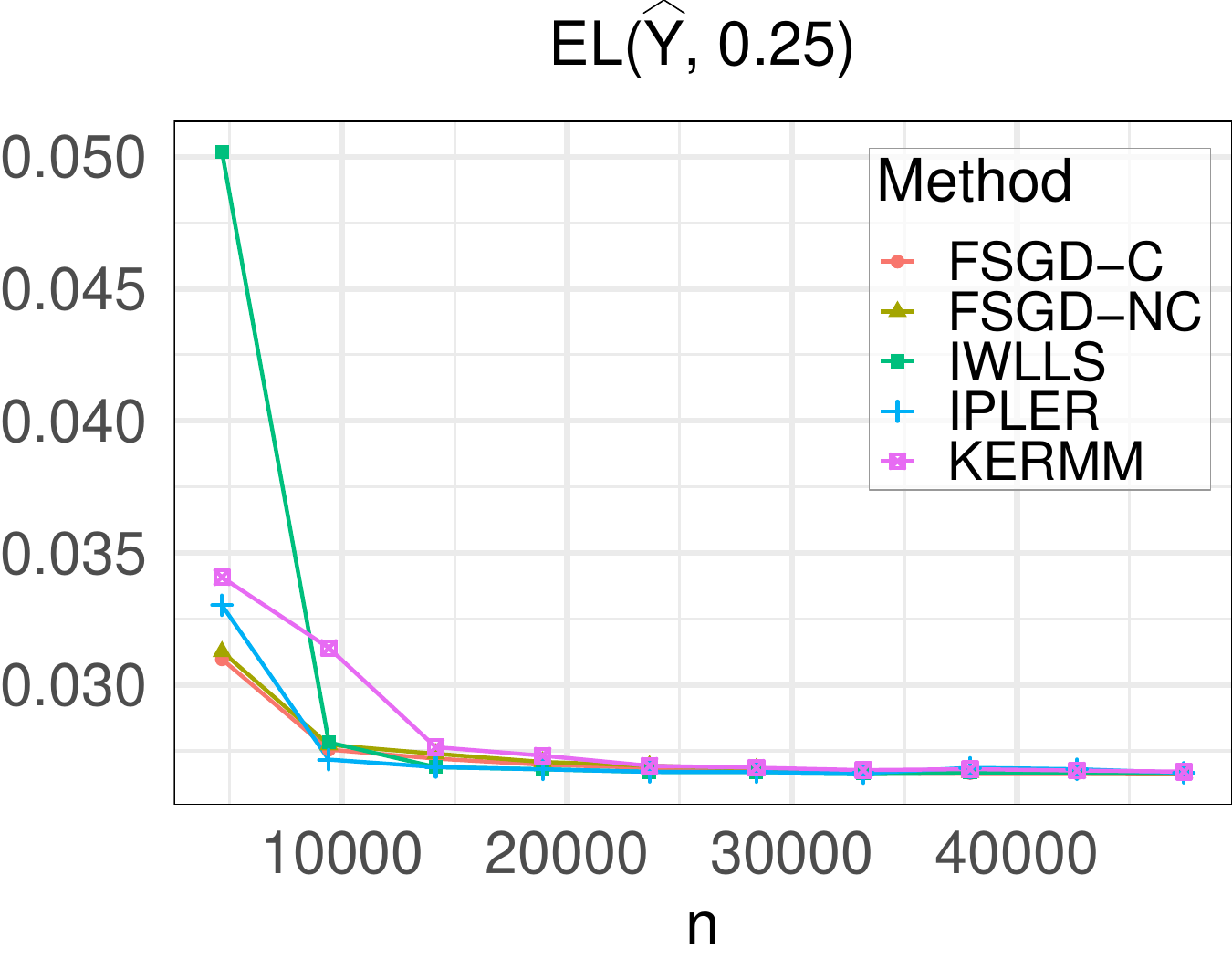}
\end{subfigure}
\begin{subfigure}{0.23\textwidth}
  \includegraphics[width=\textwidth]{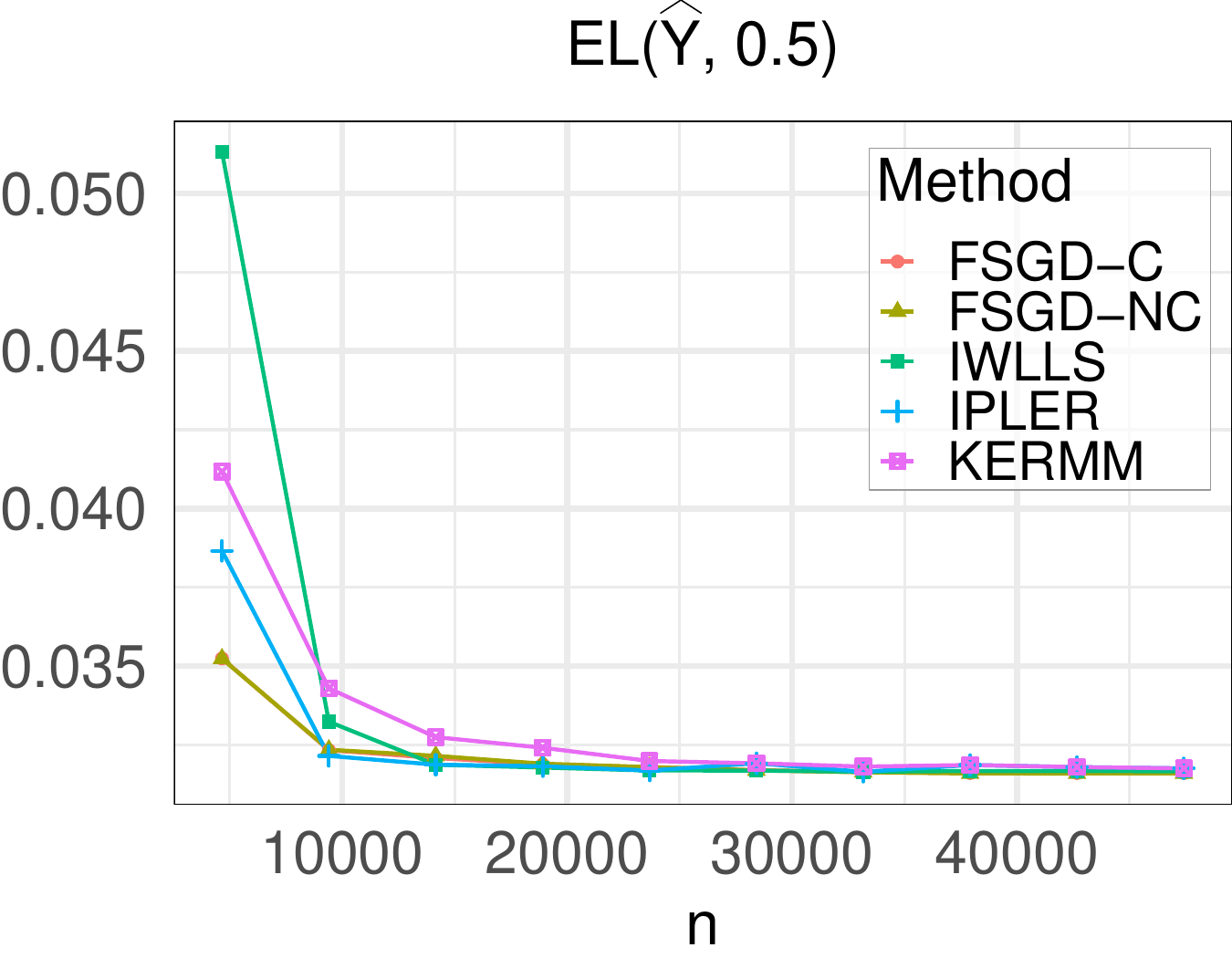}
\end{subfigure}
\begin{subfigure}{0.23\textwidth}
  \includegraphics[width=\textwidth]{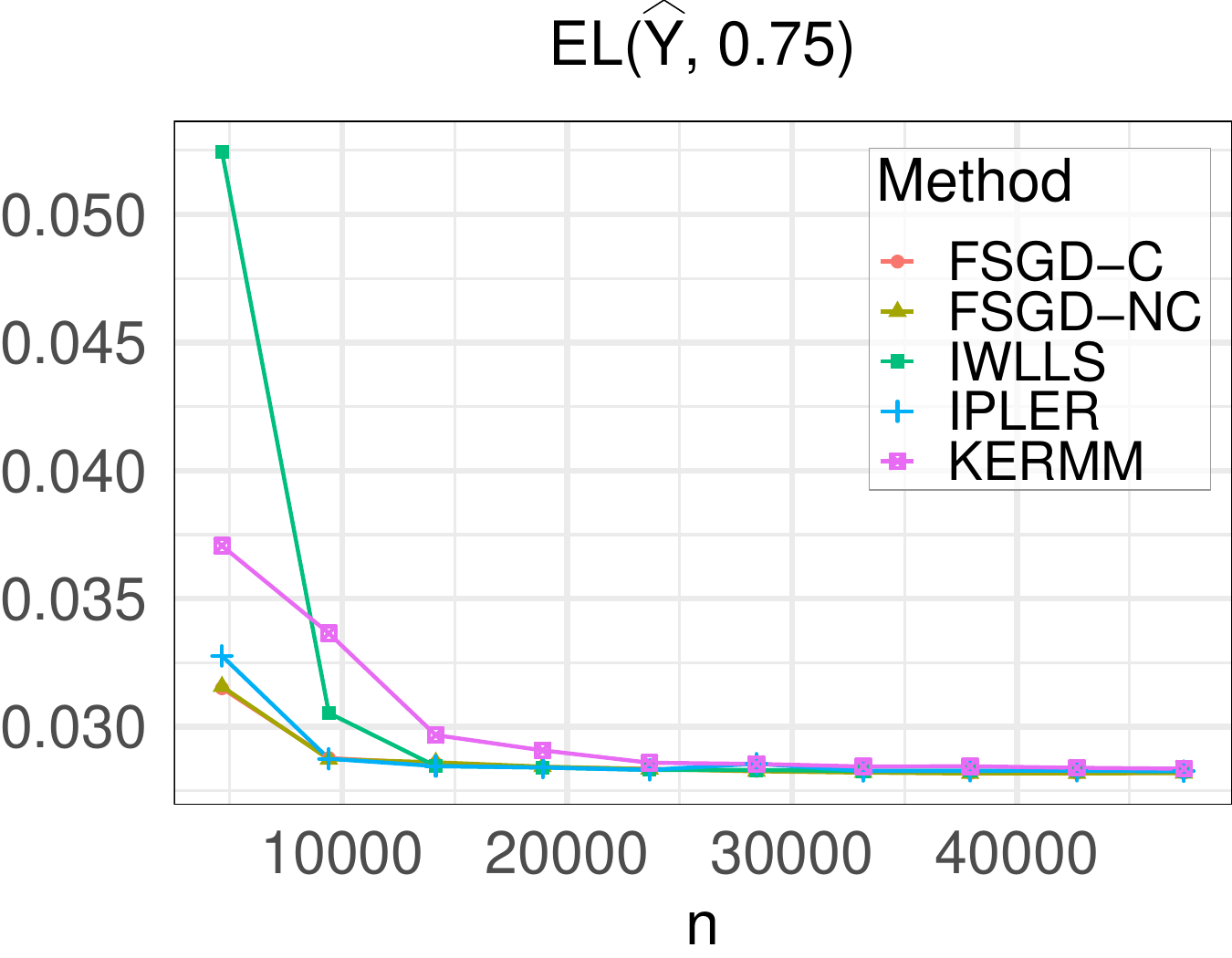}
\end{subfigure}
\begin{subfigure}{0.23\textwidth}
  \includegraphics[width=\textwidth]{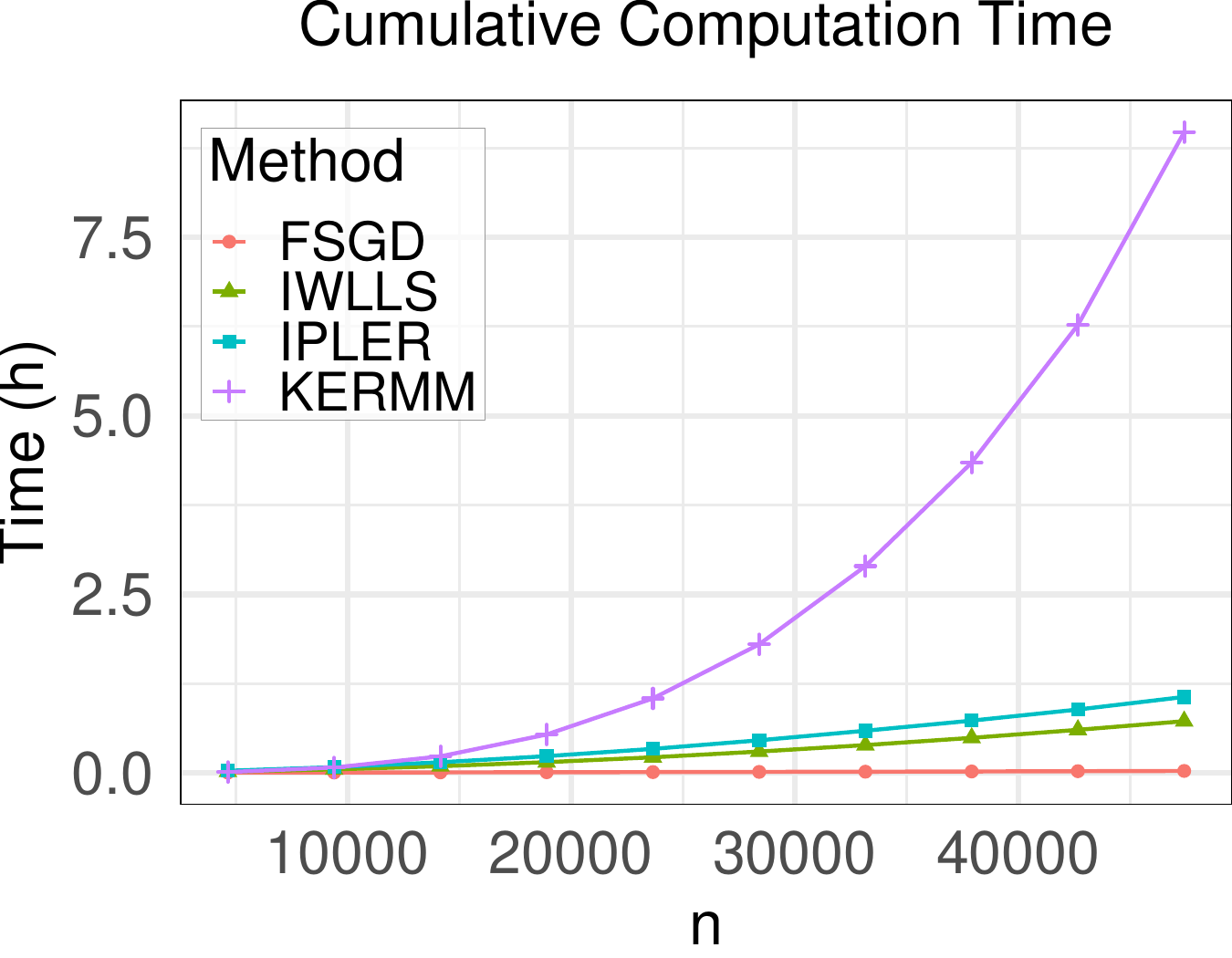}
\end{subfigure}
\caption{Comparison of empirical EL and the cumulative computation time (hours).}
\label{fig:real_checkpoints}
\end{figure}

Figure \ref{fig:real_checkpoints} compares the empirical expectile loss and cumulative estimation time of all methods at $\tau=0.25$, $0.5$, and $0.75$. FSGD-C and FSGD-NC show a clear advantage early in training, but the offline methods quickly close the gap after a few training days. At the final checkpoint, after all training data are processed, the expectile loss curves of the online and offline methods almost coincide. Thus, online estimation achieves essentially the same expectile prediction performance as the offline procedures but with significantly less cumulative estimation time. Computation time is reported only for FSGD-NC.

\begin{figure}[htbp]
\def\baselinestretch{1}
\centering
\begin{subfigure}{0.23\textwidth}
  \includegraphics[width=\textwidth]{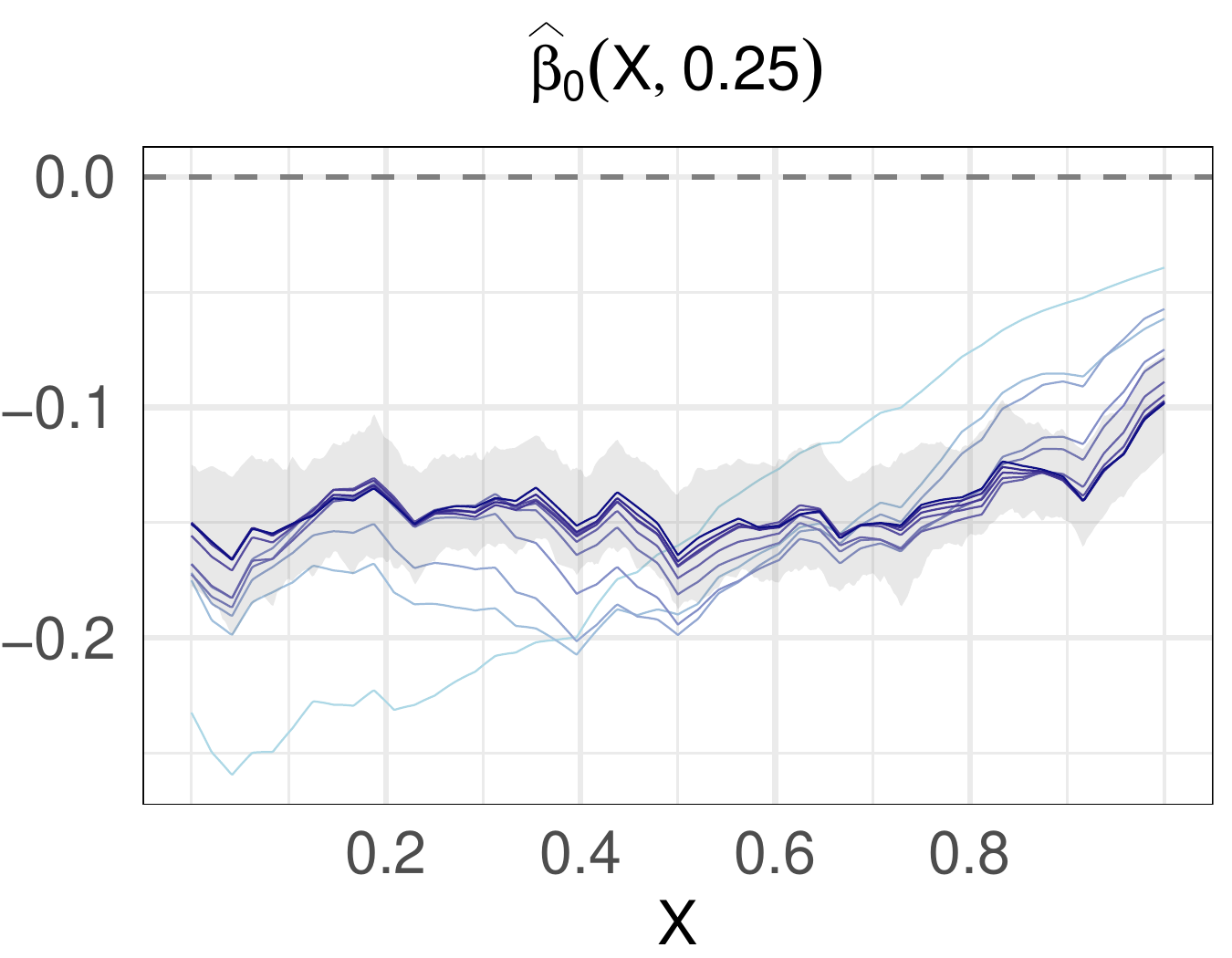}
\end{subfigure}
\begin{subfigure}{0.23\textwidth}
  \includegraphics[width=\textwidth]{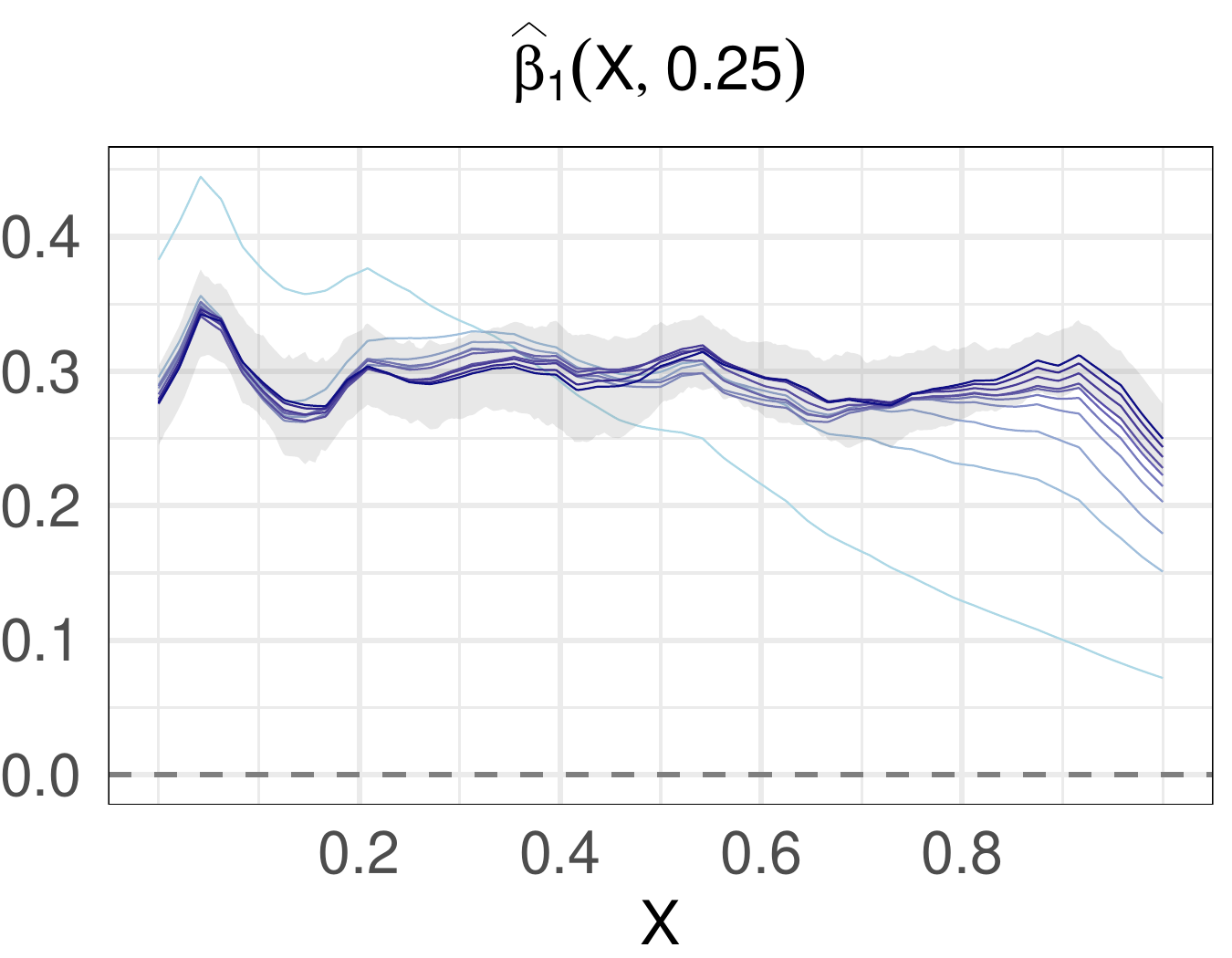}
\end{subfigure}
\begin{subfigure}{0.23\textwidth}
  \includegraphics[width=\textwidth]{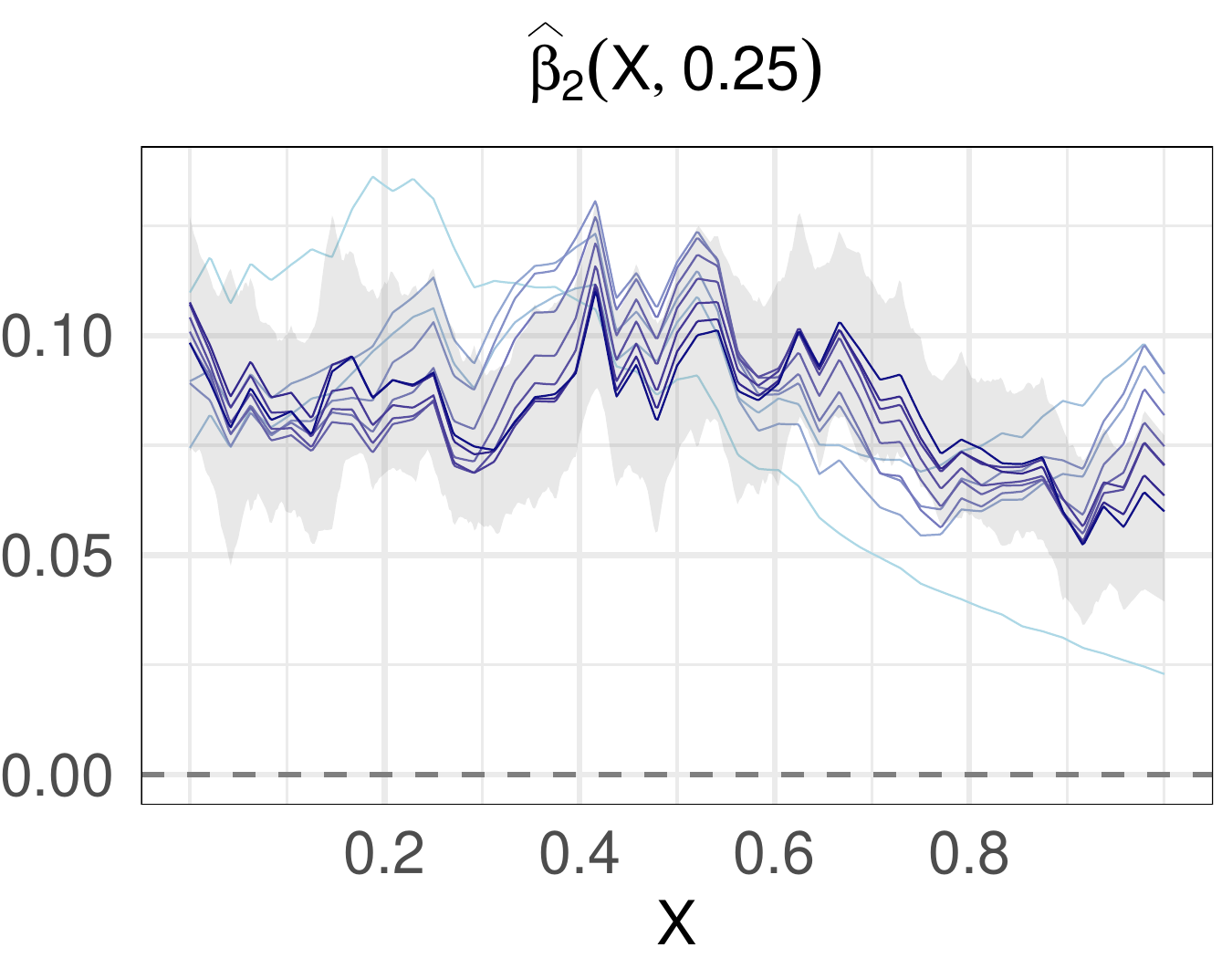}
\end{subfigure}
\begin{subfigure}{0.23\textwidth}
  \includegraphics[width=\textwidth]{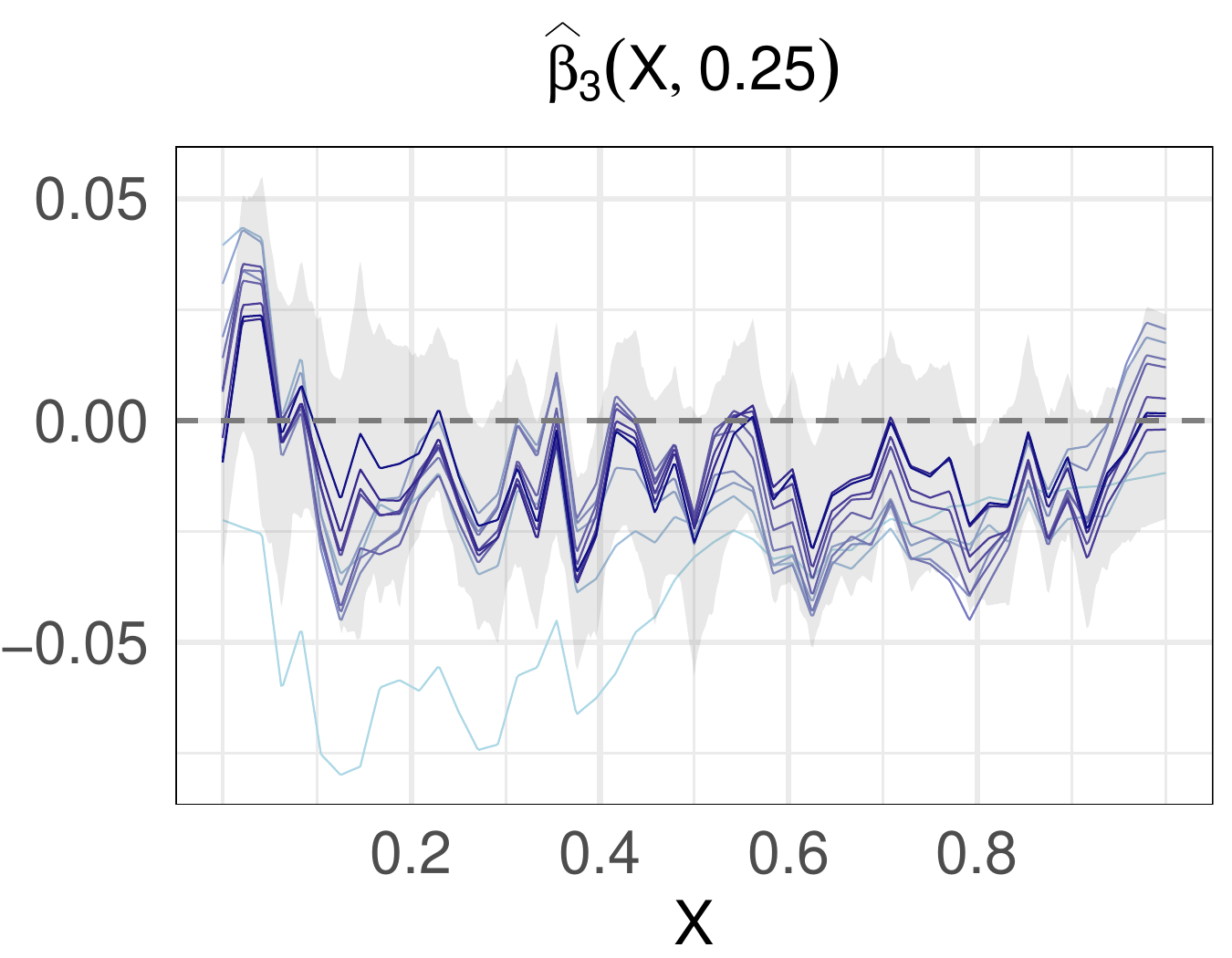}
\end{subfigure}

\begin{subfigure}{0.23\textwidth}
  \includegraphics[width=\textwidth]{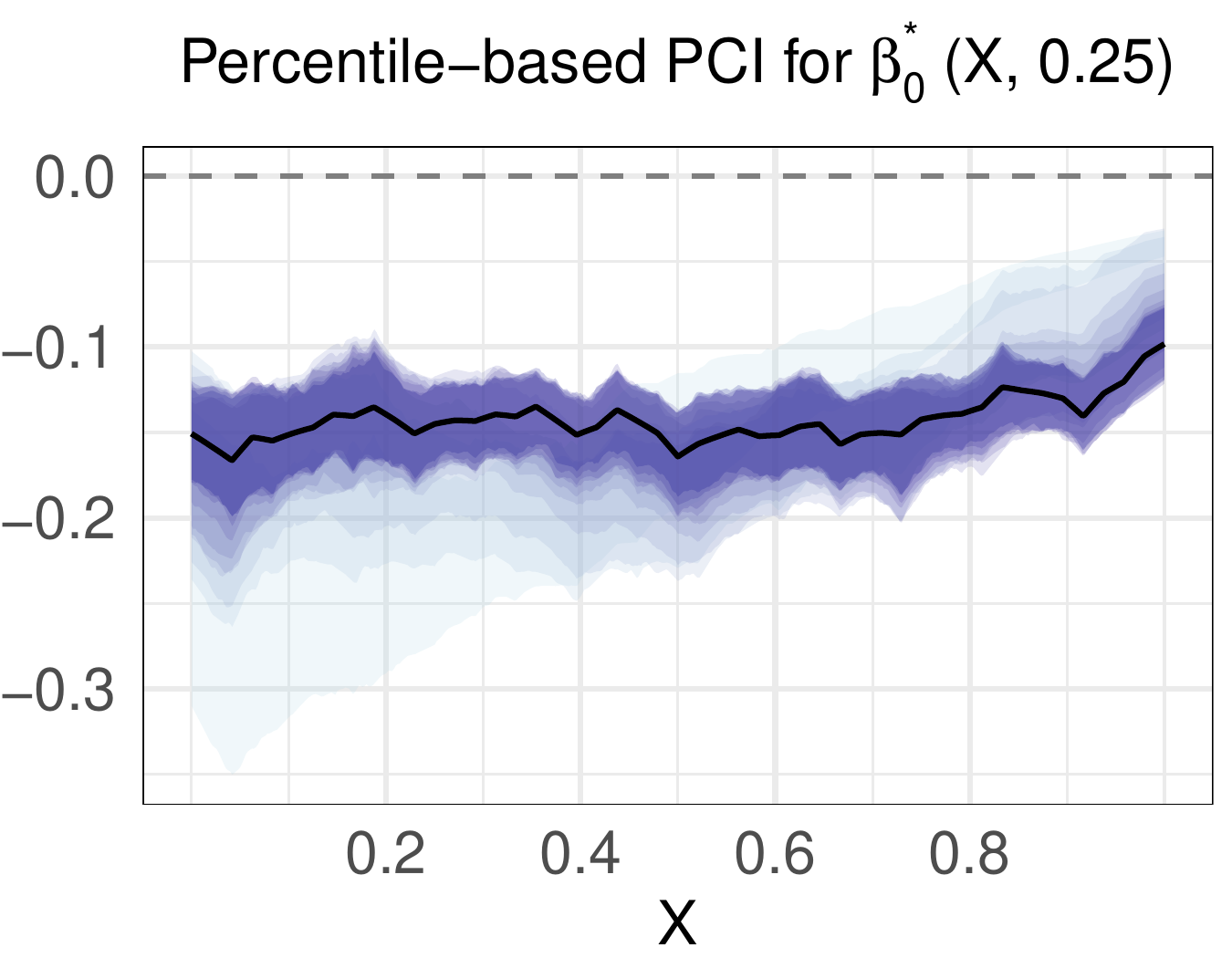}
\end{subfigure}
\begin{subfigure}{0.23\textwidth}
  \includegraphics[width=\textwidth]{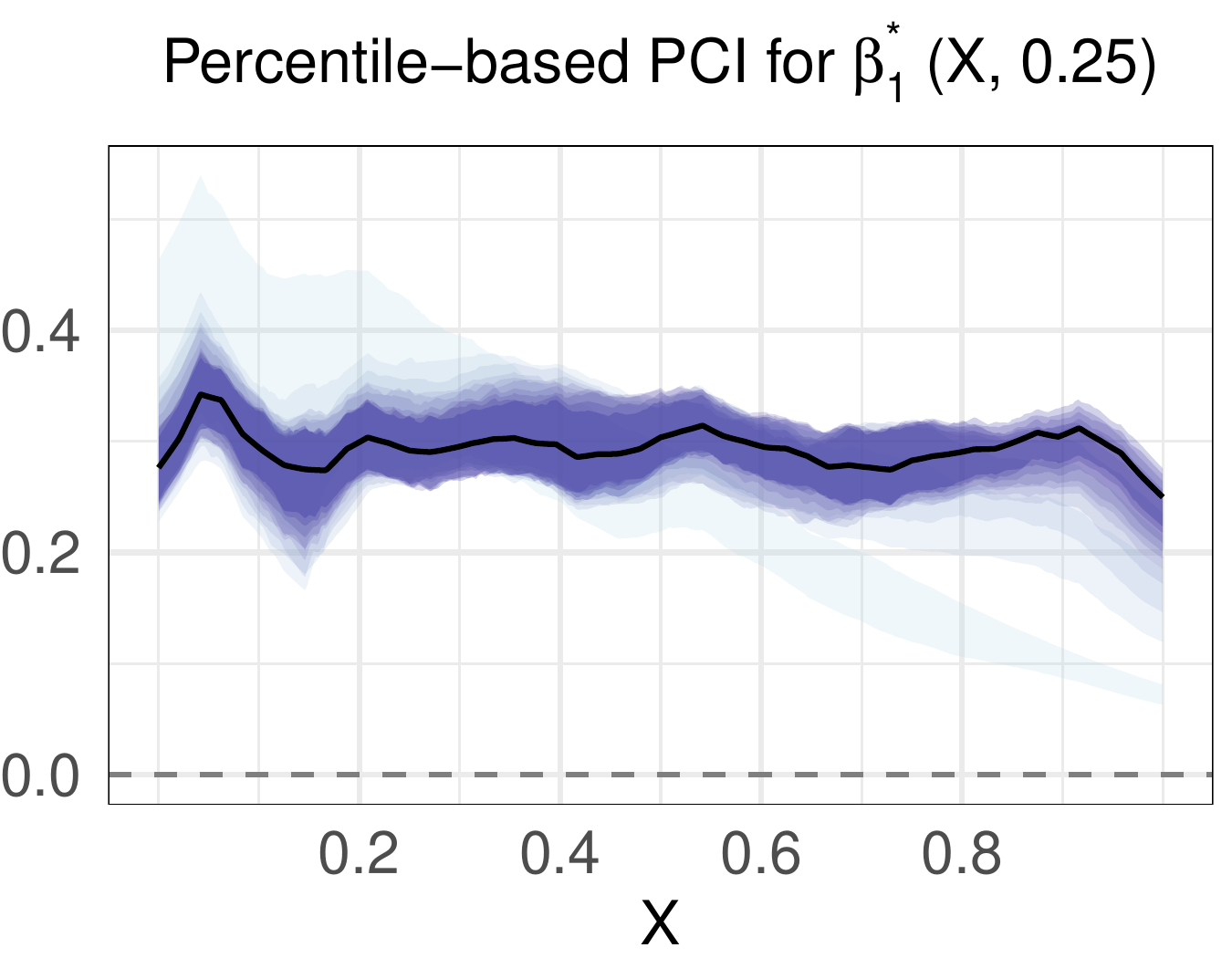}
\end{subfigure}
\begin{subfigure}{0.23\textwidth}
  \includegraphics[width=\textwidth]{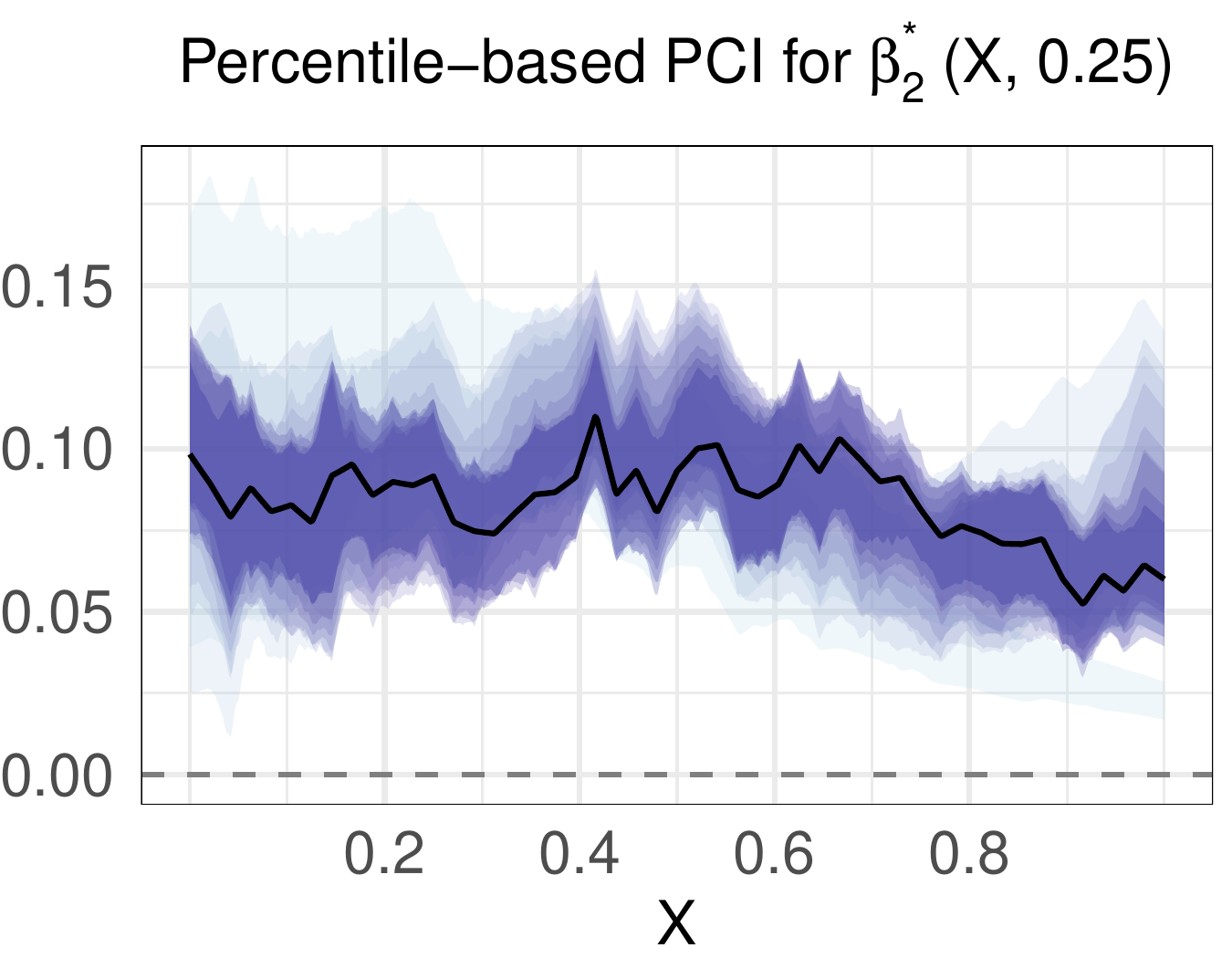}
\end{subfigure}
\begin{subfigure}{0.23\textwidth}
  \includegraphics[width=\textwidth]{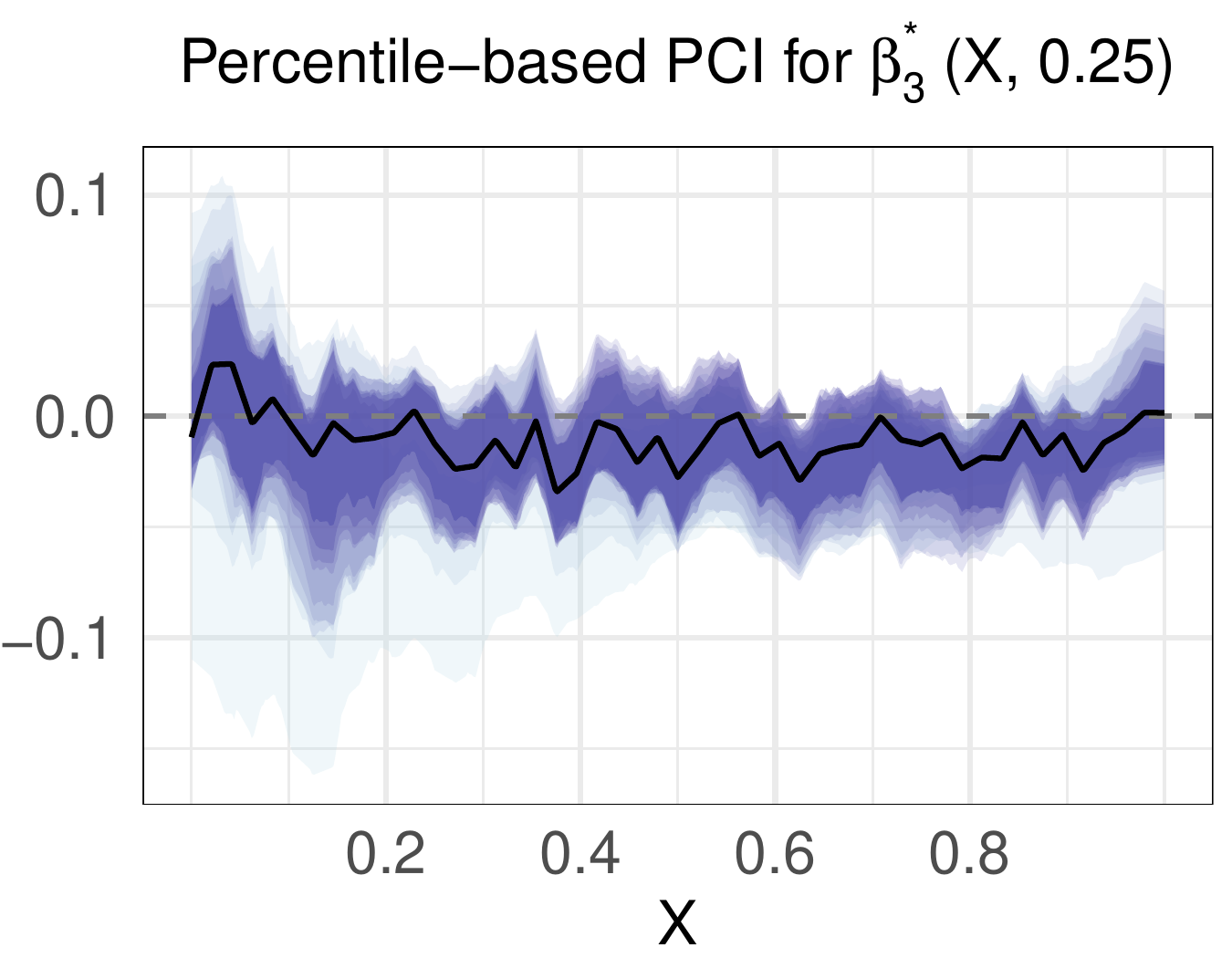}
\end{subfigure}

\begin{subfigure}{0.23\textwidth}
  \includegraphics[width=\textwidth]{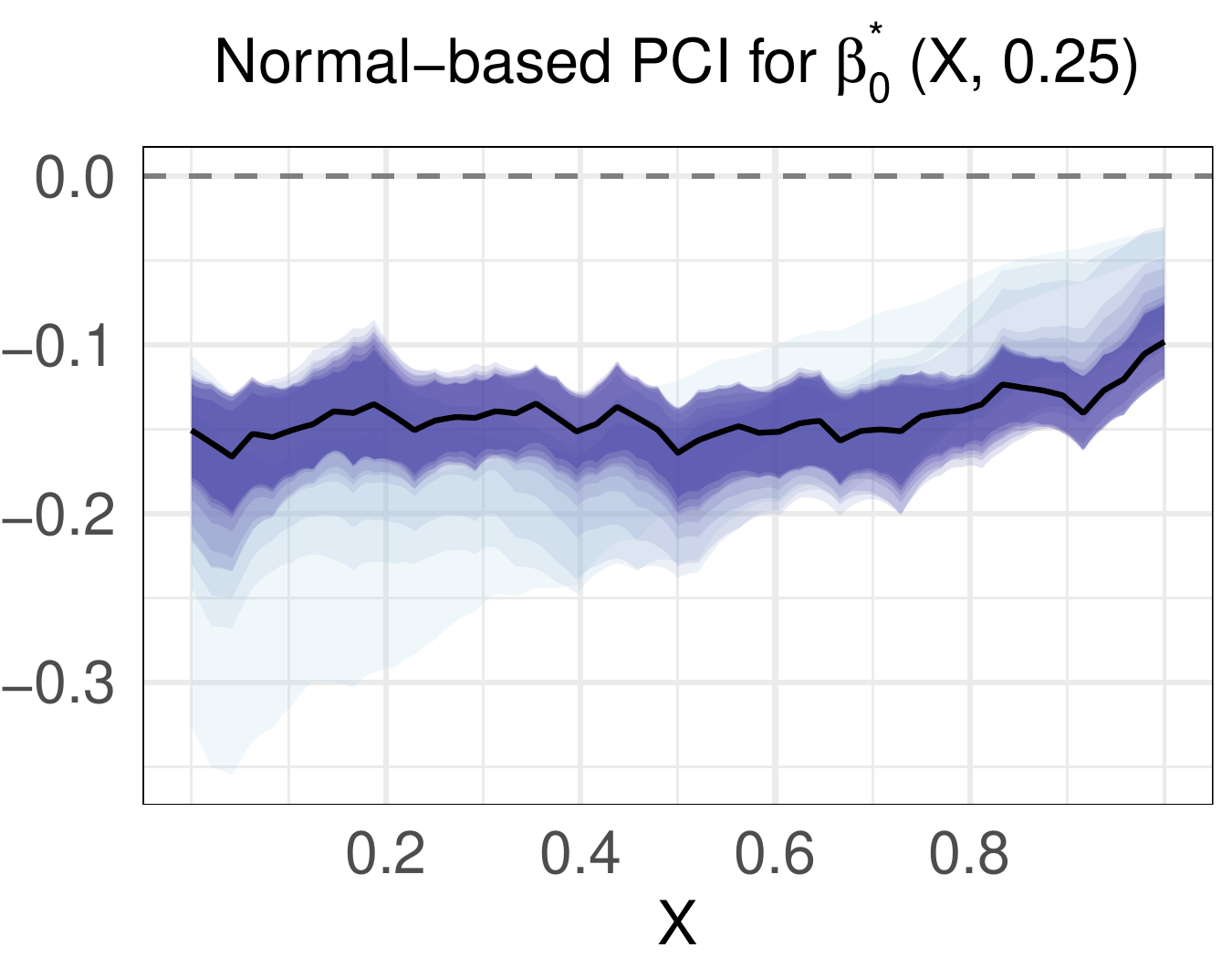}
\end{subfigure}
\begin{subfigure}{0.23\textwidth}
  \includegraphics[width=\textwidth]{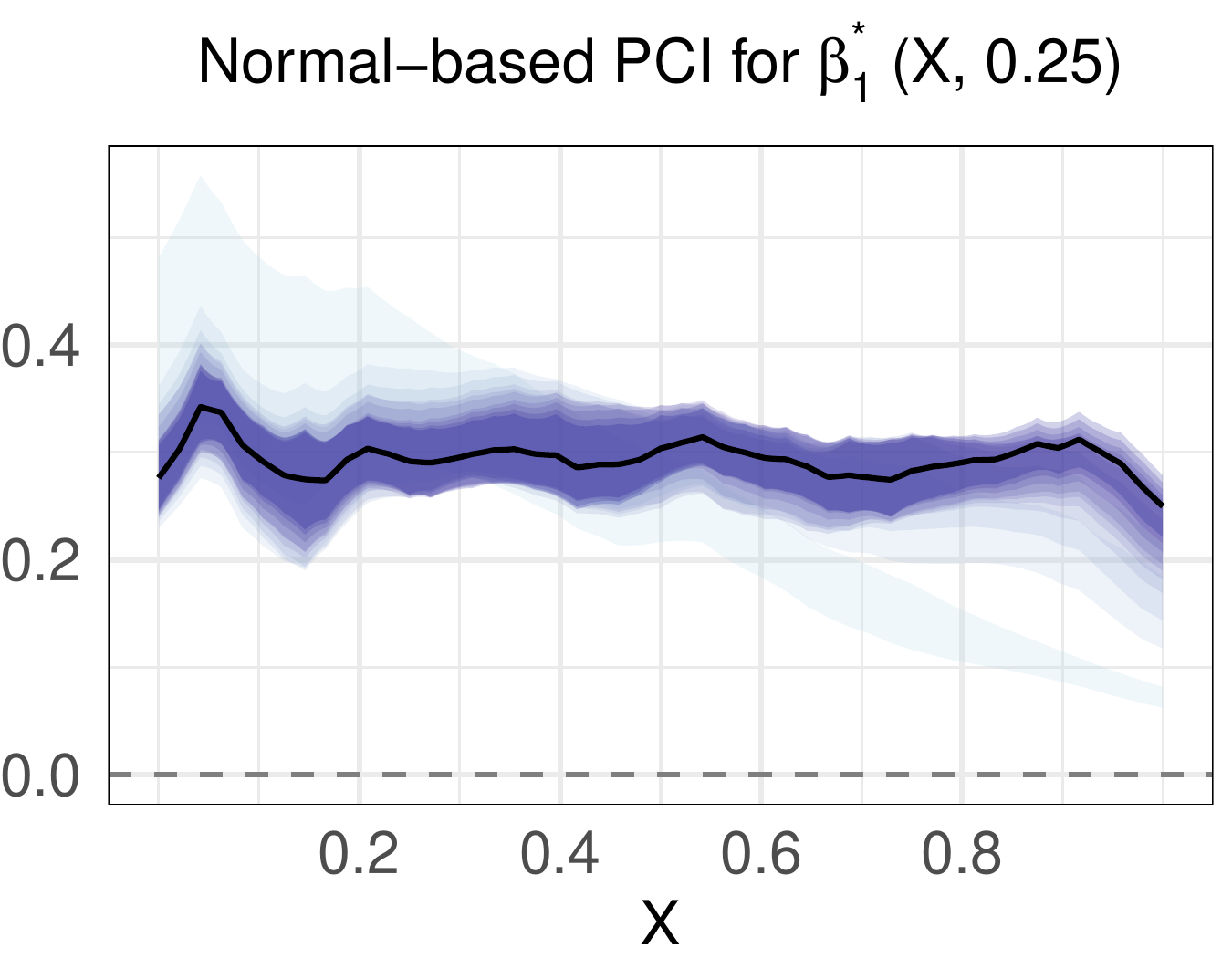}
\end{subfigure}
\begin{subfigure}{0.23\textwidth}
  \includegraphics[width=\textwidth]{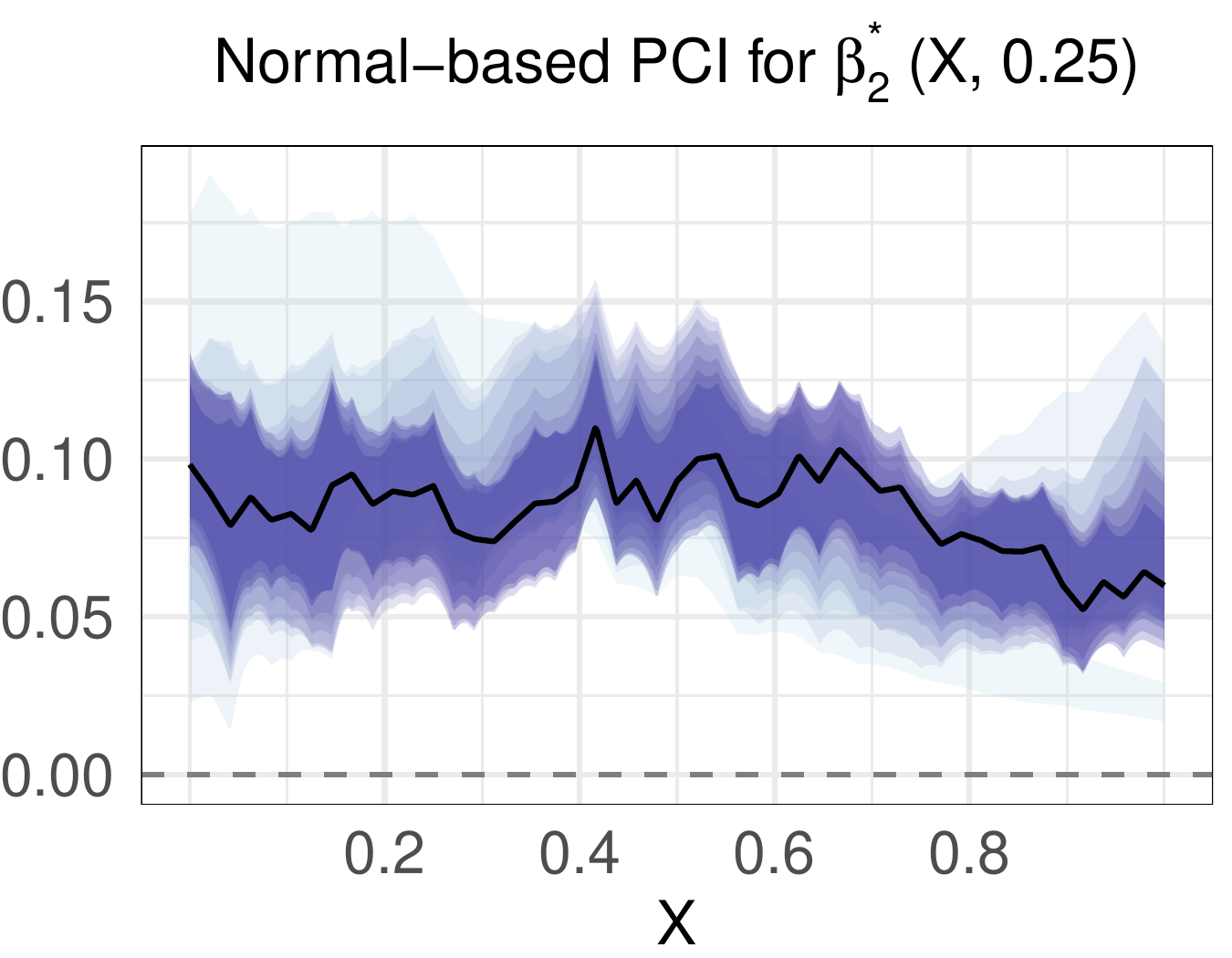}
\end{subfigure}
\begin{subfigure}{0.23\textwidth}
  \includegraphics[width=\textwidth]{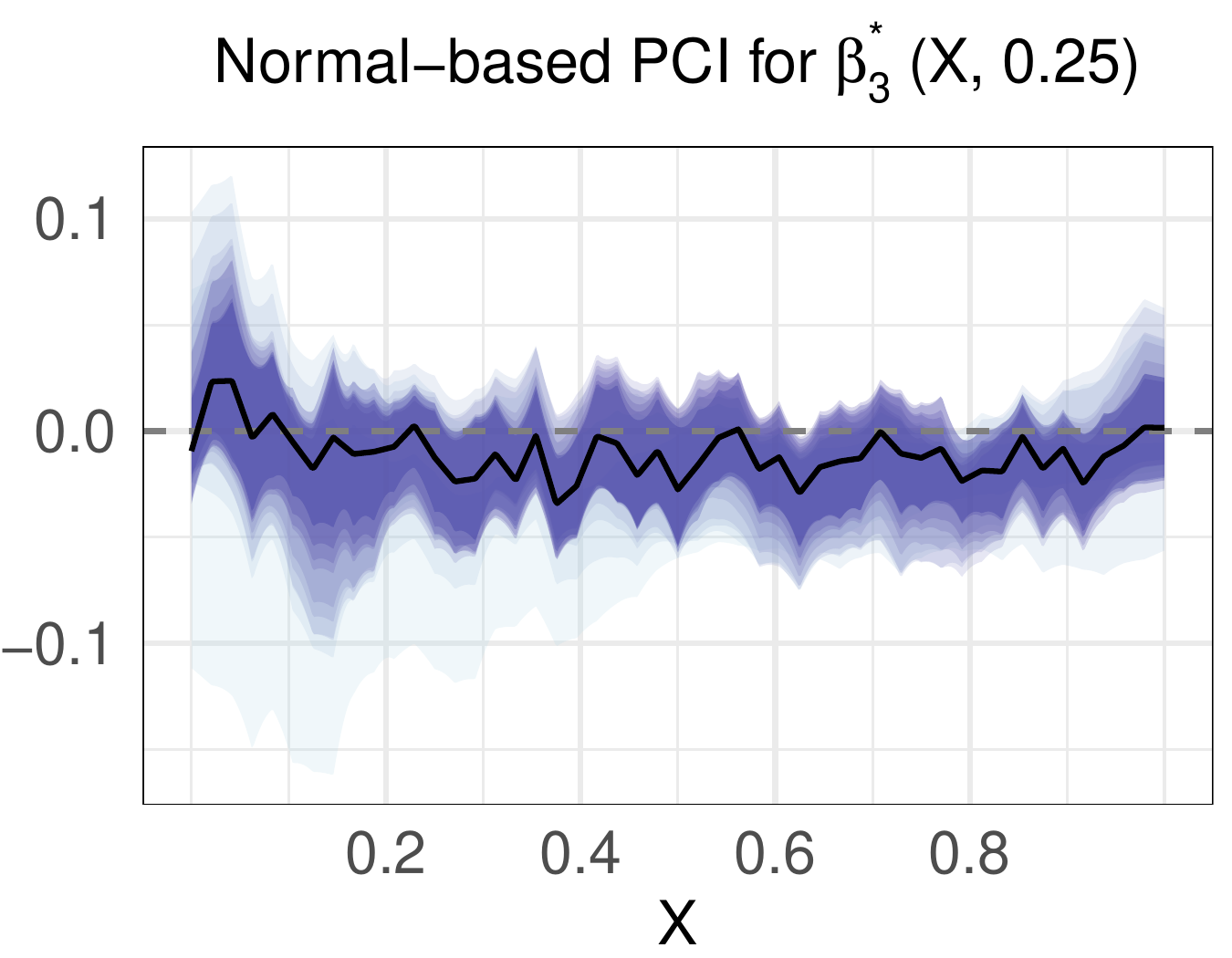}
\end{subfigure}

\begin{subfigure}{\textwidth}
  \centering
  \includegraphics[width=0.8\textwidth]{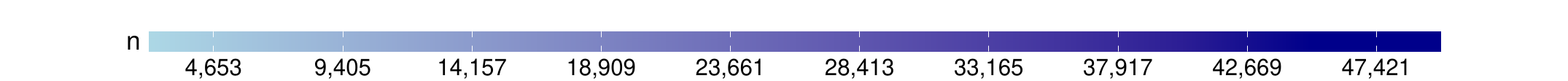}
\end{subfigure}
\caption{Online estimated curves and PCI trajectories via FSGD-NC at $\tau=0.25$. }
\label{fig:NC_traj_0.25}
\end{figure}

\begin{figure}[htbp]
\def\baselinestretch{1}
\centering
\begin{subfigure}{0.23\textwidth}
  \includegraphics[width=\textwidth]{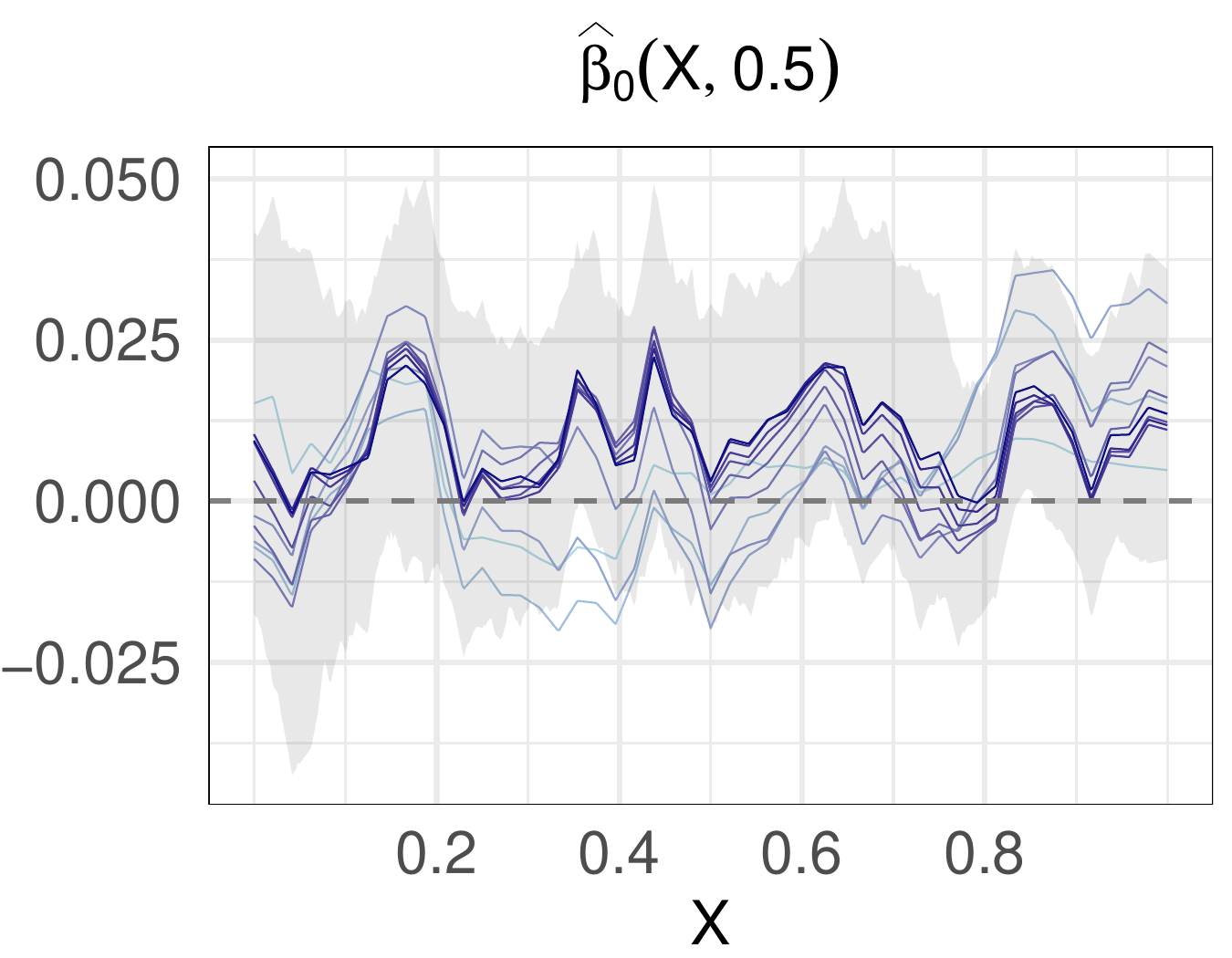}
\end{subfigure}
\begin{subfigure}{0.23\textwidth}
  \includegraphics[width=\textwidth]{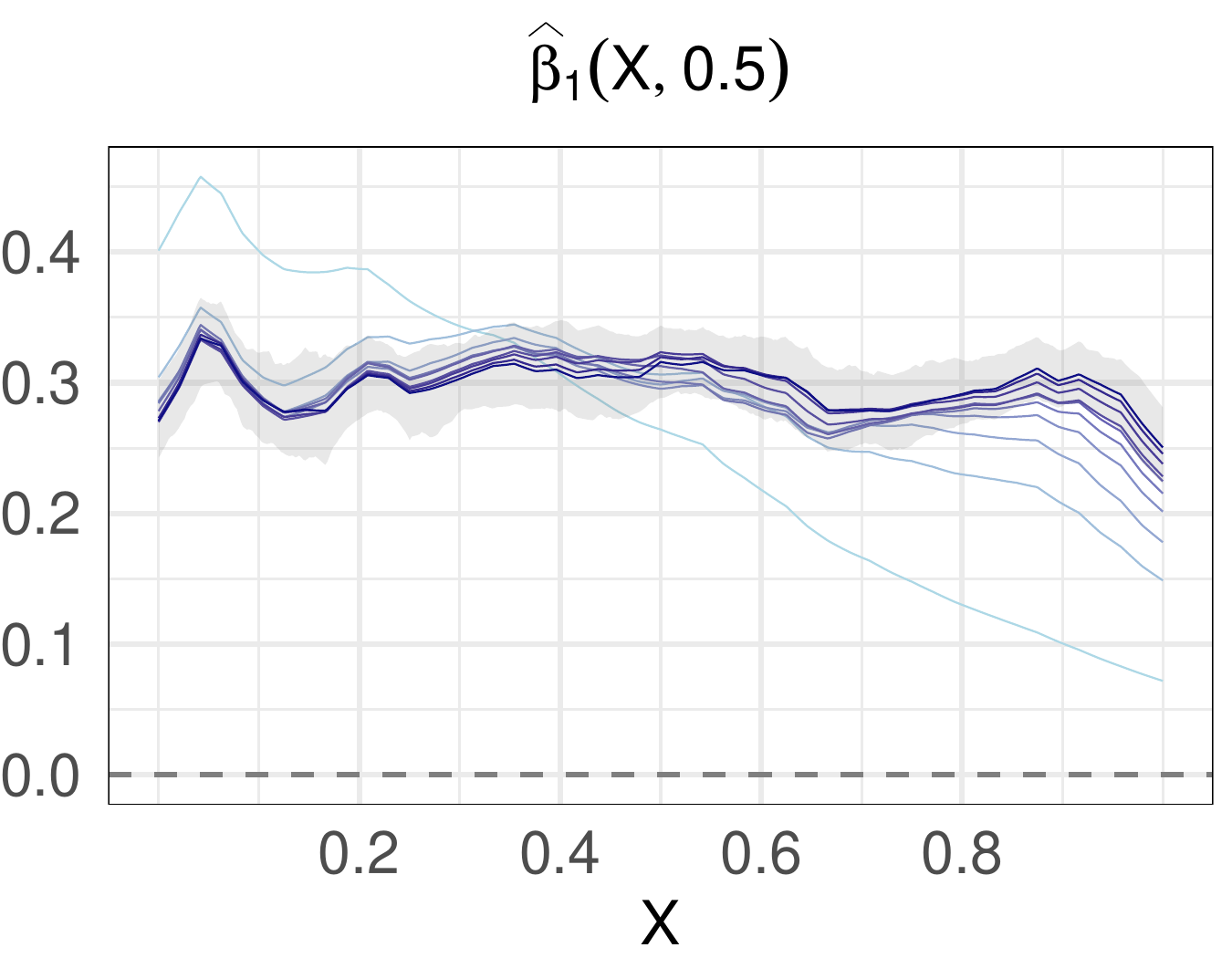}
\end{subfigure}
\begin{subfigure}{0.23\textwidth}
  \includegraphics[width=\textwidth]{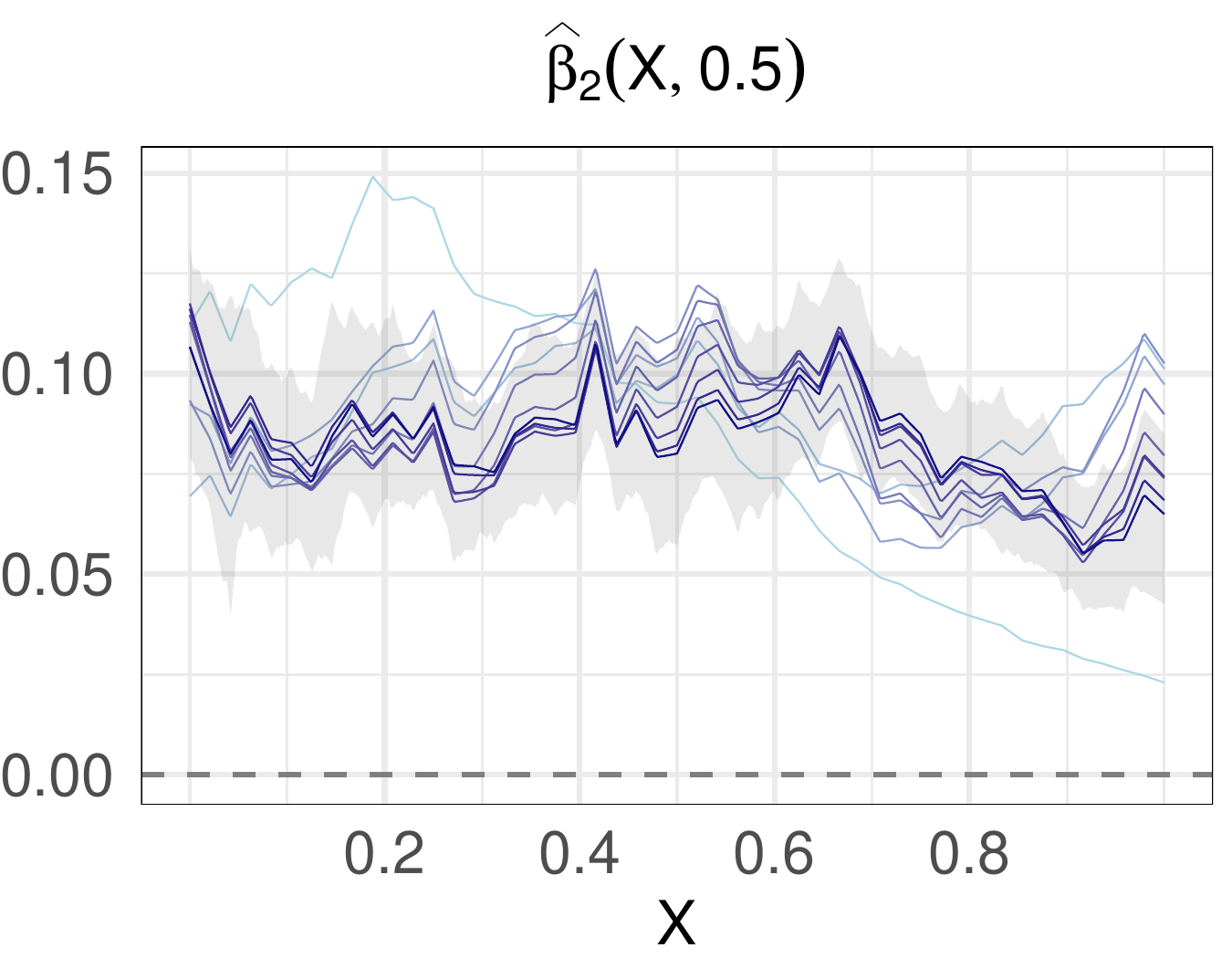}
\end{subfigure}
\begin{subfigure}{0.23\textwidth}
  \includegraphics[width=\textwidth]{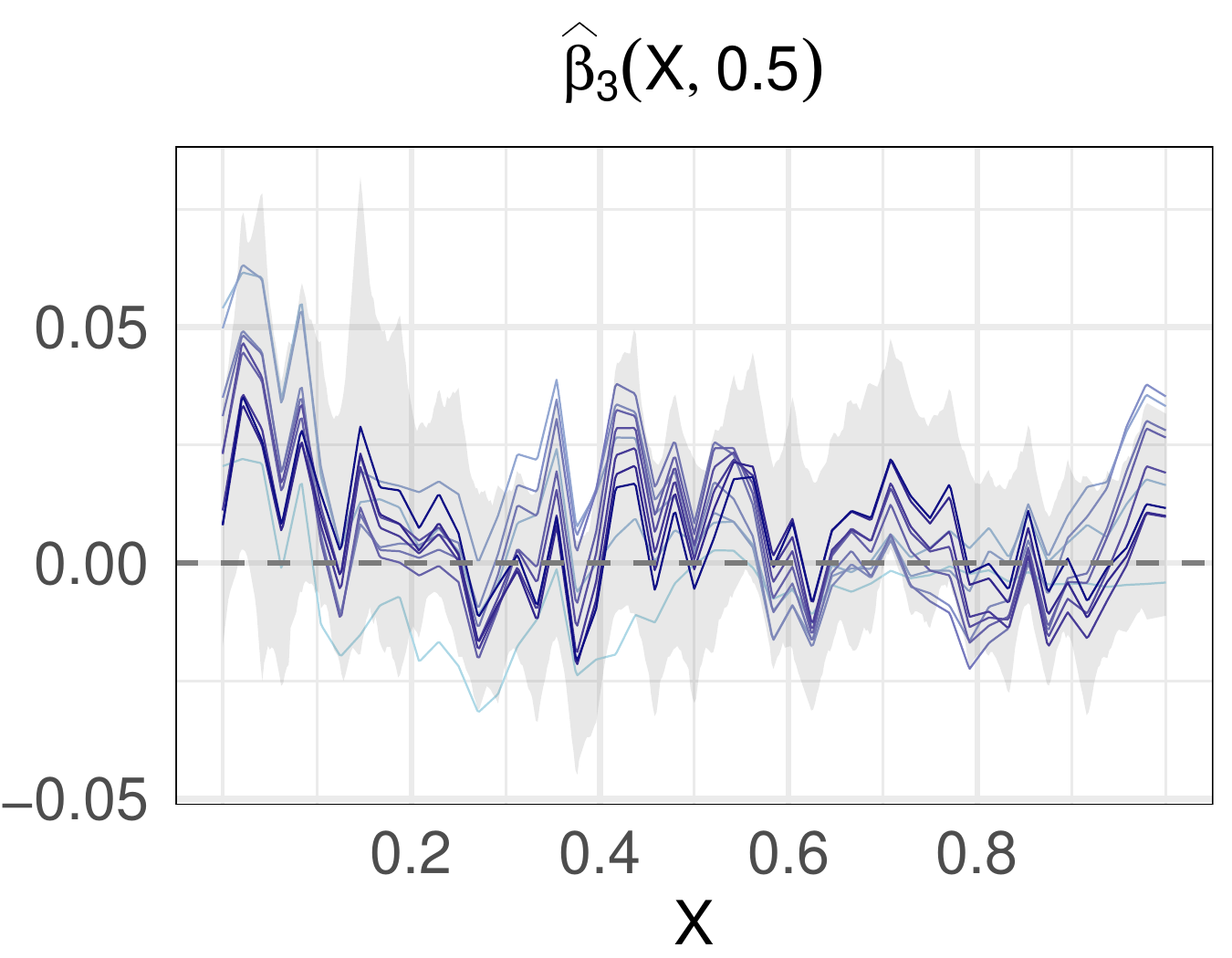}
\end{subfigure}

\begin{subfigure}{0.23\textwidth}
  \includegraphics[width=\textwidth]{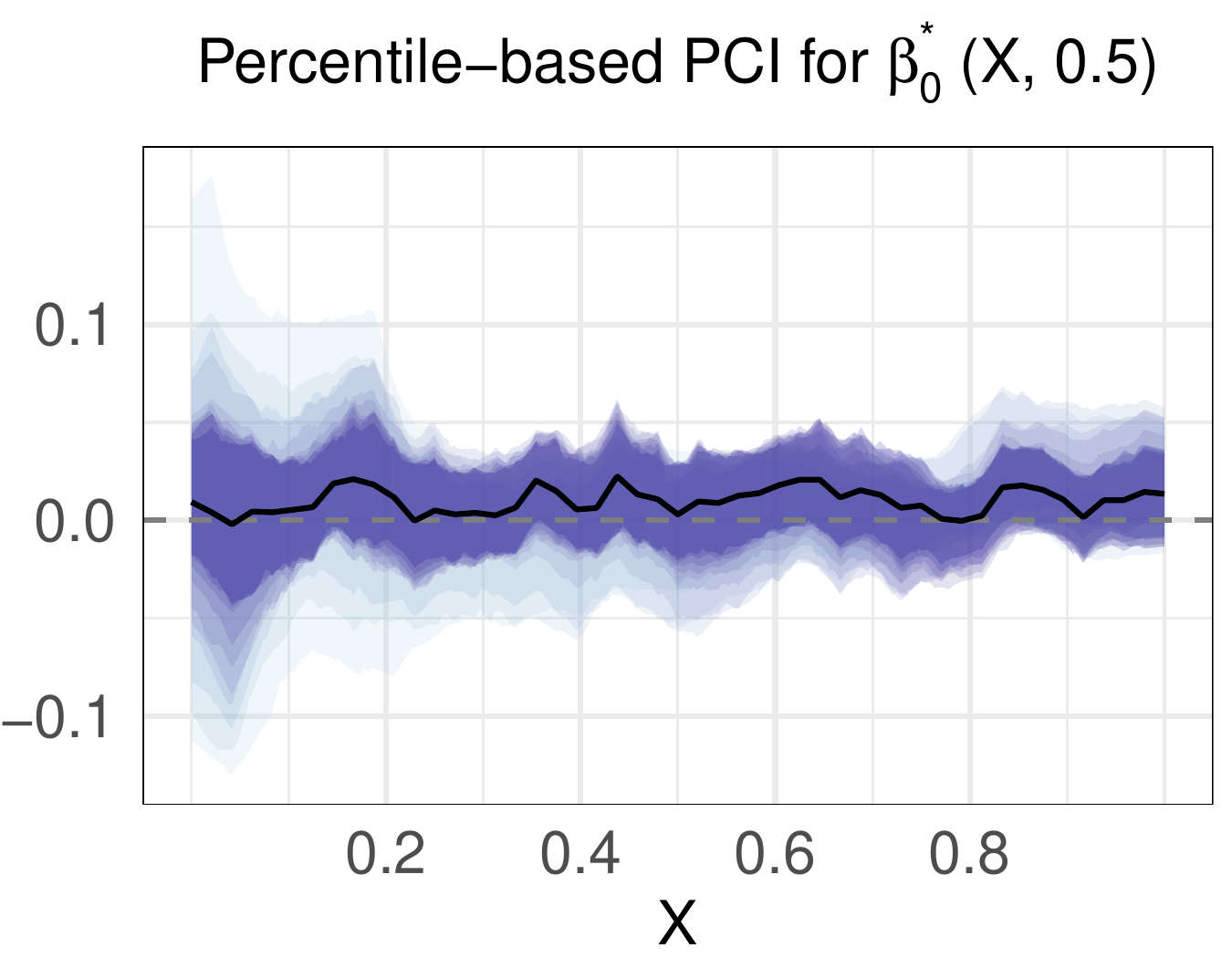}
\end{subfigure}
\begin{subfigure}{0.23\textwidth}
  \includegraphics[width=\textwidth]{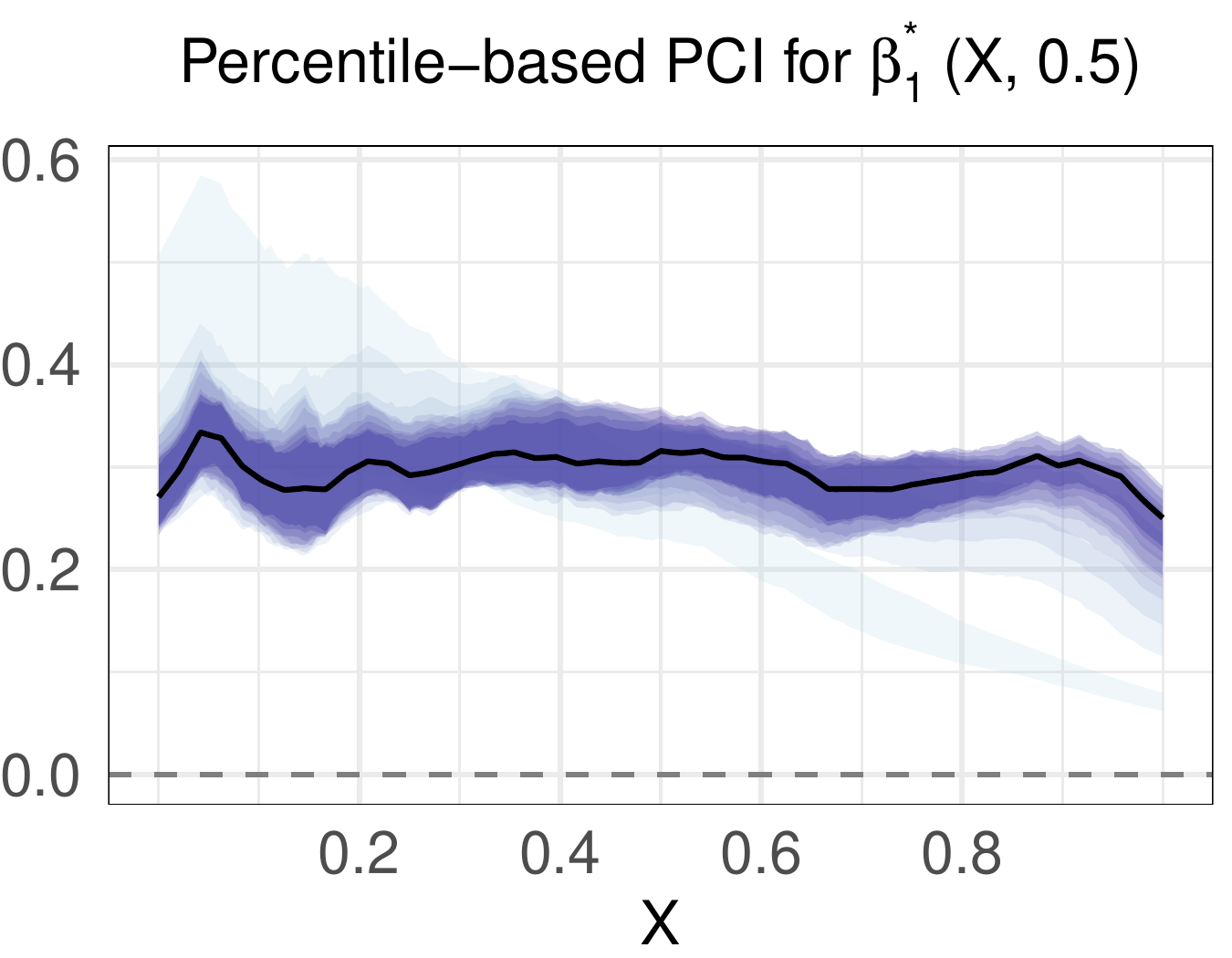}
\end{subfigure}
\begin{subfigure}{0.23\textwidth}
  \includegraphics[width=\textwidth]{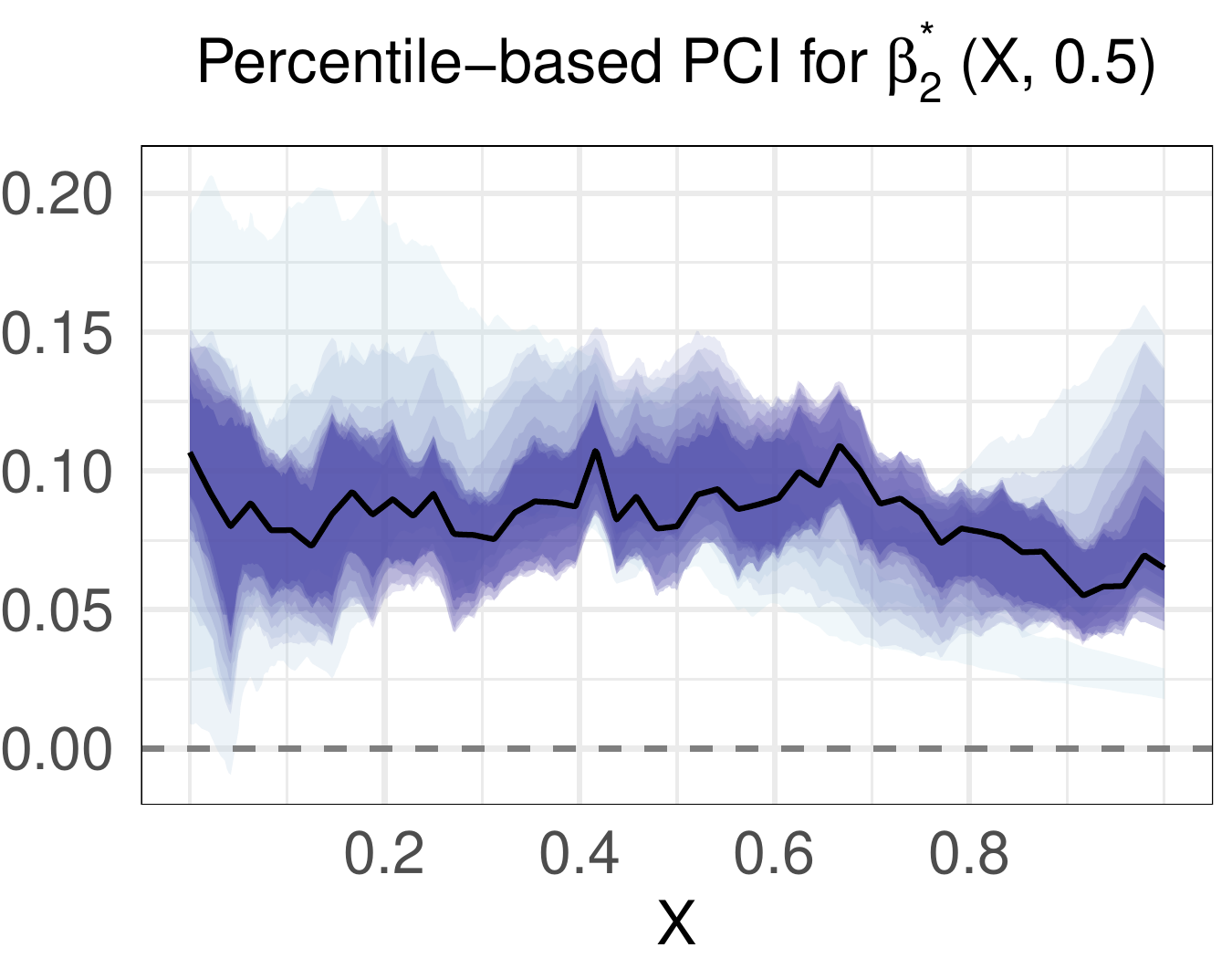}
\end{subfigure}
\begin{subfigure}{0.23\textwidth}
  \includegraphics[width=\textwidth]{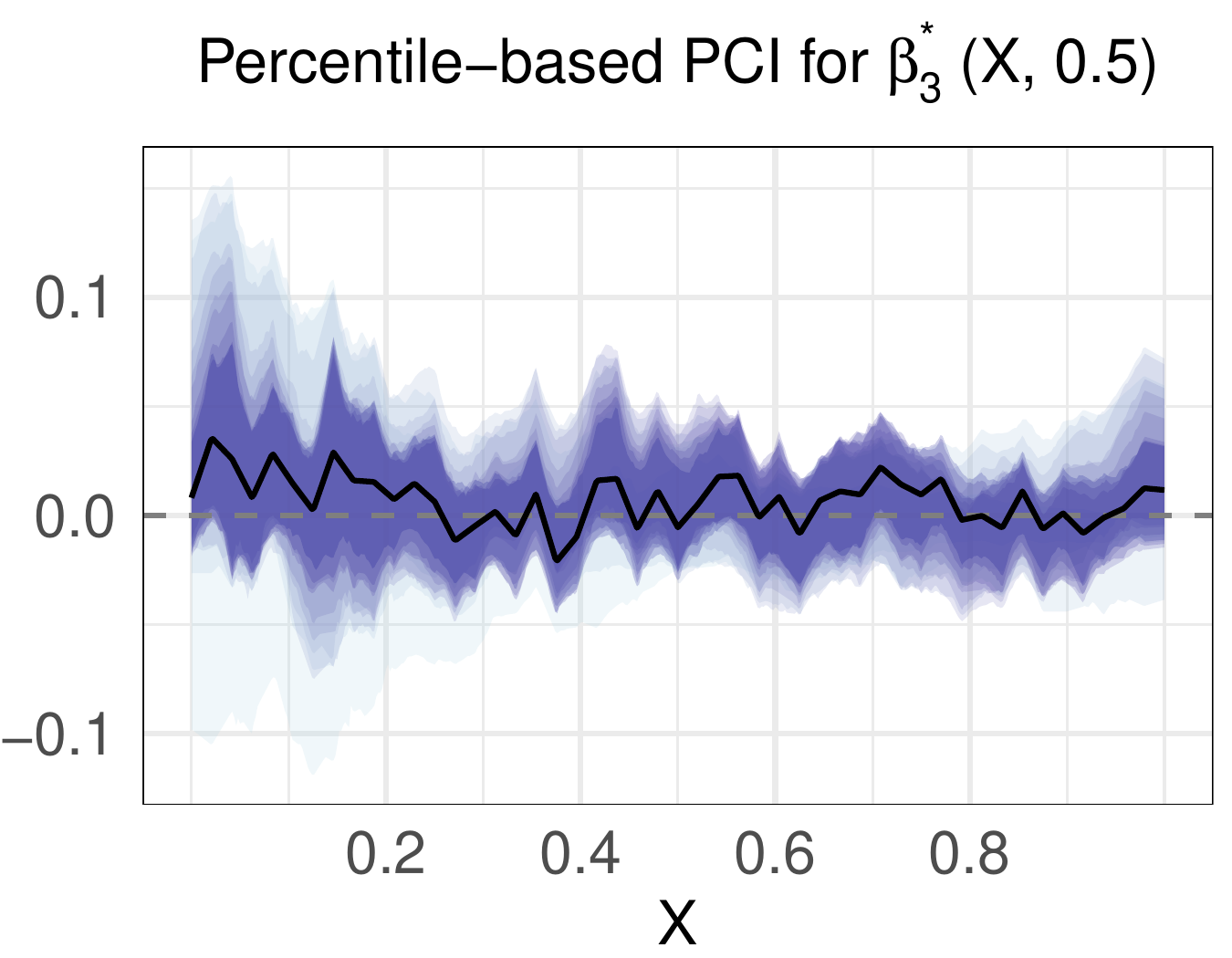}
\end{subfigure}

\begin{subfigure}{0.23\textwidth}
  \includegraphics[width=\textwidth]{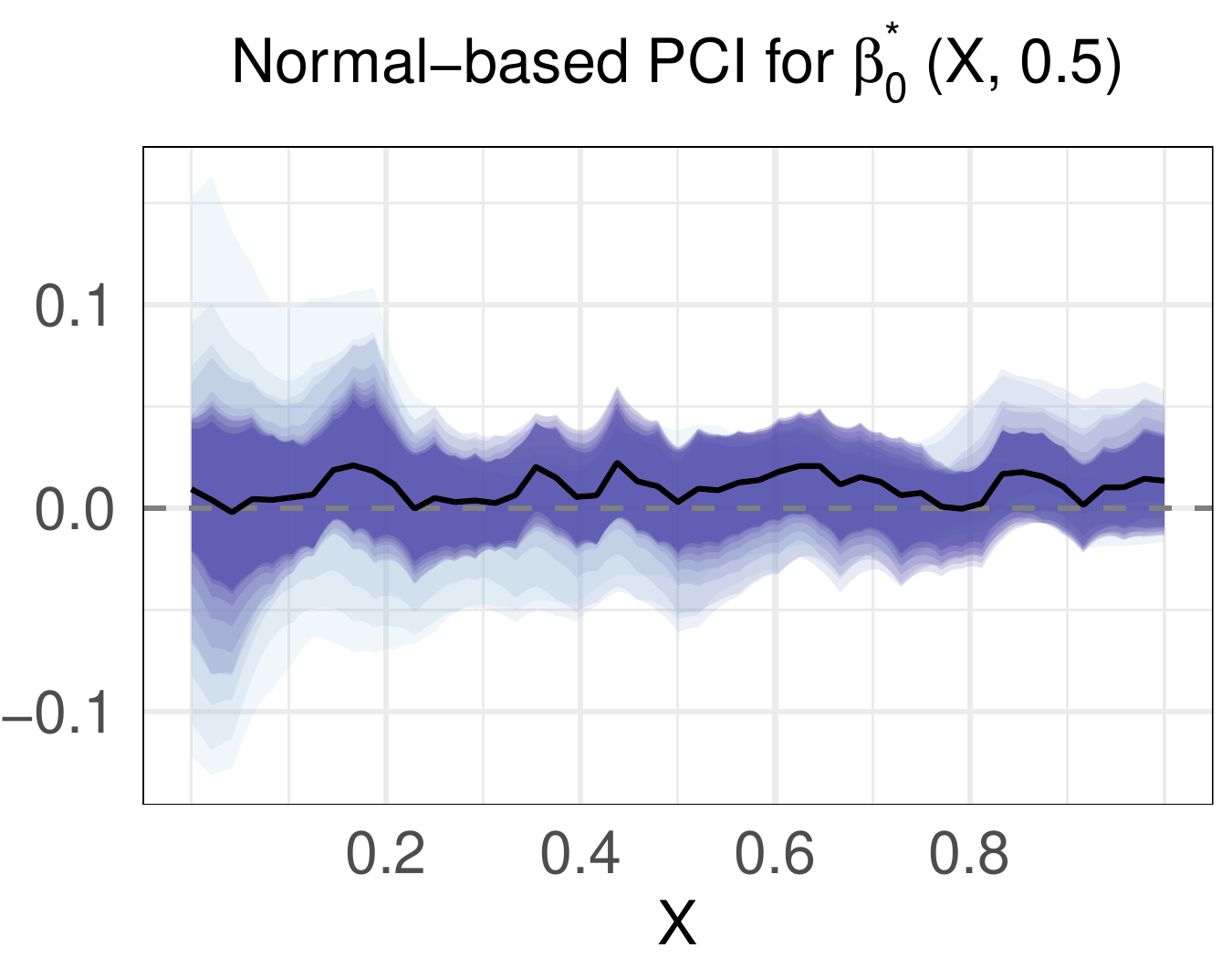}
\end{subfigure}
\begin{subfigure}{0.23\textwidth}
  \includegraphics[width=\textwidth]{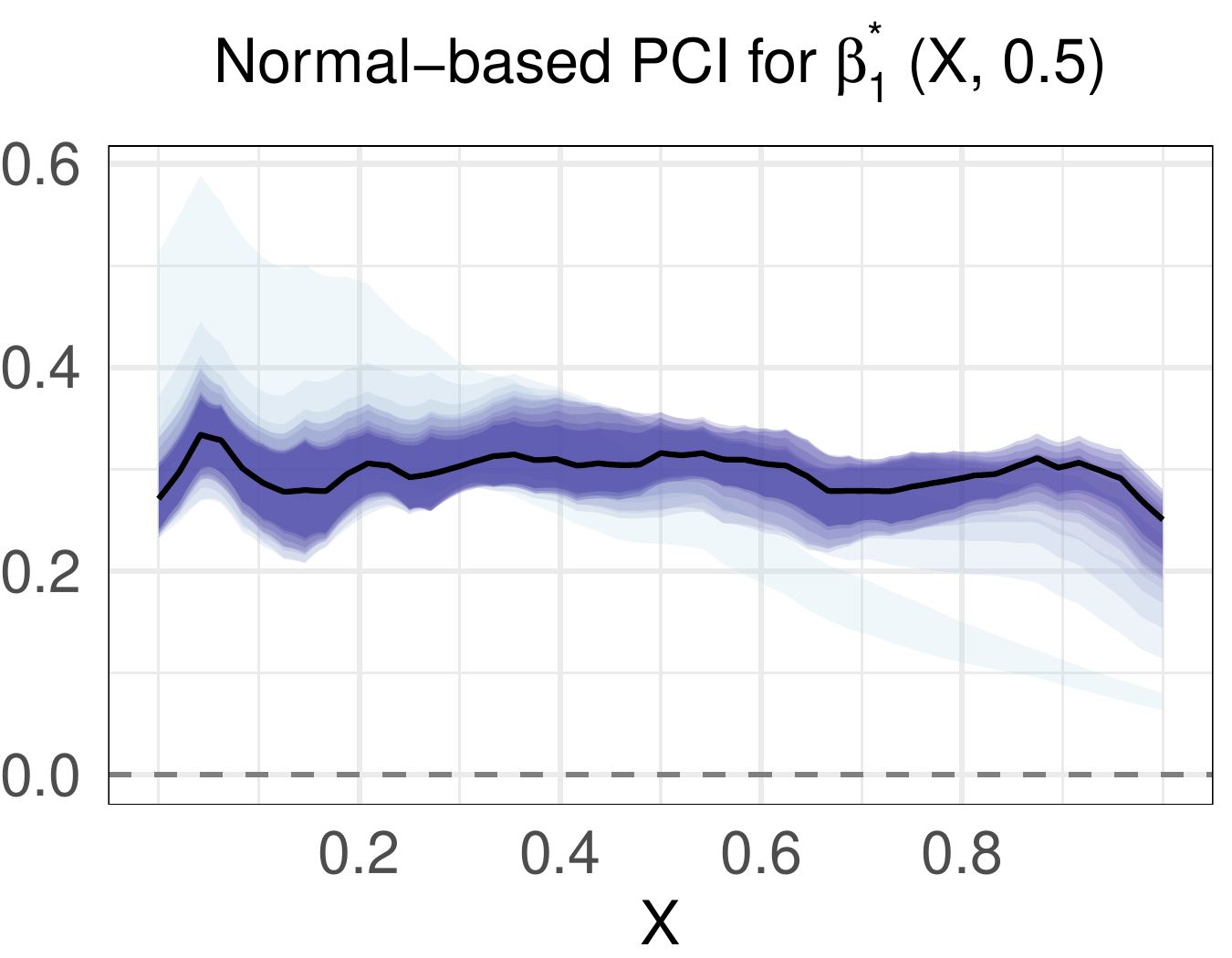}
\end{subfigure}
\begin{subfigure}{0.23\textwidth}
  \includegraphics[width=\textwidth]{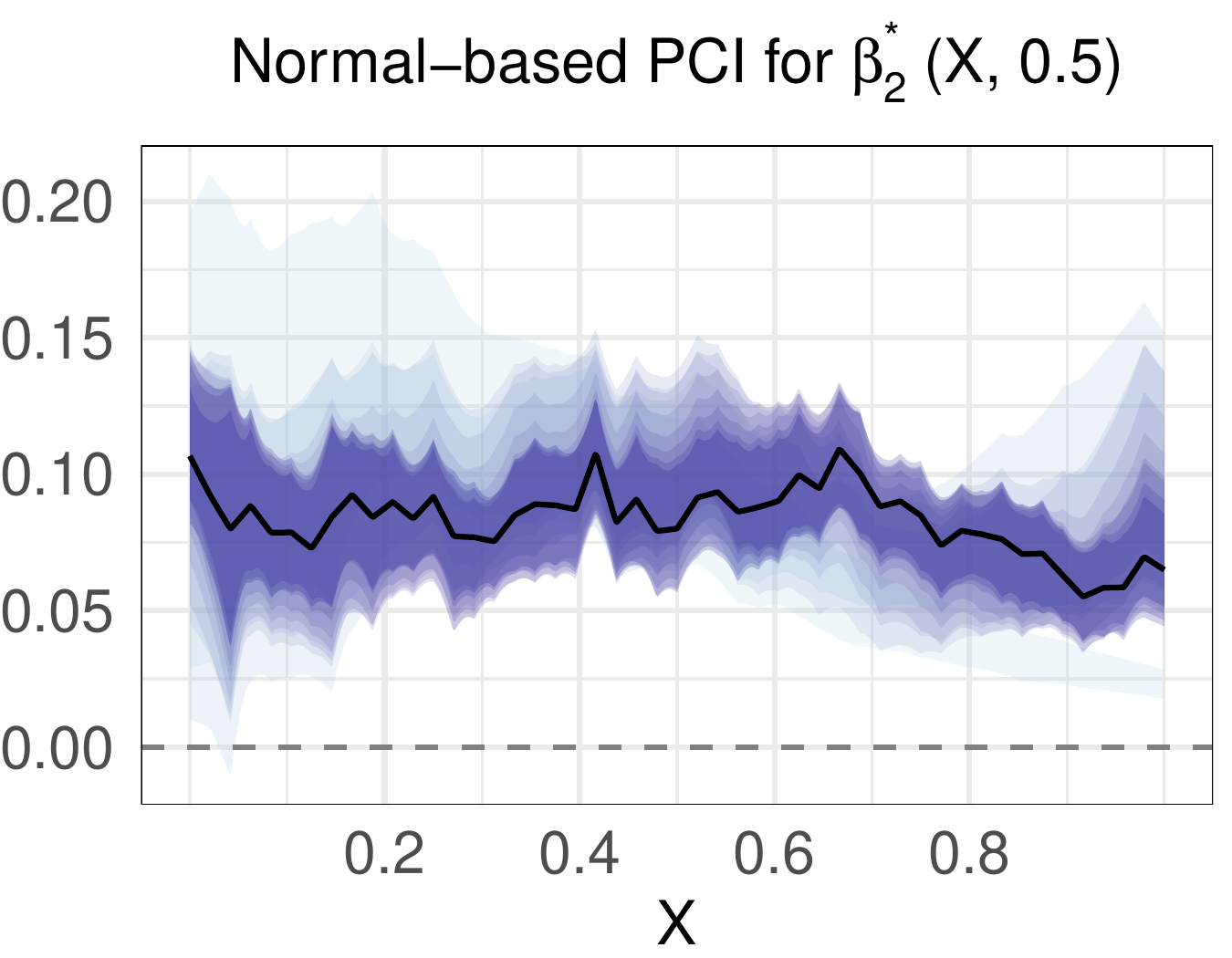}
\end{subfigure}
\begin{subfigure}{0.23\textwidth}
  \includegraphics[width=\textwidth]{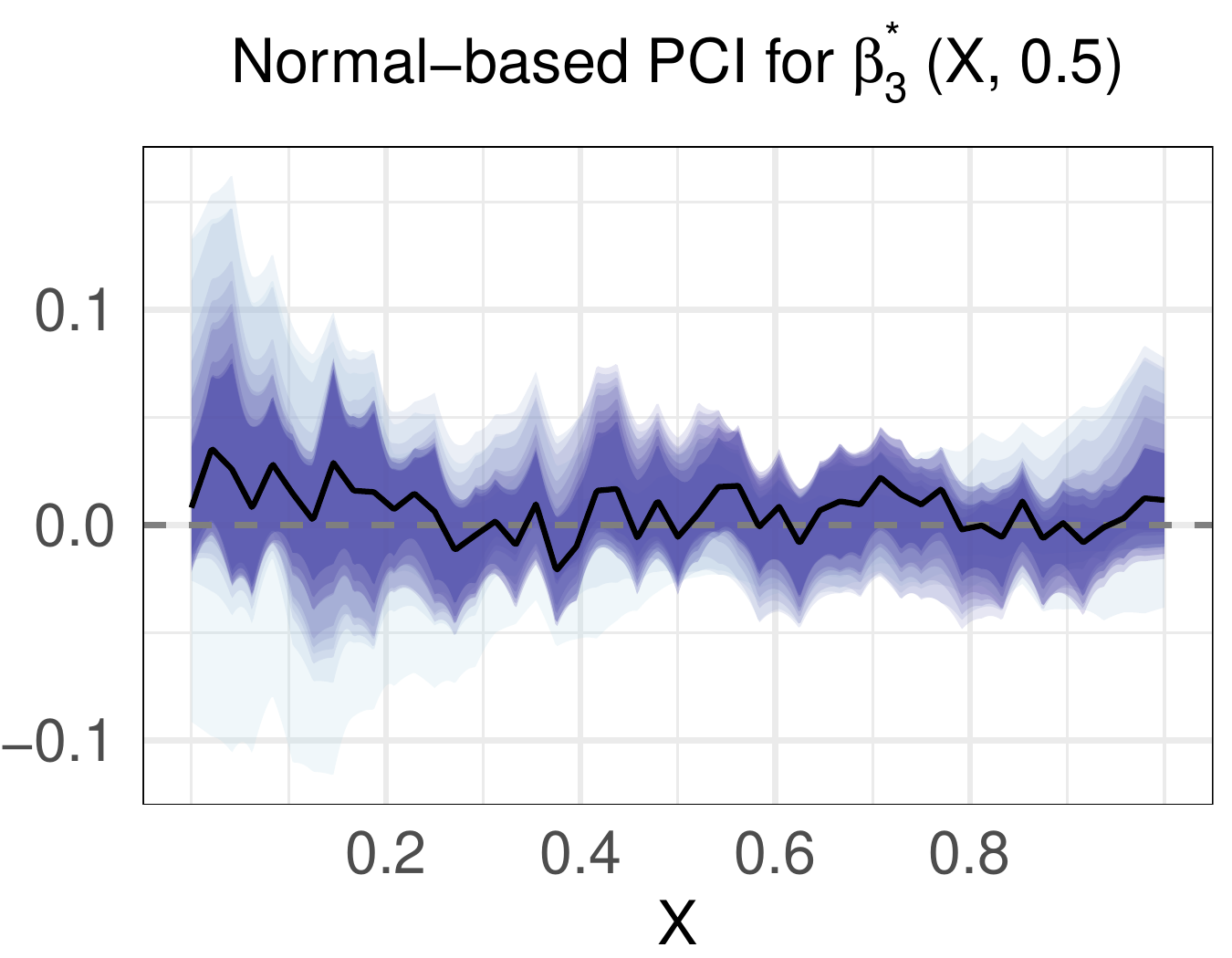}
\end{subfigure}

\begin{subfigure}{\textwidth}
  \centering
  \includegraphics[width=0.8\textwidth]{Rdfigures/FSGD_n_legend.pdf}
\end{subfigure}
\caption{Online estimated curves and PCI trajectories via FSGD-NC at $\tau=0.5$.}
\label{fig:NC_traj_0.5}
\end{figure}

\begin{figure}[htbp]
\def\baselinestretch{1}
\centering
\begin{subfigure}{0.23\textwidth}
  \includegraphics[width=\textwidth]{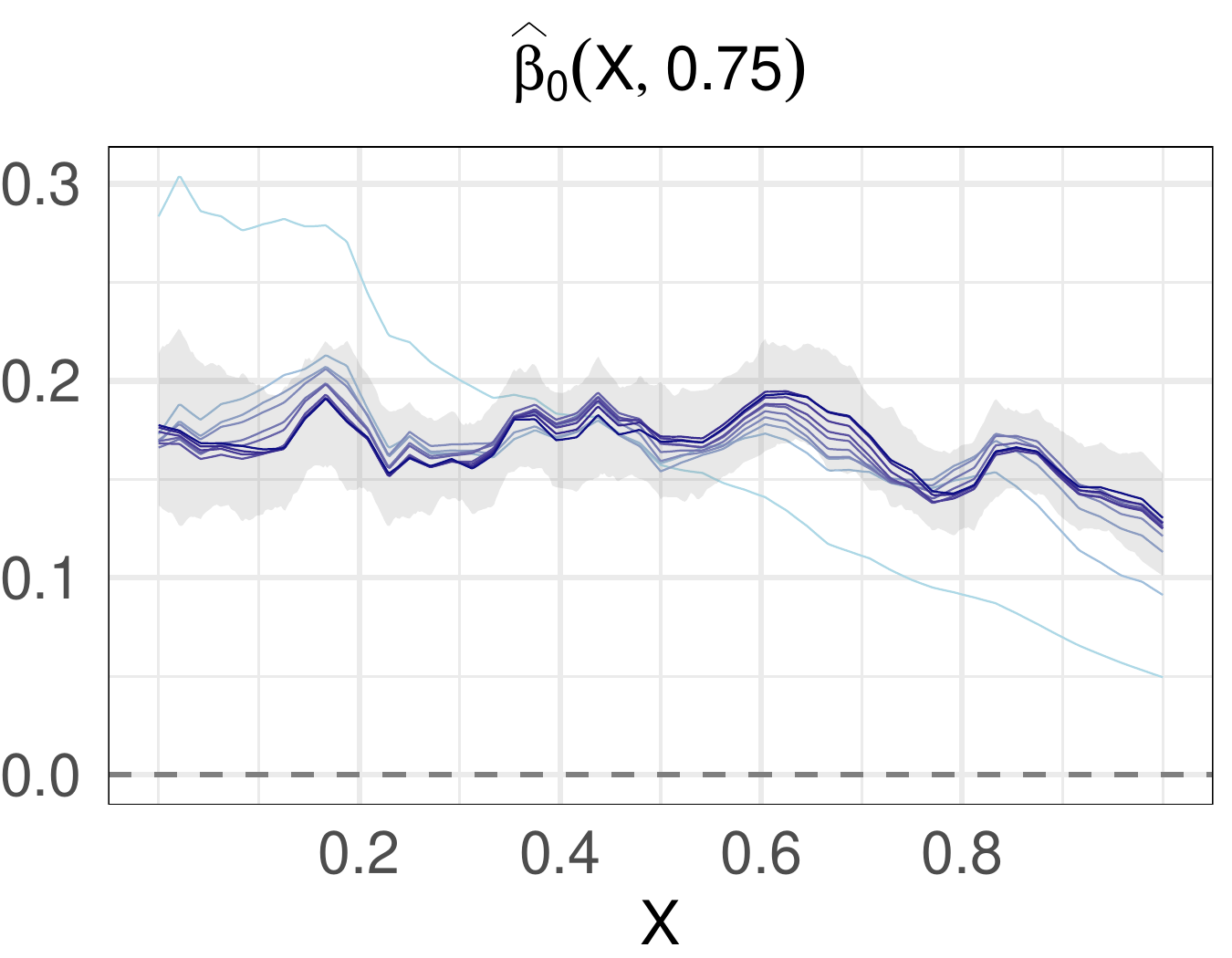}
\end{subfigure}
\begin{subfigure}{0.23\textwidth}
  \includegraphics[width=\textwidth]{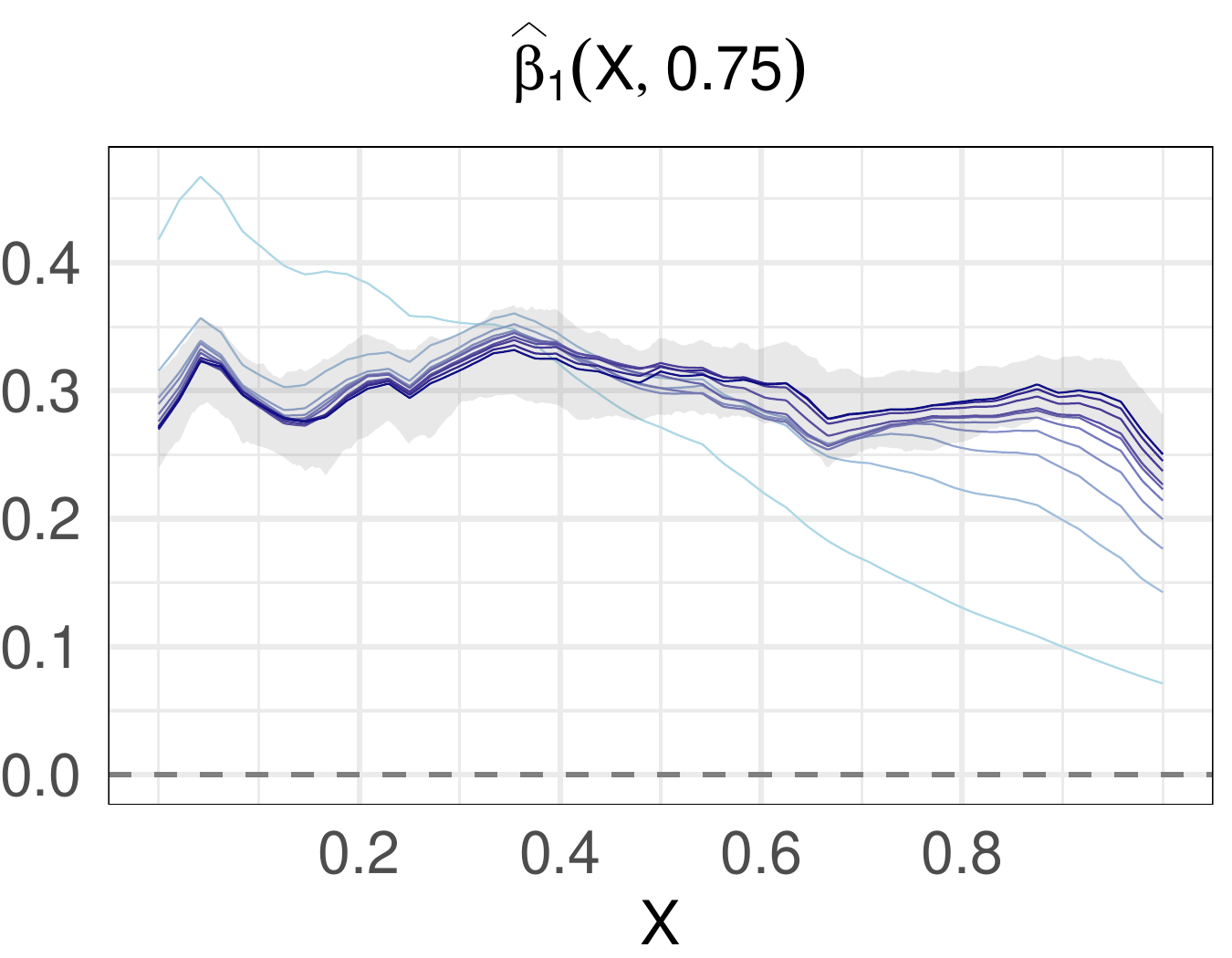}
\end{subfigure}
\begin{subfigure}{0.23\textwidth}
  \includegraphics[width=\textwidth]{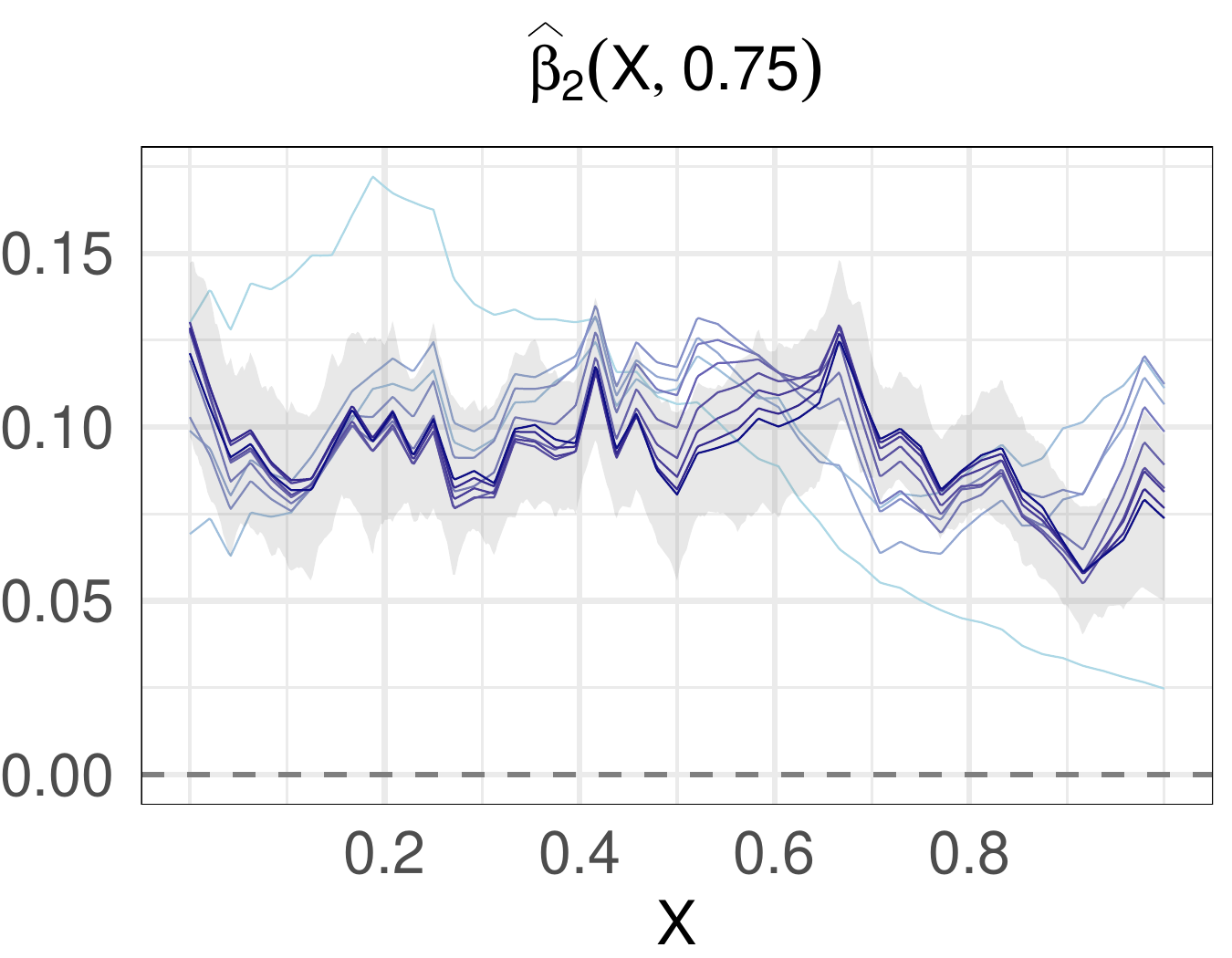}
\end{subfigure}
\begin{subfigure}{0.23\textwidth}
  \includegraphics[width=\textwidth]{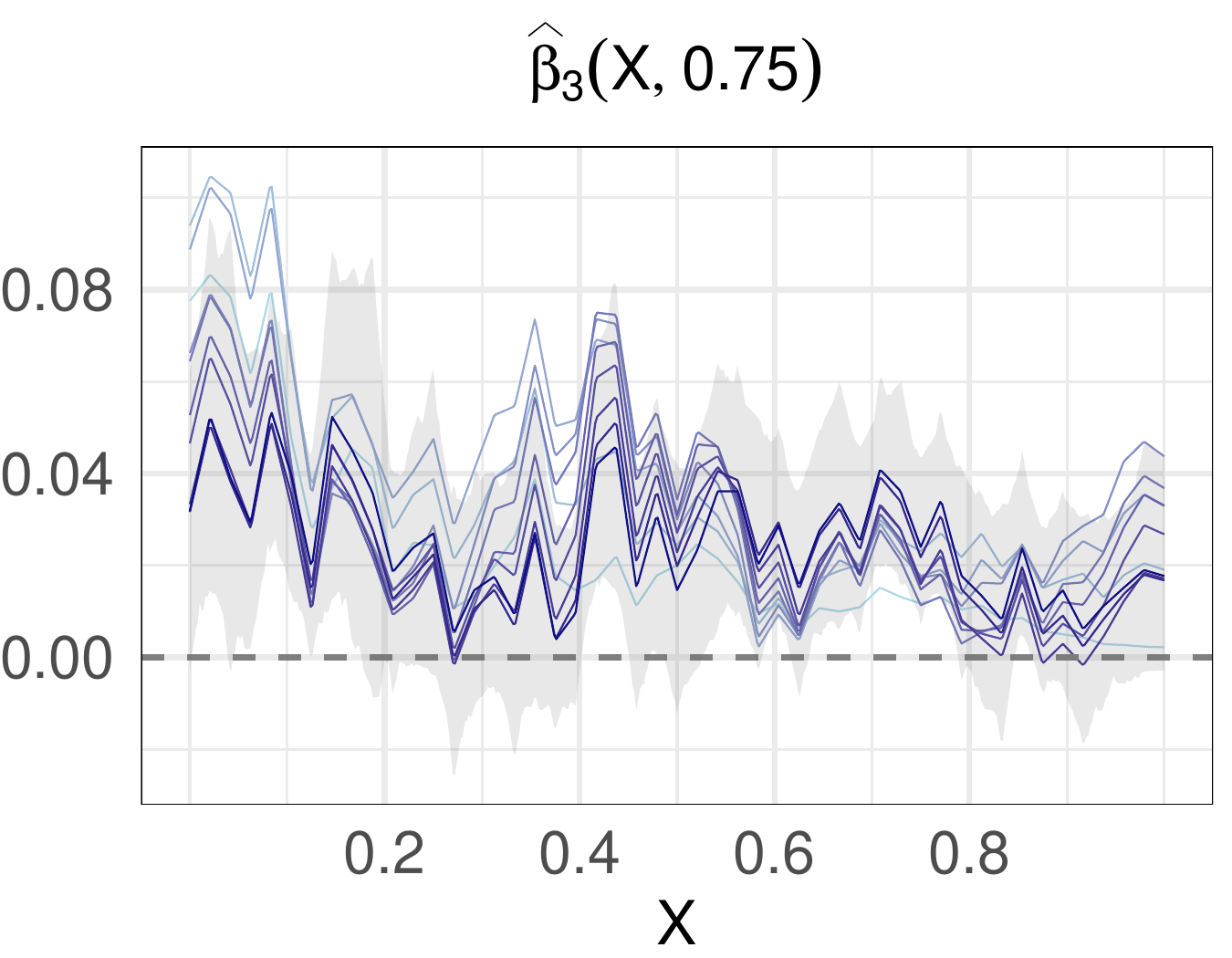}
\end{subfigure}

\begin{subfigure}{0.23\textwidth}
  \includegraphics[width=\textwidth]{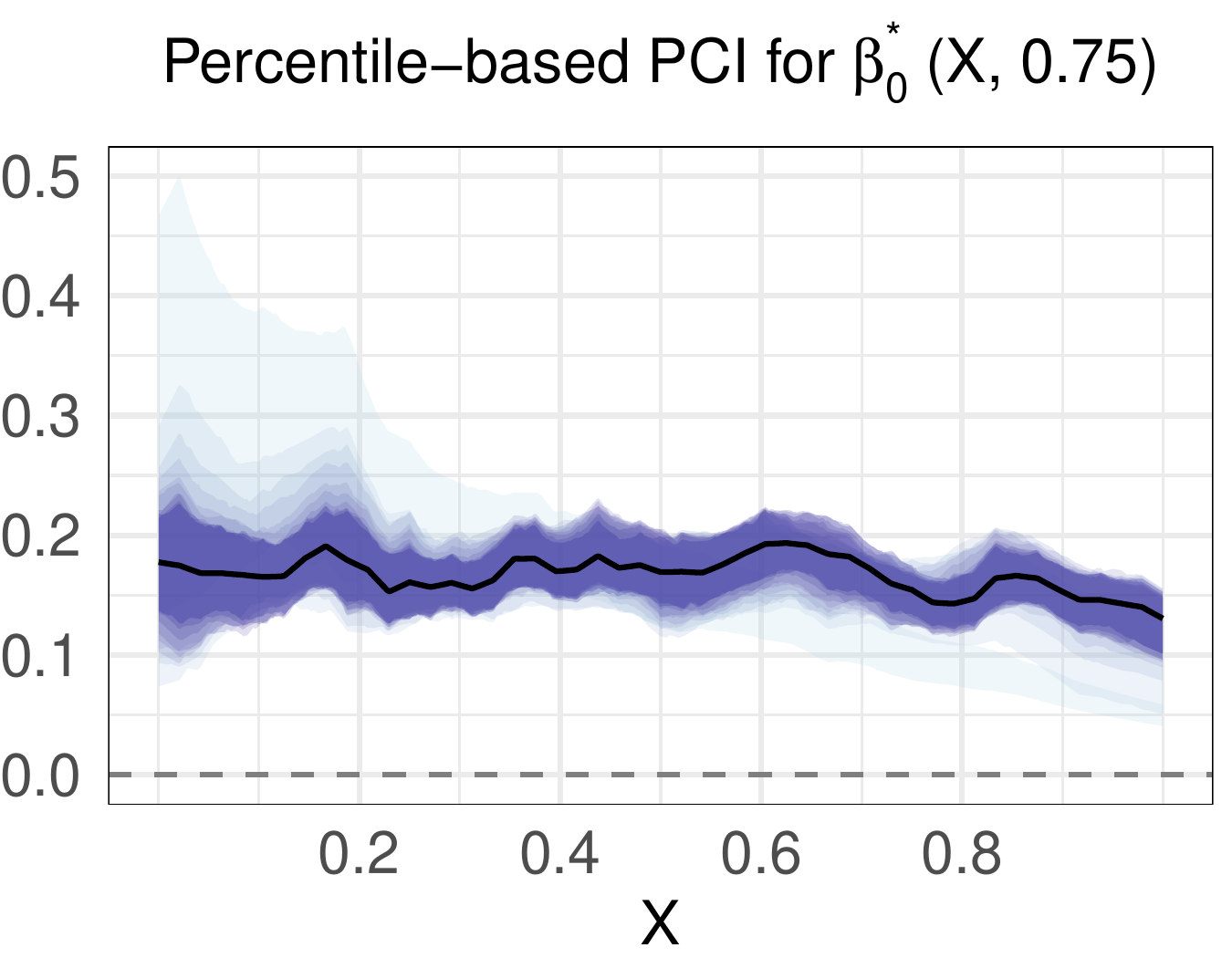}
\end{subfigure}
\begin{subfigure}{0.23\textwidth}
  \includegraphics[width=\textwidth]{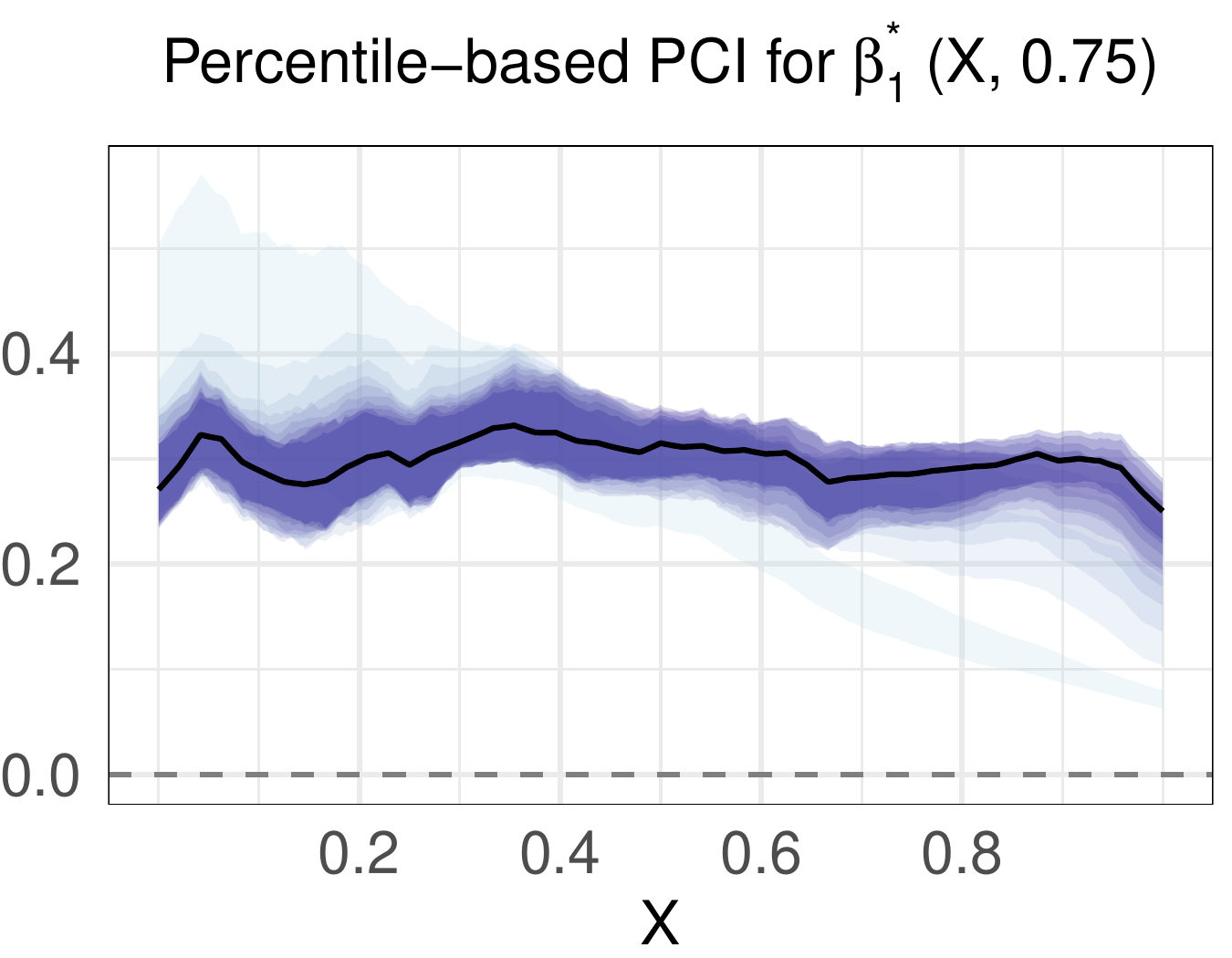}
\end{subfigure}
\begin{subfigure}{0.23\textwidth}
  \includegraphics[width=\textwidth]{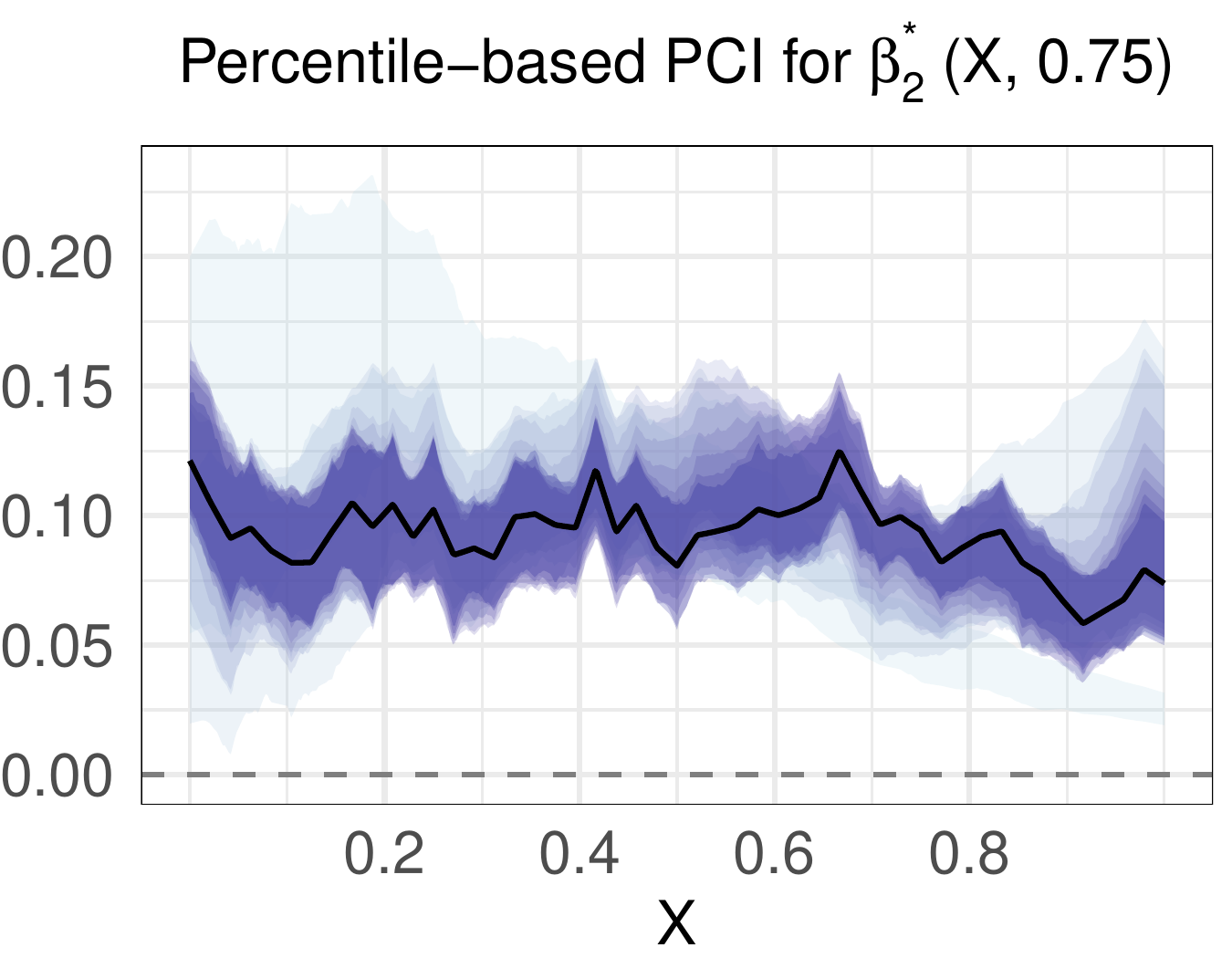}
\end{subfigure}
\begin{subfigure}{0.23\textwidth}
  \includegraphics[width=\textwidth]{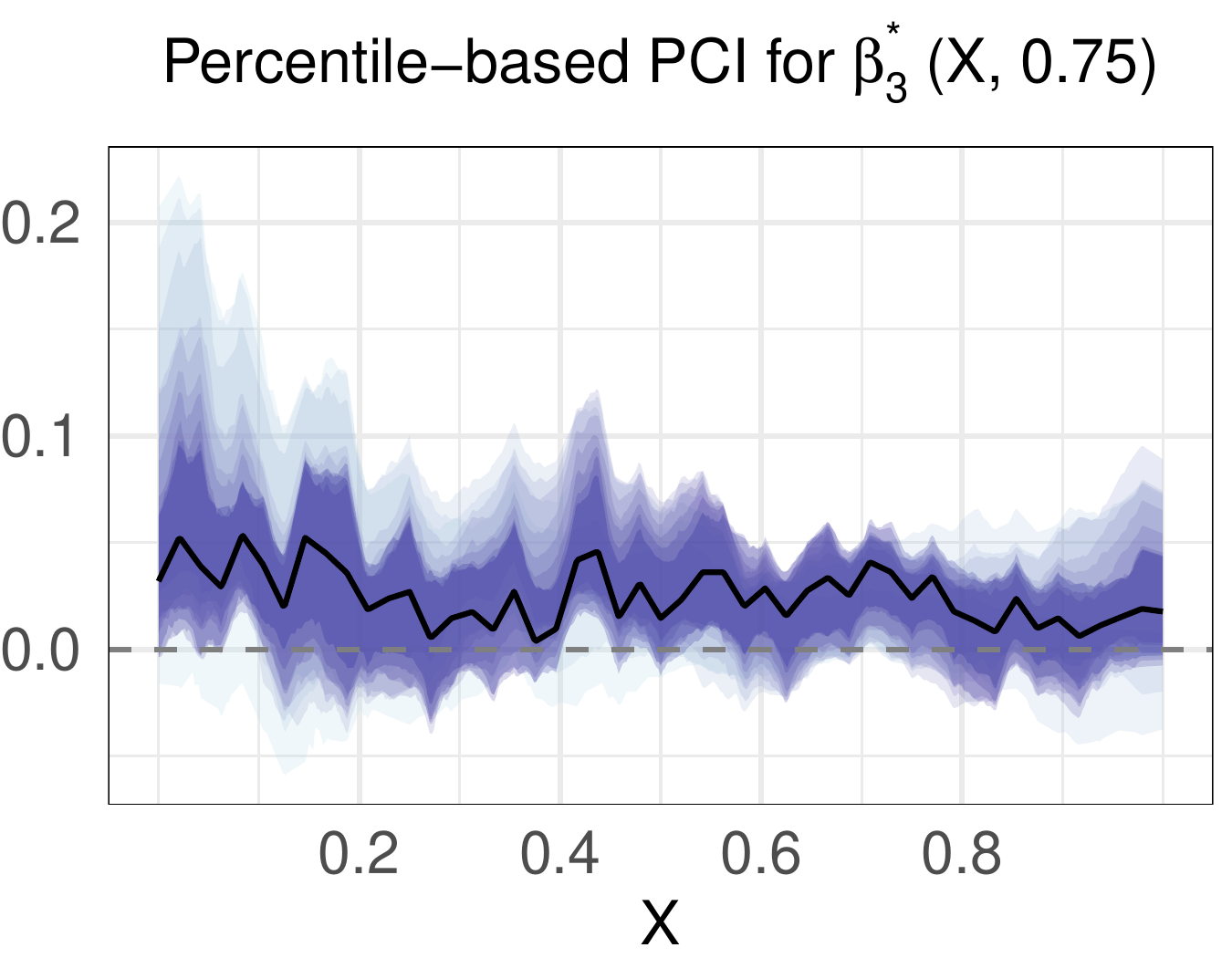}
\end{subfigure}

\begin{subfigure}{0.23\textwidth}
  \includegraphics[width=\textwidth]{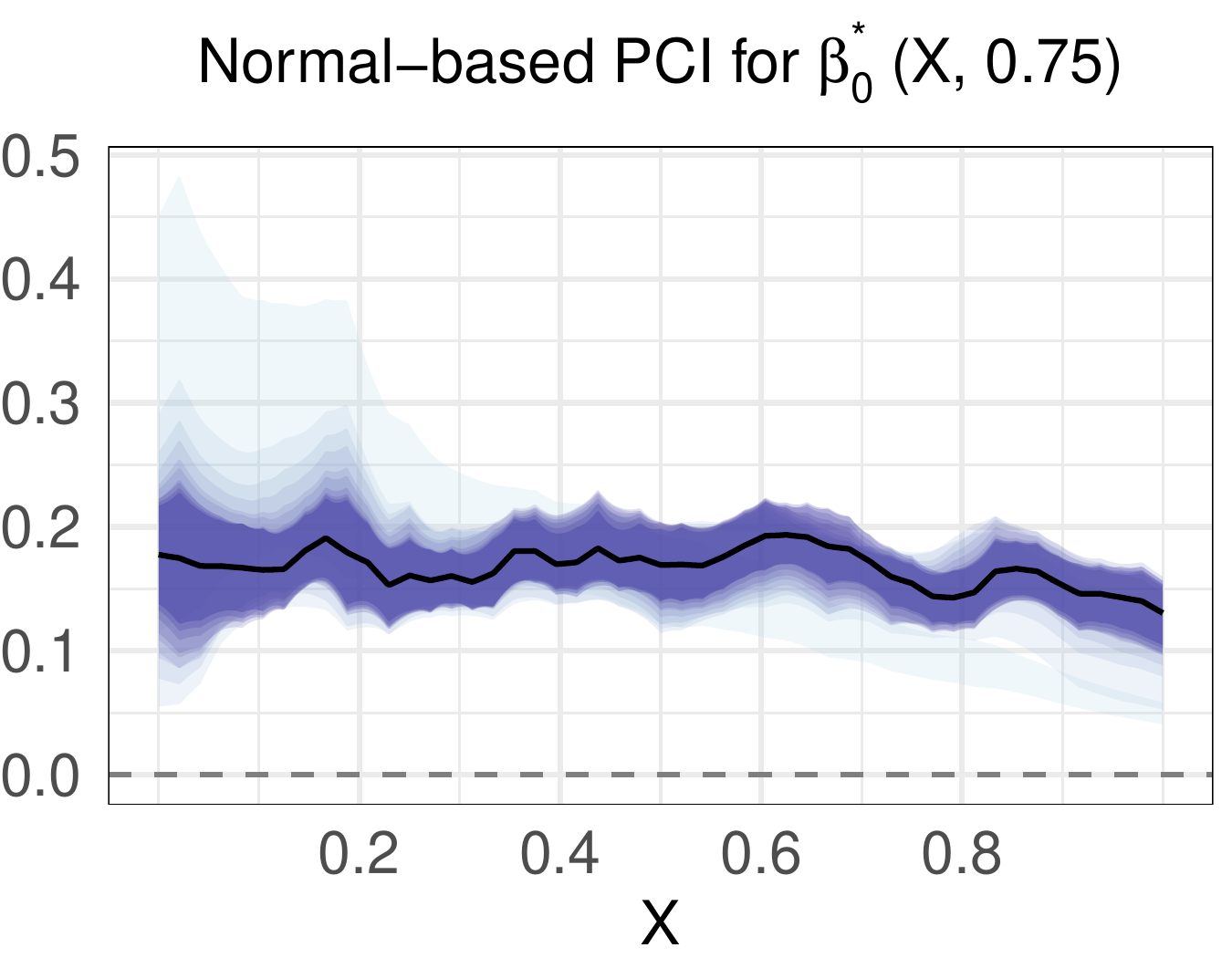}
\end{subfigure}
\begin{subfigure}{0.23\textwidth}
  \includegraphics[width=\textwidth]{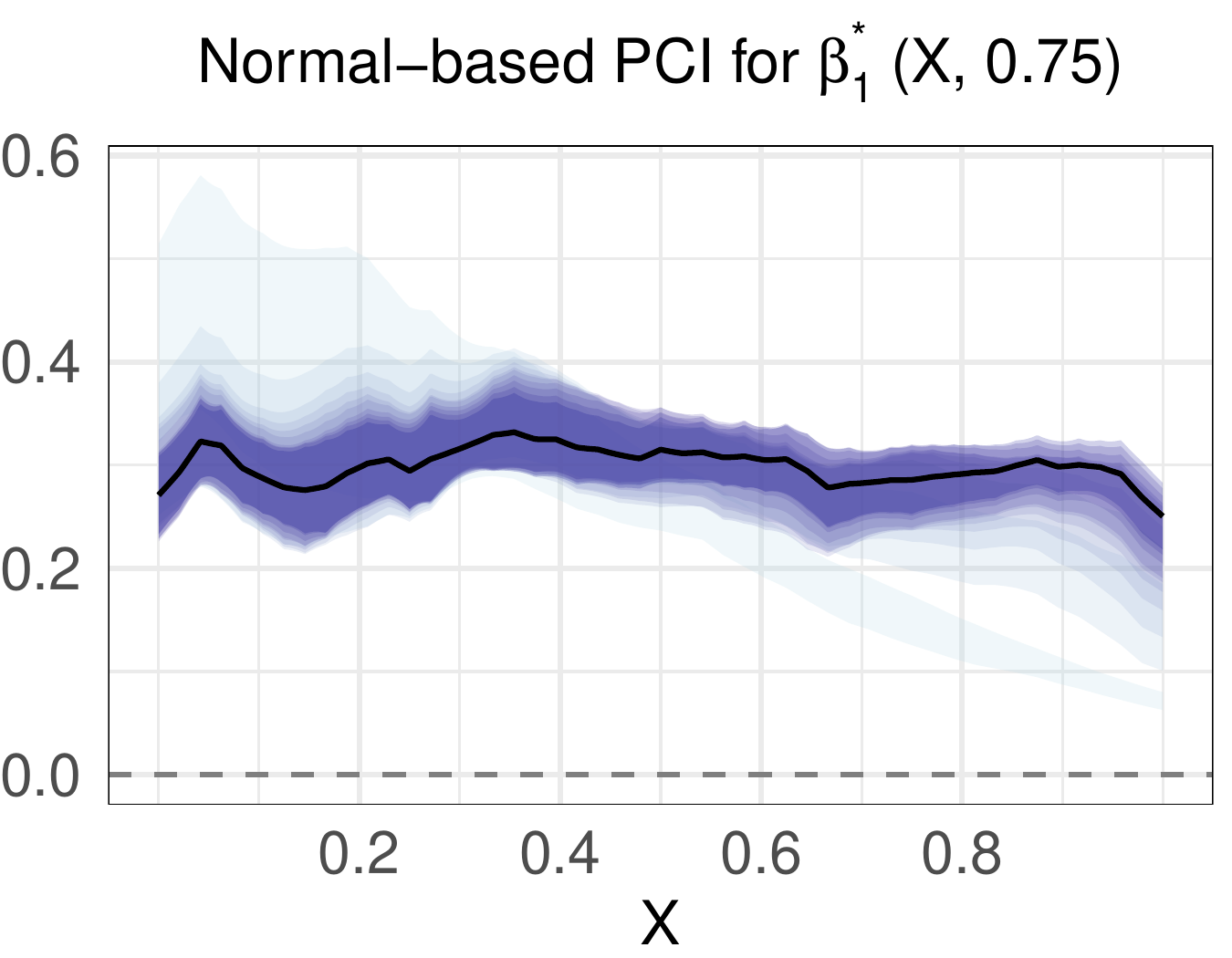}
\end{subfigure}
\begin{subfigure}{0.23\textwidth}
  \includegraphics[width=\textwidth]{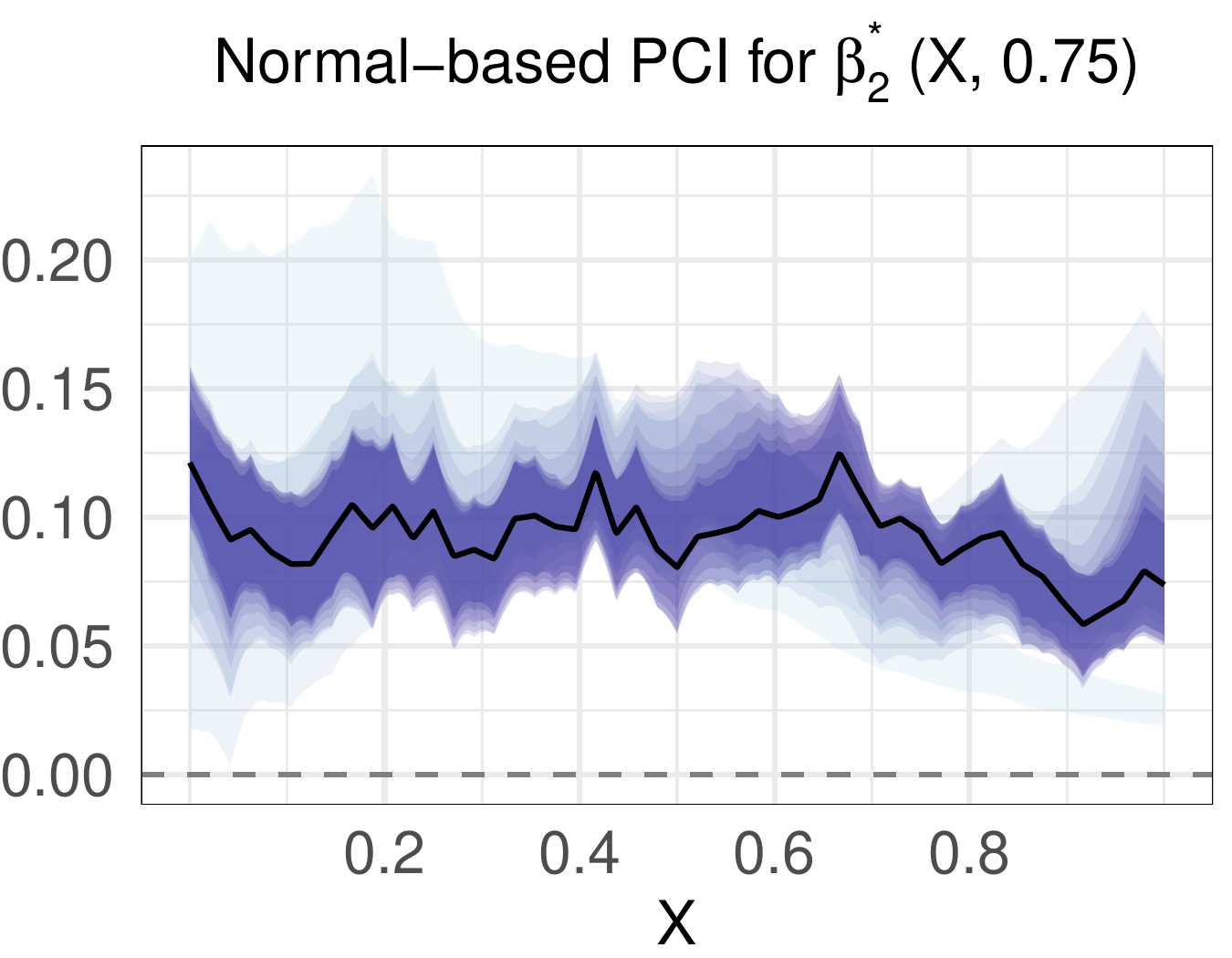}
\end{subfigure}
\begin{subfigure}{0.23\textwidth}
  \includegraphics[width=\textwidth]{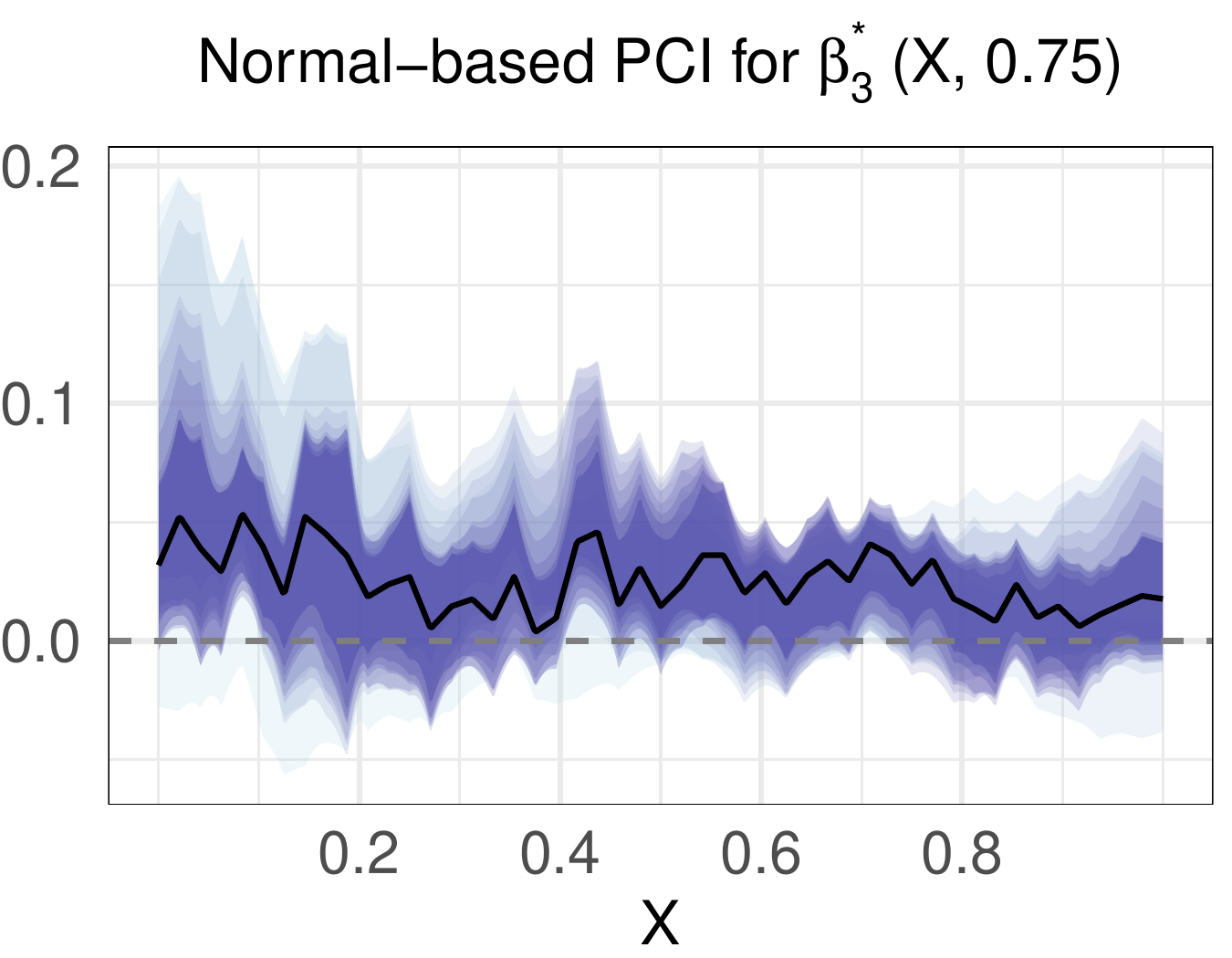}
\end{subfigure}

\begin{subfigure}{\textwidth}
  \centering
  \includegraphics[width=0.8\textwidth]{Rdfigures/FSGD_n_legend.pdf}
\end{subfigure}
\caption{Online estimated curves and PCI trajectories via FSGD-NC at $\tau=0.75$.}
\label{fig:NC_traj_0.75}
\end{figure}

Figures \ref{fig:NC_traj_0.25}--\ref{fig:NC_traj_0.75} show the online evolution of the FSGD-NC estimated curves and their $95\%$ PCIs after each training day. The color represents the cumulative training sample size $n$. In the estimation row, the gray shaded region is the final PCI; in the PCI rows, the black solid curve is the final estimated coefficient curve. FSGD-C trajectories lead to essentially the same conclusions as FSGD-NC. For coefficient estimation, the early curves differ noticeably from their later counterparts. After about seven training days, they stabilize and essentially coincide with their final versions. The percentile-based and normal-based PCIs show a similar pattern: they gradually narrow as the data stream grows and remain close to each other. Thus, the online procedure provides reliable inference at each stage of data stream processing.

We next examine the signal patterns of the final estimated coefficient functions across expectile levels. The intercept function, $\widehat\beta_0(X,\tau)$, is negative at $\tau=0.25$, close to zero at $\tau=0.5$, and positive at $\tau=0.75$; this sign ordering holds throughout the trading day. The market-return coefficient $\widehat\beta_1(X,\tau)$ and the order-flow imbalance coefficient $\widehat\beta_2(X,\tau)$ remain positive throughout the trading day and vary little across expectile levels, indicating broadly consistent effects on the conditional return distribution. In contrast, the scaled log quote volume coefficient $\widehat\beta_3(X,\tau)$ varies substantially across expectile levels: it is negative at $\tau=0.25$, suggesting that higher trading volume lowers the lower tail; close to zero at $\tau=0.5$, indicating a negligible effect on the central part; and positive at $\tau=0.75$, implying that higher trading volume raises the upper tail. Thus, this application demonstrates the practical utility of the proposed online method in uncovering heterogeneous covariate effects across the conditional return distribution.

\section{Discussion}\label{Sec6}

In this paper, we develop an online nonparametric framework for the varying coefficient ER model. By embedding the coefficient functions in a product RKHS and combining FSGD with Polyak--Ruppert averaging, the proposed estimator processes streaming data in one pass. We establish non-asymptotic error bounds, show that the online estimator attains the optimal nonparametric convergence rate under suitable step-size schemes, and derive its Berry--Esseen bound for pointwise asymptotic normality. For uncertainty quantification, we propose an online multiplier bootstrap that recursively maintains perturbed FSGD trajectories and constructs asymptotically valid PCIs. Simulation studies and a cryptocurrency application demonstrate that the online method is statistically competitive with offline procedures while substantially reducing computational costs.

Several limitations point to future research. First, our analysis fixes the expectile level $\tau\in(0,1)$. The error bounds, normal approximation, and bootstrap consistency involve constants that grow polynomially in $1/\min\{\tau,1-\tau\}$, and thus deteriorate rapidly as $\tau\to 0$ or $1$. Extending the framework to online nonparametric inference for extreme expectiles would benefit real-time tail-risk monitoring in finance and energy. Second, we assume i.i.d. streaming observations. This is common in the multiplier bootstrap literature and is essential for the conditional unbiasedness of the stochastic gradient. Under temporal dependence, the stochastic gradient is no longer conditionally unbiased. Independent bootstrap weights also cannot reproduce serial correlation. Recursive nonparametric estimation for time series has been studied under dependence (\cite{huang2013recursive}). Nevertheless, online inference for nonparametric ER with dependent data remains challenging. Third, the proposed inference provides PCIs at a fixed point, whereas many applications require simultaneous confidence bands over the entire domain. The multiplier bootstrap can be adapted to supremum-norm critical values, but valid online simultaneous inference also requires a supremum-norm Gaussian approximation and consistency of the corresponding bootstrap quantile. Such bands have been constructed for online nonparametric least squares by combining a supremum-norm Gaussian approximation with the multiplier bootstrap \cite{liu2026scalable}. Extending these bands to varying coefficient ER under the asymmetric squared loss remains an open problem.

\bibliographystyle{apalike}
\bibliography{ref}

\end{document}